\documentclass[12pt]{article}

\usepackage{bbm,latexsym}
\usepackage{amssymb,amsmath}
\usepackage{graphicx}
\usepackage[font=small,labelfont=bf]{caption}
\usepackage[font=small,labelfont=small]{subcaption}
\usepackage{slashed}
\usepackage{feynmp-auto}
\usepackage[hidelinks,plainpages=false]{hyperref}
\newcommand{\mc}{\mathcal{M}}
\newcommand{\diag}{\mbox{diag}}
\newcommand{\bone}{\mathbbm{1}}

\newcommand{\dd}{\mathrm{d}}
\newcommand{\mt}[1]{m_{\mathrm{tot},#1}}
\newcommand{\ca}{\mathcal{A}}
\newcommand{\um}{\underline{\hat m}}
\newcommand{\cf}{\mathcal{F}}
\newcommand{\defotimes}{%
	\fmfcmd{
	path quadrant, q[], otimes;
	quadrant = (0, 0) -- (0.5, 0) & quartercircle & (0, 0.5) -- (0, 0);
	for i=1 upto 4: q[i] = quadrant rotated (45 + 90*i); endfor
	otimes = q[1] & q[2] & q[3] & q[4] -- cycle;
	}
	\fmfwizard}
\allowdisplaybreaks
\begin{document}

\title{
\normalsize \hfill UWThPh-2018-17 \\[10mm]
\LARGE Renormalization of the multi-Higgs-doublet Standard Model 
and one-loop lepton mass corrections} 

\author{
W.~Grimus\thanks{E-mail: walter.grimus@univie.ac.at}\;
\addtocounter{footnote}{1}
and M.~L\"oschner\thanks{E-mail: maximilian.loeschner@univie.ac.at}
\\[5mm]
\small University of Vienna, Faculty of Physics \\
\small Boltzmanngasse 5, A--1090 Vienna, Austria
}

%%%%
\date{November 5, 2018}

\maketitle

\begin{abstract}
Motivated by models for neutrino masses and lepton mixing, 
we consider the renormalization of the lepton sector of a general 
multi-Higgs-doublet Standard Model 
with an arbitrary number of right-handed neutrino singlets.
We propose to make the theory finite by $\overline{\mbox{MS}}$ renormalization
of the parameters of the unbroken theory. However, 
using a general $R_\xi$ gauge, in the explicit one-loop 
computations of one-point and two-point functions it becomes clear 
that---in addition---a renormalization of the vacuum expectation values 
(VEVs) is necessary. Moreover, in order 
to ensure vanishing one-point functions of the
physical scalar mass eigenfields, finite shifts of the tree-level VEVs, 
induced by the finite parts of the tadpole diagrams, are required. 
As a consequence of our renormalization scheme, physical masses are 
functions of the renormalized parameters and VEVs and thus 
derived quantities. Applying our scheme to one-loop corrections of 
lepton masses, we perform a thorough discussion of finiteness and 
$\xi$-independence. In the latter context, the tadpole
contributions figure prominently. 
\end{abstract}

\newpage

\section{Introduction}
\label{introduction}
In this paper we propose a renormalization scheme for the 
multi-Higgs-doublet Standard Model (mHDSM). We are motivated by  
models for neutrino masses and lepton mixing, which all have an extended 
scalar sector. In the simplest cases, they have several Higgs doublets and 
a number of right-handed neutrino gauge singlet fields and permit in this 
way to incorporate the 
seesaw mechanism~\cite{minkowski,yanagida,glashow,gell-mann,mohapatra}. 

Our framework is the following. We consider the lepton sector\footnote{We 
 do not consider quarks in our discussion though they could be included 
 in a straightforward way.} 
of an 
extended Standard Model (SM), comprising $n_L = 3$ left-handed lepton 
gauge doublets,\footnote{Actually, the value $n_L = 3$ comes solely from 
 the known three families of fermions, but has otherwise no bearing 
 on our discussion.} 
$n_R$ right-handed neutrino gauge singlet fields and $n_H$ 
Higgs doublets. We assume
\begin{equation}
n_R \geq 1 \quad \mbox{and} \quad n_H \geq 1,
\end{equation}
but otherwise these numbers are arbitrary. We furthermore postulate that 
spontaneous gauge-symmetry breaking in the mHDSM happens 
in the same way as in the SM, \textit{i.e.}\ the SM gauge group is 
broken down to $U(1)_\mathrm{em}$. We do not discuss conditions on the scalar 
potential $V(\phi)$ which make this symmetry breaking possible. Though above we 
have mentioned the seesaw mechanism, in the renormalization of the 
mHDSM we do not assume anything about the scale of the right-handed 
neutrino masses; our setting is completely general with respect to fermion 
mass scales, but the seesaw mechanism is included. 
One-loop corrections to the seesaw mechanism have been
computed earlier in~\cite{neufeld,pilaftsis,lavoura} 
(see also~\cite{dev,fink})
and at the end of the present 
paper we will comment on the relationship between our renormalization scheme 
here and the radiative corrections of ref.~\cite{lavoura}.

Before we lay out the renormalization scheme, we have to discuss some of 
our notation.
For the precise definitions of the parameters of the Lagrangian we refer the 
reader to section~\ref{lagrangians}. We choose positive gauge coupling 
constants $g$ and $g'$ of $SU(2)_L$ and $U(1)_Y$, respectively, and thus 
the sine and cosine of the Weinberg angle, $s_w$ and $c_w$, 
respectively, are positive 
as well.\footnote{Note the sign difference to~\cite{grimus,osland,branco} 
 where $-g$ occurs in the covariant derivative.}
In the discussion in the present paper we do not need to renormalize $g$ and 
$g'$. We will always use renormalized parameters of the Lagrangian; these 
include the Yukawa coupling matrices $\Delta_k$ and $\Gamma_k$ 
($k=1,\ldots,n_H$), the parameters of the scalar potential $V(\phi)$, 
$\mu^2_{ij}$ and $\lambda_{ijkl}$, and the Majorana mass matrix $M_R$ 
of the right-handed neutrino singlets.
The corresponding counterterm 
parameters are denoted by $\delta\Delta_k$, $\delta\Gamma_k$,
$\delta\mu^2_{ij}$, $\delta\lambda_{ijkl}$ and $\delta M_R$.
The vacuum expectation values $v_k$ (VEVs) are in the notation of 
our paper pure tree-level quantities, in principle expressible in 
terms of $\mu^2_{ij}$ and $\lambda_{ijkl}$ by finding the minimum of the 
(tree level) scalar potential $V(\phi)$, written in terms of the renormalized 
parameters.

%%%%
We will work in the $R_\xi$ or 't~Hooft gauge~\cite{thooft,fujikawa,yao} 
with general parameters $\xi$ 
in almost all our computations\footnote{When we write $\xi$ 
 we mean $\xi_W$, $\xi_Z$ and 
 $\xi_A$, which occur in the propagators of the $W^\pm$, $Z$ and photon, 
 respectively.} and use dimensional regularization in the one-loop 
computations. We will not go beyond the one-loop level.

Physically, only the spontaneously broken mHDSM makes sense, because otherwise 
all fermions and vector bosons would be massless.
The renormalization scheme for the \emph{broken} theory 
we propose consists of $\overline{\mbox{MS}}$
renormalization of the parameters of the \emph{unbroken} theory
plus a VEV renormalization with renormalization parameters $\delta v_k$. 
That the latter is necessary in a gauge theory quantized in an 
$R_\xi$ gauge with $\xi \neq 0$ has been proven in~\cite{sperling}.
Note that this $\delta v_k$ is a
renormalization in addition to the scalar wave-function renormalization 
already included in the VEVs $v_k$.
Complying with our proposed renormalization scheme, we use  
dimensional regularization in $d = 4 - \varepsilon$ dimensions. 
Therefore, at one-loop order the
renormalization parameters $\delta\Delta_k$, $\delta\Gamma_k$,
$\delta\mu^2_{ij}$, $\delta\lambda_{ijkl}$, $\delta M_R$ and $\delta v_k$
are all proportional to 
\begin{equation}\label{cinfty}
%%%%
c_\infty = \frac{2}{\varepsilon} - \gamma_E + \ln (4\pi).
\end{equation}
(Note that in the present paper we use the symbol ``$\delta$'' solely for the 
purpose of indicating quantities proportional to $c_\infty$.)
We will show that the proposed scheme, with counterterms induced by
renormalization parameters listed here, 
allows to remove all divergences at the one-loop level 
and that the divergences uniquely determine the counterterm parameters.

Previously, avoiding the intricacies of gauge theories, this very fact 
has been demonstrated in~\cite{fox} for a general Yukawa model with an 
arbitrary number of real scalars. (For an early attempt with only one 
scalar see~\cite{ludl}.)

In detail, we proceed as follows:
\begin{enumerate}
\item
We determine
$\delta\lambda_{ijkl} + \delta\lambda_{ilkj}$
from the divergence of the neutral-scalar four-point 
function  of the unbroken theory.\footnote{This combination is 
sufficient for our purposes---see section~\ref{cts for one and two}.}
\item
Plugging $\delta\lambda_{ijkl} + \delta\lambda_{ilkj}$ into the 
counterterm of the scalar two-point function of the broken phase, 
the remaining divergencies uniquely 
fix $\delta\mu^2_{ij}$ and $\delta v_k$.
\item
With the so far obtained renormalization parameters we compute the 
counterterm for the scalar one-point function and, 
as a check, we prove its finiteness.
\item
We determine $\delta\Delta_k$ and $\delta\Gamma_k$ from the divergencies 
of vertex corrections of the neutral-scalar couplings to neutrinos and 
charged leptons, respectively. For simplicity, this is also done in the 
unbroken theory.
\item
Having obtained $\delta\Delta_k$, $\delta\Gamma_k$, $\delta v_k$ and 
the counterterm of the scalar one-point function, all 
ingredients required for the counterterms of the fermion self-energies 
are at hand and can be determined. We demonstrate that these make indeed 
the neutrino self-energy $\Sigma_\nu$ and the charged-lepton 
self-energy $\Sigma_\ell$ finite.\footnote{This constitutes another
independent cross check of our renormalization scheme.}
\item
Finally, having in mind formulas for the extraction of corrections to 
the tree-level pole masses from $\Sigma_\nu$ and $\Sigma_\ell$---see for
instance~\cite{aoki,loeschner}, we discuss radiative corrections 
to the tree-level physical neutrino and charged-lepton masses.
In particular, we carefully examine the $\xi$-independence of 
these physical quantities.
\end{enumerate}
We emphasize that in our renormalization scheme there is no mass 
renormalization because both scalar and fermion masses are derived 
quantities and the mass counterterms are, therefore, derived quantities 
as well. This is a consequence of renormalizing the parameters of the 
\emph{unbroken} theory, which is owing to the fact that, for an arbitrary
number of Higgs doublets, the number of Yukawa coupling constants is 
in general much larger than the number of fermion masses.

Since we are discussing the lepton sector of the mHDSM, we have both 
Dirac and Majorana fermions in the theory. When we deem it helpful for the
reader, we stress the differences 
in the treatment of both types of fermions and 
dwell on the field-theoretical specifics for Majorana neutrinos.

The $2n_H$ neutral scalar mass eigenfields 
have by
definition vanishing VEVs. These have to be re-adjusted, order by order, 
by finite VEV shifts $\Delta v_k$ such that the scalar one-point functions
vanish~\cite{fleischer,jegerlehner}.\footnote{We emphasize once more 
 that, in our notation, $\delta v_k$ is infinite while $\Delta v_k$ is finite.
 In the rest of the paper we always reserve the symbol ``$\Delta$'' for 
 finite quantities.}
In an $n$-point function with $n \geq 2$ 
one can either take into account these VEV shifts or, equivalently, 
include all tadpole diagrams instead, as shown for the SM
in~\cite{fleischer,jegerlehner}. We show this explicitly in the mHDSM at the
one-loop level for the neutrino and charged-lepton self-energies.
Moreover, tadpole diagrams play an important role with respect to
$\xi$-independence of physical observables~\cite{weinberg}. 
We present a thorough discussion of this role in the context of radiative
fermion mass corrections.
Other methods for the treatment of tadpole contributions 
are carried out in~\cite{Denner:2016etu, Krause:2016oke, Krause:2017mal} for
variants of the two-Higgs doublet model.

Concerning the notation, we have already explained that the parameters
of the Lagrangian are always considered as being renormalized quantities.
Furthermore, $-i\Sigma_\nu(p)$, $-i\Sigma_\ell(p)$ and $-i\Pi(p^2)$ 
denote the two-point functions of neutrinos, charged fermions and neutral
scalars, respectively, 
as obtained in perturbation theory, including all counterterms.
Thus the corresponding quantities with $(-i)$ removed denote the 
\emph{renormalized} self-energies. 

In order to enhance legibility of the paper we list here the definition and 
notation of all masses which occur in the paper:
\begin{itemize}
\item
The tree-level neutrino masses are $m_i$ ($i=1,2,\ldots,n_L+n_R$).
\item
The tree-level charged-lepton masses are $m_\alpha$ ($\alpha = e,\mu,\tau$).
\item
The \emph{finite} radiative corrections to tree-level fermion 
masses are denoted by $\Delta m_i$ and $\Delta m_\alpha$.
\item
Then the total fermion masses are given by $\mt{i} = m_i + \Delta m_i$ 
for neutrinos and $\mt{\alpha} = m_\alpha + \Delta m_\alpha$ for charged leptons.
\item
The scalar masses are denoted by $M_{+a}$ ($a=1,\ldots,n_H$) and 
$M_b$ ($b=1,\ldots,2n_H$) for charged and neutral scalars, respectively.
\item
The vector boson masses are denoted by $m_W$ and $m_Z$ for $W^\pm$ and $Z$ 
boson, respectively.
\item
Charged and neutral Goldstone bosons correspond to the indices $a=1$ and $b=$1,
respectively, with eigenvalues $M_{+1}^2 = M_1^2 = 0$ of the respective 
mass matrices.
\item
However, due to the $R_\xi$ gauge the Goldstone bosons have squared masses 
$\xi_W m_W^2$ and $\xi_Z m_Z^2$ in the respective propagators. 
\end{itemize}
\emph{All} boson masses are tree-level masses. If in an expression 
several summations 
occur referring to charged-scalar mass eigenfields or masses, then the indices 
$a,a'$ or $a_1,a_2,\ldots$ are used. In the case of neutral scalars, 
$b,b'$ or $b_1,b_2,\ldots$ is utilized.

The paper is organized as follows. 
In section~\ref{lagrangians} we write down all interaction Lagrangians of the
mHDSM needed for the computation of one- and two-point scalar functions and the
self-energies of the charged leptons and Majorana neutrinos. 
This section also includes important relations
concerning the diagonalization of the $(n_L + n_R) \times (n_L + n_R)$ 
neutrino mass matrix. In section~\ref{scalar sector} we discuss the
counterterms of the scalar one- and two-point functions and determine all
counterterm parameters, including $\delta v_k$ of the scalar
sector. Section~\ref{gauge-invariance} is devoted to a thorough examination of
the $\xi$-independence of the one-loop fermion masses. In
section~\ref{finiteness} we 
prove 
the finiteness of the fermion self-energies in our renormalization scheme. 
In section~\ref{one-loop} we present formulas for these 
self-energies in Feynman gauge, 
by listing the individual contributions originating from 
charged-scalar, neutral-scalar, $W$ and $Z$ exchange,
%%%%
and discuss the special case of the seesaw mechanism.
Finally, our conclusions are found in section~\ref{conclusions}. 
In appendix~\ref{app-MM} we show how the charged and
neutral-scalar mass matrices are obtained from the scalar potential, while in
appendix~\ref{app-UV} we discuss properties of the diagonalization matrices of
the scalar mass terms. A short consideration of on-shell contributions to 
fermion self-energies is found in appendix~\ref{app-on-shell}.
Lastly, in appendix~\ref{conversion} we convert the loop functions
that we use in section~\ref{one-loop} to other functions commonly used 
in the literature.

\section{Lagrangians}
\label{lagrangians}
The formalism for the mHDSM has been developed in~\cite{grimus,osland,bento} 
(see also~\cite{lavoura}).

\subsection{Yukawa Lagrangian and lepton mass matrices}
\label{scalar interactions}
In this subsection we follow the notation of~\cite{grimus} and 
repeat some material from this paper. 
As mentioned in the introduction, we assume that the electric charge 
remains conserved after spontaneous symmetry breaking. Therefore, we can 
parameterize the Higgs doublets and their VEVs as
\begin{equation}
\phi_k = \left( \begin{array}{c}
\varphi_k^+ \\ \varphi_k^0 
\end{array} \right), \quad
\tilde\phi_k = \left( \begin{array}{c}
{\varphi_k^0}^\ast \\ -\varphi_k^- 
\end{array} \right), \quad
\langle \phi_k \rangle_0 = \frac{v_k}{\sqrt{2}} 
\left( \begin{array}{c}  0 \\ 1 
\end{array} \right)
\end{equation}
with
\begin{equation}\label{v}
v = \sqrt{\sum_k |v_k|^2} \simeq 246\,\mbox{GeV}.
\end{equation}
The Yukawa Lagrangian may be written as 
\begin{equation}
{\cal L}_\mathrm{Y} = 
- \sum_{k=1}^{n_H}
\left[ \left( \begin{array}{cc} \varphi_k^-, & {\varphi_k^0}^\ast
\end{array} \right) \bar e_R\, \Gamma_k
+ \left( \begin{array}{cc} \varphi_k^0, & - \varphi_k^+
\end{array} \right) \bar \nu_R\,
\Delta_k \right]
\left( \begin{array}{c} \nu_L \\ e_L
\end{array} \right) + \mathrm{H.c.},
\label{Yukawa}
\end{equation}
where the Yukawa coupling matrices $\Gamma_k$ and $\Delta_k$ 
are $n_L \times n_L$ and $n_R \times n_L$, respectively.
The lepton mass terms are given by
\begin{equation}\label{Lmass}
\mathcal{L}_\mathrm{mass} = -\bar e_R M_\ell e_L - \bar\nu_R M_D \nu_L +
\frac{1}{2} \nu_R^T C^{-1} M_R^* \nu_R + \mbox{H.c.}
\end{equation}
with\footnote{Here and in the following we use the summation convention.}
\begin{equation}
M_\ell = \frac{1}{\sqrt{2}}\, v_k^\ast \Gamma_k
\quad \mathrm{and} \quad
M_D = \frac{1}{\sqrt{2}}\, v_k \Delta_k.
\end{equation}
The $n_R \times n_R$ matrix $M_R$ is in general complex and symmetric.
With the chiral projectors
\begin{equation}
\gamma_{L} = \frac{1}{2}\left( \bone - \gamma_5 \right)
\quad \mbox{and} \quad
\gamma_{R} = \frac{1}{2}\left( \bone + \gamma_5 \right),
\end{equation}
the fermion mass eigenfields $\ell_\alpha$ ($\alpha = e,\mu,\tau$) and 
$\chi_i$ ($i=1,\ldots,n_L + n_R$) are obtained from the 
weak chiral eigenfields $e_{L,R}$ and $\nu_{L,R}$ by the transformations
\begin{equation}\label{mass-eigen}
e_L = W_L \gamma_L \ell, \quad
e_R = W_R \gamma_R \ell, \quad
\nu_L = U_L \gamma_L \chi, \quad
\nu_R = U_R \gamma_R \chi.
\end{equation}
The matrices $W_L$ and $W_R$ are unitary $n_L \times n_L$ matrices 
such that\footnote{We deviate slightly in notation from that 
of~\cite{grimus} where a basis has been assumed with $W_L = W_R = \bone$. 
In the present paper, for the sake of clarity, we stick to general 
unitary matrices $W_L$ and $W_R$.} 
\begin{equation}
W_R^\dagger M_\ell W_L \equiv \hat m_\ell = 
\diag \left( m_e, m_\mu, m_\tau \right).
\end{equation}
The matrices $U_L$ and $U_R$ are $n_L \times (n_L + n_R)$ and 
$n_R \times (n_L + n_R)$, respectively, such that
the matrix 
\begin{equation}
{\cal U} \equiv \left( \begin{array}{c} U_L \\ U_R^\ast \end{array} \right)
\end{equation}
is $(n_L + n_R) \times (n_L + n_R)$ unitary. The unitarity of ${\cal U}$ is
expressed as
\begin{subequations}
\begin{eqnarray}
U_L U_L^\dagger &=& \bone_{n_L}\, , \label{uni1}
\\
U_R U_R^\dagger &=& \bone_{n_R}\, , \label{uni2}
\\
U_L U_R^T &=& 0_{n_L \times n_R}\, , \label{uni3}
\end{eqnarray}
\end{subequations}
and
\begin{equation}
U_L^\dagger U_L + U_R^T U_R^\ast = \bone_{n_L + n_R}.
\end{equation}
${\cal U}$ diagonalizes the $(n_L + n_R) \times (n_L + n_R)$ Majorana neutrino 
mass matrix, \textit{i.e.}\ 
\begin{equation}
{\cal U}^T \left( \begin{array}{cc} 0 & M_D^T \\
M_D & M_R \end{array} \right) {\cal U} \equiv \hat m_\nu
= \mathrm{diag} \left( m_1, m_2, \ldots, m_{n_L + n_R} \right),
\label{rgets}
\end{equation}
with real and non-negative $m_i$~\cite{schur}.
Therefore,
\begin{subequations}
\begin{eqnarray}
U_L^\ast \hat m_\nu U_L^\dagger &=& 0_{n_L \times n_L}\, , \label{d1} \\
U_R \hat m_\nu U_L^\dagger &=& M_D\, , \label{d2} \\
U_R \hat m_\nu U_R^T &=& M_R\, . \label{d3}
\end{eqnarray}
\end{subequations}
A further relation is given by~\cite{lavoura}
\begin{equation}\label{URLM}
U_R^\dagger M_D = \hat m_\nu U_L^\dagger.
\end{equation}

Now we turn to the scalar mass eigenfields $S^+_a$ ($a=1,\ldots,n_H$)
and $S^0_b$ ($b=1,\ldots,2n_H$), related to $\varphi_k^+$ and 
$\varphi^0_k$ by 
\begin{equation}
\varphi^+_k = U_{ka} S^+_a 
\quad \mbox{and} \quad
\varphi^0_k = \frac{1}{\sqrt{2}} \left( v_k + V_{kb} S^0_b \right),
\end{equation}
respectively.
For the definition and properties of the matrices $U$
and $V$ we refer the reader to appendix~\ref{app-UV}.

Now we are in a position to formulate the Yukawa interactions in 
terms of mass eigenfields. Since we perform computations 
with Majorana neutrinos, \textit{i.e.}\ $\chi^c = \chi$, 
it is useful to have at hand both the 
interaction Lagrangians 
of the charged-lepton fields $\ell$ and of the charge-conjugated 
fields $\ell^c$~\cite{denner}. 
The neutral-scalar Yukawa interaction Lagrangian may be written as 
\begin{equation}\label{LS0}
\mathcal{L}_\mathrm{Y}(S^0) =
-\frac{1}{\sqrt{2}} \, S^0_b \left\{ 
\bar\chi \left[ F_b \gamma_L + F_b^\dagger \gamma_R \right] \chi 
+ \bar\ell \left[ G_b \gamma_L + G_b^\dagger \gamma_R \right] \ell \right\}.
\end{equation}
Note that
\begin{equation}\label{LS0c}
\bar\ell \left[ G_b \gamma_L + G_b^\dagger \gamma_R \right] \ell = 
\overline{\ell^c} \left[ G_b^T \gamma_L + G_b^* \gamma_R \right] \ell^c.
\end{equation}
The charged-scalar Yukawa interaction Lagrangian can be formulated as
\begin{subequations}\label{LS+}
\begin{eqnarray}
\mathcal{L}_\mathrm{Y}(S^\pm) &=&
S_a^- \bar \ell \left[ R_a \gamma_R
- L_a \gamma_L \right] \chi
+ S_a^+ \bar \chi \left[ R_a^\dagger \gamma_L
- L_a^\dagger \gamma_R \right] \ell
\label{LS+a} \\
&=&
S_a^- \bar \chi \left[ R_a^T \gamma_R
- L_a^T \gamma_L \right] \ell^c
+ S_a^+ \overline{\ell^c} \left[ R_a^* \gamma_L
- L_a^* \gamma_R \right] \chi.
\label{LS+b}
\end{eqnarray}
\end{subequations}
Then for these Lagrangians the coupling matrices are 
given by~\cite{grimus}
\begin{subequations}
\begin{eqnarray}
F_b &=& 
\frac{1}{2}
\left( U_R^\dagger \Delta_k U_L + U_L^T \Delta_k^T U_R^\ast \right) V_{kb},
\label{Fb}
\\
G_b &=& \left( W_R^\dagger \Gamma_k W_L \right) V_{kb}^*,
\label{Gb}
\\
R_a &=& \left( W_L^\dagger \Delta_k^\dagger U_R \right) U_{ka}^*,
\label{Ra}
\\
L_a &=& \left( W_R^\dagger \Gamma_k U_L \right) U_{ka}^*.
\label{La}
\end{eqnarray}
\end{subequations}

Since we identify the scalars carrying index~1 with the Goldstone bosons, 
we have $S^0_1 \equiv G^0$ and $S^+_1 \equiv G^+$.
Using the matrix elements $V_{k1}$ and $U_{k1}$, required for the Goldstone
boson couplings, of equation~(\ref{GUV}) in appendix~\ref{app-UV}, we obtain 
\begin{equation}
\Delta_k V_{k1}   = i \Delta_k U_{k1} = i \frac{\sqrt{2}}{v} M_D, 
\quad
\Gamma_k V_{k1}^* = -i \Gamma_k U_{k1}^* = -i \frac{\sqrt{2}}{v} M_\ell, 
\end{equation}
with $v$ being defined in equation~(\ref{v}). Then, 
exploiting the formulas for the diagonalization of the fermion mass matrices,
the coupling matrices of $G^0$ and $G^\pm$ can be converted into 
\begin{equation}\label{F1G1}
F_1 = \frac{i}{\sqrt{2}v} \left(  
\hat m_\nu U_L^\dagger U_L + U_L^T U_L^* \hat m_\nu \right),
\quad
G_1 = -i \frac{\sqrt{2}}{v} \hat m_\ell
\end{equation}
and~\cite{grimus}
\begin{equation}\label{R1L1}
R_1 = \frac{\sqrt{2}}{v} W_L^\dagger U_L \hat m_\nu,
\quad
L_1 = \frac{\sqrt{2}}{v} \hat m_\ell W_L^\dagger U_L,
\end{equation}
respectively.

\subsection{Charged and neutral current interactions}
\label{current interactions}
The squares of vector boson masses are 
\begin{equation}\label{mV}
m_W^2 = \frac{g^2 v^2}{4}, \quad m_Z^2 = \frac{g^2 v^2}{4 c_w^2},
\end{equation}
with $c_w = m_W/m_Z$ being the cosine of the weak mixing or Weinberg angle.

In terms of the lepton mass eigenfields, we obtain the 
charged-current Lagrangian 
\begin{subequations}\label{Lcc}
\begin{eqnarray}
{\cal L}_\mathrm{cc} &=& -\frac{g}{\sqrt{2}}\, \left[
W_\mu^- \bar\ell \left( W_L^\dagger U_L \right)
\gamma^\mu \gamma_L \chi 
+
W_\mu^+ \bar \chi \left( U_L^\dagger W_L \right)
\gamma^\mu \gamma_L \ell
\right] 
\label{Lcca} \\
&=&
+\frac{g}{\sqrt{2}}\, \left[
W_\mu^- \bar\chi \left( W_L^\dagger U_L \right)^T
\gamma^\mu \gamma_R \ell^c 
+
W_\mu^+ \overline{\ell^c} \left( U_L^\dagger W_L \right)^T
\gamma^\mu \gamma_R \chi \right],
\label{Lccb}
\end{eqnarray}
\end{subequations}
and the neutral-current Lagrangians~\cite{lavoura,grimus}
\begin{equation}\label{Lnc}
{\cal L}_\mathrm{nc} =  -\frac{g}{4 c_w}\, Z_\mu 
\bar \chi \gamma^\mu F_{LR} \chi
-\frac{g}{c_w}\, Z_\mu \bar\ell \gamma^\mu \left[ 
\left( s_w^2 - \frac{1}{2} \right) \gamma_L + s_w^2 \gamma_R \right] \ell
\end{equation}
with
\begin{equation}\label{FLR}
F_{LR} = 
\left( U_L^\dagger U_L \right) \gamma_L - \left( U_L^T U_L^* \right) \gamma_R.
\end{equation}
Finally, the electromagnetic interaction Lagrangian 
of the charged leptons with charge $-e$ is 
\begin{equation}
{\cal L}_\mathrm{em} =  e\, A_\mu \bar \ell \gamma^\mu \ell.
\end{equation}
Concerning the vector boson propagators in the $R_\xi$ gauge, they have the
form 
\begin{subequations}\label{vbp}
\begin{eqnarray}
\Delta_V^{\mu\nu}(k) &=& 
-\frac{g^{\mu\nu}}{k^2 - m_V^2 + i\epsilon} + 
\frac{k^\mu k^\nu}{m_V^2} \left( 
\frac{1}{k^2 - m_V^2 + i\epsilon} - 
\frac{1}{k^2 - \xi_V m_V^2 + i\epsilon} \right) 
\label{vbp1}
\\
&=&
-\frac{g^{\mu\nu}}{k^2 - m_V^2 + i\epsilon} + \left( 1-\xi_V \right)
\frac{k^\mu k^\nu}{(k^2 - m_V^2 + i\epsilon) (k^2 - \xi_V m_V^2 + i\epsilon)}
\label{vbp2}
\end{eqnarray}
\end{subequations}
with $V = Z, W, A$.
For photons only the second form of the propagator is meaningful.

\subsection{Vector boson--scalar interactions}
Here we only display those interaction Lagrangians which we need in the 
present paper. For the complete set of vector boson--scalar interaction 
Lagrangians see~\cite{osland,bento}.
Derivative couplings of the vector bosons to scalars are given by
\begin{subequations}\label{vb-s-der}
\begin{eqnarray}
\mathcal{L}_\partial &=& \hphantom{+}
\frac{ig}{2} V_{kb} U_{ka}^* W^+_\mu 
\left( S^0_b \partial^\mu S^-_a - S^-_a \partial^\mu S^0_b \right) 
\\ && + 
\frac{ig}{2} V_{kb}^* U_{ka} W^-_\mu 
\left( S^+_a \partial^\mu S^0_b - S^0_b \partial^\mu S^+_a \right) 
\\ && -
\frac{g}{4c_w}\, \mbox{Im} \left( V_{kb}^* V_{kb'} \right) Z_\mu 
\left( S^0_b \partial^\mu S^0_{b'} - S^0_{b'} \partial^\mu S^0_b \right). 
\end{eqnarray}
\end{subequations}
In the case of non-derivative couplings, we will only need those to the
neutral scalars:
\begin{equation}\label{vb-s-non}
\mathcal{L}_{\mathrm{non-}\partial} = 
\left( \frac{g^2}{4} W^+_\mu W^{-\,\mu} + \frac{g^2}{8c_w^2} Z_\mu Z^\mu \right) 
\left[ \left( v_k^* V_{kb} + V_{kb}^* v_k \right) S^0_b + 
V_{kb}^* V_{kb'} S^0_b S^0_{b'} \right].
\end{equation}

\subsection{Scalar--ghost interactions}
Defining 
\begin{equation}\label{omega}
\omega_k = \frac{v_k}{v},
\end{equation}
we can write the interaction of the scalar mass eigenfields $S^0_b$ with 
the ghost fields $c^+$, $c^-$ and $c_Z$ as
\begin{subequations}
\label{ghost-L}
\begin{eqnarray}
\mathcal{L}(S^0 \bar c c) &=& 
-\frac{gm_W \xi_W}{2}\, \sum_{b=2}^\infty S^0_b
\left( \omega_k^* V_{kb}\, \overline{c^+} c^+ +  
\omega_k V_{kb}^*\, \overline{c^-} c^- \right)
\label{c+-} \\ &&
-\frac{g m_Z \xi_Z}{2 c_w}\, \sum_{b=2}^\infty S^0_b
\, \mbox{Re} \left( \omega_k^* V_{kb} \right) \bar c_Z c_Z.
\label{cz}
\end{eqnarray}
\end{subequations}

\subsection{Triple scalar interactions}
The triple-scalar interactions~\cite{bento} follow straightforwardly from the 
scalar potential, equation~(\ref{V}):
\begin{eqnarray}
\label{sss0}
\mathcal{L}(S^0 S^0 S^0) &=& -\frac{1}{2}\, \lambda_{ijkl}
\left( v_i^* V_{jb_1} + V_{ib_1}^* v_j \right) V_{kb_2}^* V_{lb_3}
S^0_{b_1} S^0_{b_2} S^0_{b_3},
\\ \label{sss+-}
\mathcal{L}(S^0 S^- S^+) &=& -\lambda_{ijkl}
\left( v_i^* V_{jb} + V_{ib}^* v_j \right) U_{ka_1}^* U_{la_2}
S^0_{b} S^-_{a_1} S^+_{a_2}.
\end{eqnarray}
For the properties of $\lambda_{ijkl}$ see equation~(\ref{indices}).
If one of the scalars is a Goldstone boson, then the difference of the 
squares of the masses of 
the other two scalars is involved in the 
triple scalar coupling~\cite{grimus}. Specializing to the coupling of 
$S^0$ to the
Goldstone bosons leads to~\cite{bento}
\begin{equation}\label{sgg}
\mathcal{L}(S^0 GG) = \frac{1}{v}\, \sum_{b=2}^{2n_H} M_b^2 \,
\mbox{Im} \left( V^\dagger V \right)_{1b} S^0_b 
\left( G^+ G^- + \frac{1}{2}\, G^0 G^0 \right).
\end{equation}

\subsection{Quartic scalar interactions}
The quartic scalar couplings 
which we need in the following
are given by~\cite{bento}
\begin{eqnarray}
\mathcal{L}(S^0 S^0 S^0 S^0) &=& -\frac{1}{4}\, \lambda_{ijkl}
V_{ib_1}^* V_{jb_2} V_{kb_3}^* V_{lb_4} 
S^0_{b_1} S^0_{b_2} S^0_{b_3} S^0_{b_4},
\\ 
\mathcal{L}(S^0 S^0 S^- S^+) &=& -\lambda_{ijkl}
V_{ib_1}^* V_{jb_2} U_{ka_1}^* U_{la_2} 
S^0_{b_1} S^0_{b_2} S^-_{a_1} S^+_{a_2}. 
\end{eqnarray}

\subsection{Scale factors in dimensional regularization}
As mentioned in the introduction, we 
will be using dimensional regularization for the one-loop integrals 
in $d = 4 - \varepsilon$ dimensions. Introducing the mass scale $\mc$,
in order to keep the coupling constants dimensionless in $d$ dimensions, 
we have to make the replacements
\begin{equation}
g \to \mc^{\varepsilon/2} g, \quad
\Delta_k \to \mc^{\varepsilon/2} \Delta_k, \quad
\Gamma_k \to \mc^{\varepsilon/2} \Gamma_k, \quad
\lambda_{ijkl} \to \mc^\varepsilon \lambda_{ijkl}.
\end{equation}
Similarly, the VEVs have to be scaled by
\begin{equation}
v_i \to \mc^{-\varepsilon/2} v_i,
\end{equation}
so that they have the dimension of a mass.

\section{The scalar sector}
\label{scalar sector}
In this section we discuss the one- and two-point functions of the 
\emph{neutral} scalars.

\subsection{The counterterms for the one- and two-point scalar functions}
\label{cts for one and two}
The scalar potential is defined in equation~(\ref{V}).
The Lagrangian of the scalar potential plus its counterterm
parameters is given by 
\begin{equation}\label{V+cts}
-V(\phi) - \delta V(\phi) =
-\left( \mu^2_{ij} + \delta{\hat\mu}^2_{ij} \right) \phi_i^\dagger \phi_j
-\mc^\varepsilon \left( \lambda_{ijkl} + \delta\hat\lambda_{ijkl} \right)
\phi_i^\dagger \phi_j \phi_k^\dagger \phi_l.
\end{equation}
Here the components of the Higgs doublets $\phi_i$ are meant to be bare
fields. For the neutral components of the Higgs doublets we make the 
ansatz\footnote{Note that the symbol ``$\Delta v_j$'' used 
 in~\cite{Denner:2016etu} 
 refers to the total VEV shifts of the bare scalar fields and has, therefore, 
 a meaning different from our $\Delta v_j$.}
\begin{equation}
\varphi^0_i = \frac{1}{\sqrt{2}} \left( Z^{(1/2)}_\varphi \right)_{ij} 
\left[ \mc^{-\varepsilon/2} \left( v_j + \Delta v_j + \delta v_j \right) + 
V_{jb} S^0_b \right].
\end{equation}
The VEVs $v_j$ are in our notation pure tree-level quantities defined 
as the solution of the set of $n_H$ equations
\begin{equation}
\left( \mu^2_{ij} + \lambda_{ijkl} v_k^* v_l \right) v_j = 0. 
\end{equation}
By definition, the fields $S^0_b$ have vanishing VEVs, which is guaranteed
beyond tree level by the finite VEV shifts $\Delta v_j$. In addition, 
the VEV renormalization $\delta v_j$ is needed in the $R_\xi$ gauge 
in the case of $\xi \neq 0$~\cite{sperling}. The complex 
$n_H \times 2n_H$ matrix  $\left( V_{jb} \right)$ is connected to 
the orthogonal $2n_H \times 2n_H$ diagonalization matrix of the 
mass matrix of the neutral scalars. For its definition and properties we 
refer the reader to appendix~\ref{app-UV}.

Since we only perform one-loop computations, we write \begin{equation}
\left( Z^{(1/2)}_\varphi \right)_{ij} = \delta_{ij} + 
\frac{1}{2}\,z^{(\varphi)}_{ij} 
\end{equation}
for the wave-function renormalization of the neutral scalars.  
It is convenient to absorb the wave-function renormalization into 
the counterterm parameters of equation~(\ref{V+cts}).
Thus we define 
\begin{equation}
\delta\mu^2_{ij} = \delta{\hat\mu}^2_{ij} + \frac{1}{2}
\left( \mu^2_{i'j} \left( z^{(\varphi)}_{i'i} \right)^* + 
\mu^2_{ij'}\, z^{(\varphi)}_{j'j} \right)
\end{equation}
and
\begin{equation}
\delta\lambda_{ijkl} = \delta\hat\lambda_{ijkl} +
\frac{1}{2} \left( \lambda_{i'jkl} \left( z^{(\varphi)}_{i'i} \right)^* + 
\lambda_{ij'kl} z^{(\varphi)}_{j'j} + \cdots \right).
\end{equation}
With these definitions the counterterms for the one- and two-point functions 
are induced by $\delta\mu^2_{ij}$, $\delta\lambda_{ijkl}$ and $\delta v_j$.

In writing down the counterterm for the scalar one-point function of $S^0_b$,
we ``truncate'' it by removing $\mc^{-\varepsilon/2}i/(-M^2_b)$. 
Then the counterterm reads 
\begin{fmffile}{tadpole-ct}
\fmfset{thin}{.7pt}
\fmfset{dash_len}{1.5mm}
\begin{subequations}\label{1pf-ct}
  \begin{align}
    \begin{gathered}
      \begin{fmfgraph*}(25,25)
	\fmfbottom{b}
	\fmftop{t}
	\fmf{dashes}{b,t}
	\fmflabel{$S^0_b$}{b}
	\defotimes
	\fmfv{d.sh=otimes,d.f=empty}{t}
      \end{fmfgraph*}
    \end{gathered}
    \quad = \quad &
    -\frac{i}{2} \left[ \delta \mu^2_{ij} + \frac{1}{2}\,
    \delta\tilde\lambda_{ijkl} v_k^* v_l \right]
    \left( v_i^* V_{jb} + V_{ib}^* v_j \right)
    \label{1pf-ct1}
    \\ &
    -\frac{i}{2} \left( \delta v_i^* V_{ib} + V_{ib}^* \delta v_i \right) M_b^2.
    \label{1pf-ct2}
  \end{align}
\end{subequations}
\end{fmffile}%
Here we have introduced the definition
\begin{equation}\label{lambdatilde}
\tilde\lambda_{ijkl} \equiv \lambda_{ijkl} + \lambda_{ilkj}.
\end{equation}
and, consequently,
\begin{equation}\label{dlambdatilde}
\delta\tilde\lambda_{ijkl} \equiv \delta\lambda_{ijkl} + \delta\lambda_{ilkj}.
\end{equation}
In order to achieve the form of equation~(\ref{1pf-ct2}) with 
the neutral scalar masses $M_b$, 
we have taken into account 
equation~(\ref{columns}).
Note that, for later reference, we have split the 
counterterm of equation~(\ref{1pf-ct})
into a part induced by $\delta\mu^2_{ij}$ and $\delta\tilde\lambda_{ijkl}$
and a part induced by $\delta v_i$.

The counterterm pertaining to the scalar self-energy 
$-i \Pi_{bb'}(p^2)$ is given by
\begin{fmffile}{scalar-self-ct}
\fmfset{thin}{.7pt}
\fmfset{dash_len}{1.5mm}
\begin{subequations}\label{2pf-ct}
  \begin{align}
    \begin{gathered}
      \vspace{-4pt}
      \begin{fmfgraph*}(50,30)
	\fmfleft{i}
	\fmfright{o}
	\fmf{dashes}{i,v}
	\fmf{dashes}{o,v}
	\fmflabel{$S^0_b$}{i}
	\fmflabel{$S^0_{b'}$}{o}
	\defotimes
	\fmfv{d.sh=otimes,d.f=empty}{v}
      \end{fmfgraph*}
    \end{gathered}
    \quad \quad = \; &
    -\frac{i}{2} 
    \left[ \delta \mu^2_{ij} + \delta\tilde\lambda_{ijkl} v_k^* v_l \right]
    \left( V_{ib}^* V_{jb'} + V_{ib'}^* V_{jb} \right)
    \nonumber \\ &
    -\frac{i}{4} \delta\tilde\lambda_{ijkl} \left[ 
    v_i^* v_k^* V_{jb} V_{lb'} + v_j v_l V_{ib}^* V_{kb'}^* \right]
    \label{2pf-ct1}
    \\ &
    -\frac{i}{2} \tilde\lambda_{ijkl} 
    \left( \delta v_k^* v_l + v_k^* \delta v_l \right) 
    \left( V_{ib}^* V_{jb'} + V_{ib'}^* V_{jb} \right) 
    \nonumber \\ &
%%%%
    -\frac{i}{4} \tilde\lambda_{ijkl} \left[
    \left( \delta v_i^* v_k^* + v_i^* \delta v_k^* \right) V_{jb} V_{lb'} +
    \left( \delta v_j v_l + v_j \delta v_l  \right) V_{ib}^* V_{kb'}^*
    \right].
    \label{2pf-ct2}
    \end{align}
  \end{subequations}
\end{fmffile}%
In this counterterm we have done a splitting analogous to 
the case of the one-point function. 
Note that the second line in equation~(\ref{2pf-ct1})
comes about 
because
\begin{equation}
\delta\lambda_{ijkl} v_i^* v_k^* V_{jb} V_{lb'} = 
\frac{1}{2}\, \delta\tilde\lambda_{ijkl} v_i^* v_k^* V_{jb} V_{lb'},
\end{equation}
\textit{cf}.\ equation~(\ref{indices}).
A similar argument applies to the second line of equation~(\ref{2pf-ct2}).

We anticipate here that the counterterm of equation~(\ref{1pf-ct}) 
connects via the scalar propagator to 
a neutral or charged fermion line and contributes thus to the counterterms 
of the fermion self-energy, and the VEV renormalization $\delta v_j$ 
contributes directly via $v_j \to v_j + \delta v_j$ to the 
fermion mass counterterms---see figure~\ref{counterterms} for a graphical 
rendering. 
These counterterms will play an important role in sections~\ref{Counterterms} 
and~\ref{finiteness}.

Now we proceed as announced in the introduction in items~1-3.

\subsection{Renormalization of the quartic scalar couplings}
\label{quartic}
 \begin{figure}[ht]
  \begin{fmffile}{scalar-quartic}
   \fmfset{thin}{.7pt}
   \fmfset{dash_len}{1.5mm}
   \fmfset{wiggly_len}{2mm}
   \fmfset{wiggly_slope}{75}
   \fmfset{dot_size}{1.5thick}
  \begin{center} 
    \begin{subfigure}[t]{.3\textwidth} \centering
      \begin{fmfgraph*}(100,70)
	\fmfleft{i1,i2}
	\fmfright{o1,o2}
	\fmf{dashes}{i1,v1,i2}
	\fmf{dashes}{o1,v2,o2}
	\fmf{dashes,right=0.8,tension=.8}{v1,v2}
	\fmf{dashes,right=0.8,tension=.8}{v2,v1}
	\fmfdot{v1,v2}
      \end{fmfgraph*}
    \setlength{\abovecaptionskip}{0pt}
    \caption{}
    \end{subfigure}
    \begin{subfigure}[t]{.3\textwidth} \centering
      \begin{fmfgraph*}(100,70)
	\fmfleft{i1,i2}
	\fmfright{o1,o2}
	\fmf{dashes}{i1,v1,i2}
	\fmf{dashes}{o1,v2,o2}
	\fmf{wiggly,right=0.8,tension=.8}{v1,v2}
	\fmf{wiggly,right=0.8,tension=.8}{v2,v1}
	\fmfdot{v1,v2}
      \end{fmfgraph*}
    \setlength{\abovecaptionskip}{0pt}
    \caption{}
    \end{subfigure}
    \begin{subfigure}[t]{.3\textwidth} \centering
      \begin{fmfgraph*}(100,70)
	\fmfleft{i1,im,i2}
	\fmfright{o1,o2}
	\fmf{dashes}{vx,v1,i1}
	\fmf{dashes}{vx,v2,i2}
	\fmf{dashes}{vx,o2}
	\fmf{dashes}{vx,o1}
	\fmf{phantom}{im,vx}
	\fmffreeze
	\fmf{wiggly}{v2,v1}
	\fmfdot{v1,v2,vx}
      \end{fmfgraph*}
    \setlength{\abovecaptionskip}{0pt}
    \caption{}
    \end{subfigure}
    \\[10pt]
    \begin{subfigure}[t]{.3\textwidth} \centering
      \begin{fmfgraph*}(100,70)
	\fmfleft{i1,i2}
	\fmfright{o1,o2}
	\fmf{dashes,tension=1.}{i1,vlb,vlt,i2}
	\fmf{dashes,tension=1.}{o1,vrb,vrt,o2}
	\fmf{wiggly}{vlb,vrb}
	\fmf{wiggly}{vlt,vrt}
	\fmfdot{vlb,vlt,vrb,vrt}
      \end{fmfgraph*}
    \setlength{\abovecaptionskip}{0pt}
    \caption{}
    \end{subfigure}
    \begin{subfigure}[t]{.3\textwidth} \centering
      \begin{fmfgraph*}(100,70)
	\fmfleft{i1,im,i2}
	\fmfright{o1,o2}
	\fmf{phantom}{i1,v1,vx,o2}
	\fmf{phantom}{i2,v2,vx,o1}
	\fmf{phantom}{im,vx}
	\fmffreeze
	\fmf{dashes}{i1,v1,v2,i2}
	\fmf{dashes}{o1,vx,o2}
	\fmf{wiggly}{v1,vx}
	\fmf{wiggly}{v2,vx}
	\fmfdot{v1,v2,vx}
      \end{fmfgraph*}
    \setlength{\abovecaptionskip}{0pt}
    \caption{}
    \end{subfigure}
    \begin{subfigure}[t]{.3\textwidth} \centering
       \begin{fmfgraph*}(100,70)
	\fmfleft{i1,i2}
	\fmfright{o1,o2}
	\fmf{phantom,tension=1.}{i1,vlb,vlt,i2}
	\fmf{phantom,tension=1.}{o1,vrb,vrt,o2}
	\fmf{phantom}{vlb,vrb}
	\fmf{phantom}{vlt,vrt}
	\fmffreeze
	\fmf{plain}{vlt,vrt,vrb,vlb,vlt}
	\fmf{dashes}{vlb,i1}
	\fmf{dashes}{vlt,i2}
	\fmf{dashes}{vrb,o1}
	\fmf{dashes}{vrt,o2}
      \end{fmfgraph*}
    \setlength{\abovecaptionskip}{0pt}
    \caption{}
    \end{subfigure} 
  \end{center}
  \end{fmffile}
  \caption{The Feynman diagrams that determine $\delta\tilde{\lambda}_{ijkl}$
defined in equation~(\ref{lambdatilde}). 
The lines have the usual meaning: full, wiggly and dashed lines indicate
fermions, vector bosons and scalars, respectively.}
  \label{scalar-quartic-graphs}
\end{figure}
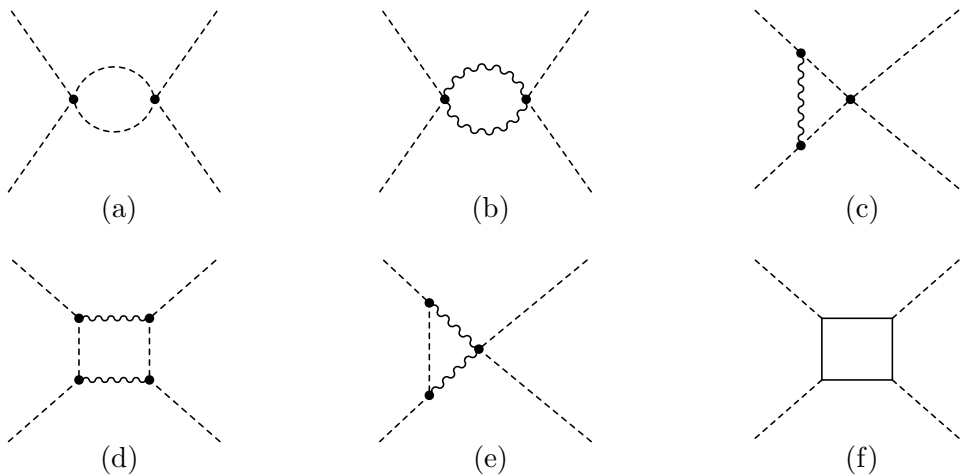
In the $\overline{\mbox{MS}}$ renormalization scheme
at one-loop order the terms proportional to $c_\infty$
of equation~(\ref{cinfty})
have to be cancelled by the respective counterterms.

Since we are only interested in the divergencies of the scalar 
four-point function, we can stick to the unbroken theory for the 
computation of $\delta\lambda_{ijkl}$. Moreover, 
as discussed in the previous subsection, it suffices to compute 
$\delta\tilde\lambda_{ijkl}$ instead of $\delta\lambda_{ijkl}$.
This leads us to consider the four-point function
\begin{equation}\label{4pf}
\langle 0 | T \varphi^0_i\, {\varphi^0_j}^* \varphi^0_k\, 
{\varphi^0_l}^* | 0 \rangle.
\end{equation}
The Feynman diagrams from which we compute the divergencies are 
displayed in figure~\ref{scalar-quartic-graphs}. There is a 
one-to-one correspondence between the labels of the subfigures and 
those of the subequations of equation~(\ref{dlambda}). Moreover, 
every subequation contains the contributions of both charged and neutral 
inner lines of the diagrams. 
We obtain the following result for the 
divergencies:\footnote{Here and in the following, a full blob means a 
sum over one-loop diagrams, whereas a circle with a cross refers to a 
counterterm. When specific parts of these entities are addressed,
they are put within parentheses with subscripts indicating the specifics.}
\vspace{10pt}
\begin{fmffile}{scalar-quartic-divs}
\fmfset{thin}{.7pt}
\fmfset{dash_len}{1.5mm}
\begin{subequations}\label{dlambda}
\begin{align}
i L_{ijkl} \equiv\,& 
    \left(\quad 
    \begin{gathered}
      \begin{fmfgraph*}(50,40)
	\fmfleft{i1,i2}
	\fmfright{o1,o2}
	\fmf{dashes}{i1,v}
	\fmf{dashes}{i2,v}
	\fmf{dashes}{o1,v}
	\fmf{dashes}{o2,v}
	\fmfv{label=\footnotesize{$\varphi^0_i$},l.d=1}{i2}
	\fmfv{label=\footnotesize{$\varphi^{0*}_j$},l.d=1}{i1}
	\fmfv{label=\footnotesize{$\varphi^0_k$},l.d=1}{o2}
	\fmfv{label=\footnotesize{$\varphi^{0*}_l$},l.d=1}{o1}
	\fmfblob{17}{v}
      \end{fmfgraph*}
    \end{gathered}
    \quad \vphantom{\begin{minipage}[t][1cm]{0pt}\end{minipage}}
    \right)_{\!\!c_\infty}
    \nonumber
    \\[1.2em]
    = & \;\frac{i}{16\pi^2}\,c_\infty \left\{
    4 \left[ \vphantom{\lambda^\dagger_j} 
    \tilde\lambda_{ijmn} \tilde\lambda_{klnm} +
    \tilde\lambda_{ilmn} \tilde\lambda_{kjnm}  \right. \right.
    \nonumber \\
    &+ \lambda_{ijmn} \lambda_{klnm} + \lambda_{ilmn} \lambda_{kjnm} 
    +\left. 
    \lambda_{imkn} \lambda_{mjnl} + \lambda_{imkn} \lambda_{mlnj} 
    \right] 
    \\
    &+
    \left[ \frac{g^4}{4} \left( 3 + \xi_W^2 \right) + 
    \frac{g^4}{8 c_w^4} \left( 3 + \xi_Z^2 \right) \right]
    \left( \delta_{ij} \delta_{kl} + \delta_{il} \delta_{kj} \right)
    \\
    &-
    \left( 2 g^2 \xi_W + \frac{g^2}{c_w^2} \xi_Z \right) \tilde\lambda_{ijkl}
    \label{Vvc}
    \\
    &+
    \left( \frac{g^4}{4} \xi_W^2 + \frac{g^4}{8 c_w^4} \xi_Z^2 \right)
    \left( \delta_{ij} \delta_{kl} + \delta_{il} \delta_{kj} \right)
    \\
    &-
    \left( \frac{g^4}{2} \xi_W^2 + \frac{g^4}{4 c_w^4} \xi_Z^2 \right)
    \left( \delta_{ij} \delta_{kl} + \delta_{il} \delta_{kj} \right)
    \\
    &- \left.
    2\, \mbox{Tr} \left( \Gamma_i \Gamma_j^\dagger \Gamma_k \Gamma_l^\dagger 
    + \Gamma_i \Gamma_l^\dagger \Gamma_k \Gamma_j^\dagger \right)
    -2\, \mbox{Tr} \left( \Delta_i^\dagger \Delta_j \Delta_k^\dagger \Delta_l
    + \Delta_i^\dagger \Delta_l \Delta_k^\dagger \Delta_j \right)
    \right\}.
    \end{align}
  \end{subequations}
\end{fmffile}%
The counterterm pertaining to the four-point function of equation~(\ref{4pf})
is given by
\begin{fmffile}{scalar-quartic-ct}
\fmfset{thin}{.7pt}
\fmfset{dash_len}{1.5mm}
  \begin{equation}
    \begin{gathered}
      \begin{fmfgraph*}(50,40)\fmfkeep{scalar-quartic-ct}
	\fmfleft{i1,i2}
	\fmfright{o1,o2}
	\fmf{dashes}{i1,v}
	\fmf{dashes}{i2,v}
	\fmf{dashes}{o1,v}
	\fmf{dashes}{o2,v}
	\defotimes
	\fmfv{d.sh=otimes,d.f=empty}{v}
      \end{fmfgraph*}
    \end{gathered}
    = \;
    -2i \left( \delta\lambda_{ijkl} + \delta\lambda_{ilkj} \right) = 
    -2i\, \delta\tilde\lambda_{ijkl}.
  \end{equation}
\end{fmffile}%
Then the $\overline{\mbox{MS}}$ condition is
\begin{fmffile}{scalar-quartic-msbar}
\fmfset{thin}{.7pt}
\fmfset{dash_len}{1.5mm}
  \begin{equation}\label{4pf-dlambda}
    \left(
    \begin{gathered}
      \begin{fmfgraph*}(50,40)
	\fmfleft{i1,i2}
	\fmfright{o1,o2}
	\fmf{dashes}{i1,v}
	\fmf{dashes}{i2,v}
	\fmf{dashes}{o1,v}
	\fmf{dashes}{o2,v}
	\fmfblob{17}{v}
      \end{fmfgraph*}
    \end{gathered}
    \;
    \right)_{\!\!c_\infty}
    +
  \begin{gathered}
    \fmfreuse{scalar-quartic-ct}
  \end{gathered}
    =
    i L_{ijkl} -2i \delta\tilde\lambda_{ijkl} = 0 
    \quad \mbox{or} \quad
    \delta\tilde\lambda_{ijkl} = \frac{1}{2} \, L_{ijkl}.
  \end{equation}
\end{fmffile}
Note that $L_{ijkl} = L_{ilkj}$, as it has to be for consistency.

For the further discussion it is convenient to decompose 
$\delta\tilde\lambda_{ijkl}$ as 
\begin{equation}\label{decomp-lambda}
\delta\tilde\lambda_{ijkl} = 
\delta\tilde\lambda_{ijkl}(S) +
\delta\tilde\lambda_{ijkl}(\ell^\pm) + \delta\tilde\lambda_{ijkl}(\chi) + 
\delta\tilde\lambda_{ijkl}(\xi^0) +
\delta\tilde\lambda_{ijkl}(\xi^1) + \delta\tilde\lambda_{ijkl}(\xi^2).
\end{equation}
The first three terms correspond to the contributions of the scalars, 
the charged leptons and the neutrinos, respectively,
\textit{i.e.} to those diagrams
which do not have a vector boson line. 
Vector boson contributions can be characterized by powers 
in the gauge parameters ($\xi^\nu$ with $\nu=0,1,2$)---see 
equation~(\ref{dlambda}). Actually, diagrams with two vector boson lines 
are proportional to $g^4$ and have parts with $\xi^0$ and $\xi^2$, 
whereas diagrams with one vector boson line are proportional to $g^2 \xi^1$ and 
a quartic scalar coupling.

However, inspection of equation~(\ref{dlambda}) reveals that the $\xi^2$-terms
cancel each other, \textit{i.e.}
\begin{equation}\label{xi2=0}
\delta\tilde\lambda_{ijkl}(\xi^2) = 0.
\end{equation}

\subsection{Divergencies of the neutral-scalar self-energy}
\label{divergencies-neutral-scalar}
\begin{figure}[ht]
 \begin{fmffile}{scalar-selfenergy}
 \fmfset{thin}{.7pt}
 \fmfset{dash_len}{1.5mm}
 \fmfset{wiggly_len}{2mm}
 \fmfset{wiggly_slope}{75}
 \fmfset{dot_len}{.8mm}
 \fmfset{dot_size}{1.5thick}
\begin{center}
 \begin{subfigure}[t]{.3\textwidth} \centering
 \begin{fmfgraph*}(100,80)
 \fmfleft{i}
 \fmfright{o}
 \fmf{dashes,tension=2}{i,v1}
 \fmf{dashes,tension=2}{v2,o}
 \fmf{dashes,right}{v1,v2}
 \fmf{dashes,right}{v2,v1}
 \fmfdot{v1,v2}
 \end{fmfgraph*}
 \setlength{\abovecaptionskip}{-12pt}
 \caption{}
 \end{subfigure}
 \begin{subfigure}[t]{.3\textwidth} \centering
 \begin{fmfgraph*}(100,80)
 \fmfleft{i}
 \fmfright{o}
 \fmftop{t}
 \fmf{dashes}{i,v,o}
 \fmffreeze
 \fmf{dashes,right,tension=2}{v,t,v}
 \fmfdot{v}
 \end{fmfgraph*}
 \setlength{\abovecaptionskip}{-12pt}
 \caption{}
 \end{subfigure}
 \begin{subfigure}[t]{.3\textwidth} \centering
 \begin{fmfgraph*}(100,80)
 \fmfleft{i}
 \fmfright{o}
 \fmftop{t}
 \fmf{dashes}{i,v,o}
 \fmffreeze
 \fmf{wiggly,right,tension=2}{v,t,v}
 \fmfdot{v}
 \end{fmfgraph*}
 \setlength{\abovecaptionskip}{-12pt}
 \caption{}
 \end{subfigure}
 \\[10pt]
 \begin{subfigure}[t]{.3\textwidth} \centering
 \begin{fmfgraph*}(100,80)
 \fmfleft{i}
 \fmfright{o}
 \fmf{dashes,tension=2}{i,v1}
 \fmf{dashes,tension=2}{v2,o}
 \fmf{wiggly,right}{v1,v2}
 \fmf{wiggly,right}{v2,v1}
 \fmfdot{v1,v2}
 \end{fmfgraph*}
 \setlength{\abovecaptionskip}{-12pt}
 \caption{}
 \end{subfigure}
 \begin{subfigure}[t]{.3\textwidth} \centering
 \begin{fmfgraph*}(100,80)
 \fmfleft{i}
 \fmfright{o}
 \fmf{dashes,tension=2}{i,v1}
 \fmf{dashes,tension=2}{v2,o}
 \fmf{dashes,right}{v1,v2}
 \fmf{wiggly,right}{v2,v1}
 \fmfdot{v1,v2}
 \end{fmfgraph*}
 \setlength{\abovecaptionskip}{-12pt}
 \caption{}
 \end{subfigure}
 \begin{subfigure}[t]{.3\textwidth} \centering
 \begin{fmfgraph*}(100,80)
 \fmfleft{i}
 \fmfright{o}
 \fmf{dashes,tension=2}{i,v1}
 \fmf{dashes,tension=2}{v2,o}
 \fmf{dots,right}{v1,v2}
 \fmf{dots,right}{v2,v1}
 \fmfdot{v1,v2}
 \end{fmfgraph*}
 \setlength{\abovecaptionskip}{-12pt}
 \caption{}
 \end{subfigure}
 \begin{subfigure}[t]{.3\textwidth} \centering
 \begin{fmfgraph*}(100,80)
 \fmfleft{i}
 \fmfright{o}
 \fmf{dashes,tension=2}{i,v1}
 \fmf{dashes,tension=2}{v2,o}
 \fmf{plain,right}{v1,v2}
 \fmf{plain,right}{v2,v1}
 \fmfdot{v1,v2}
 \end{fmfgraph*}
 \setlength{\abovecaptionskip}{-12pt}
 \caption{}
 \end{subfigure}
\end{center}
 \end{fmffile}
 \caption{The Feynman diagrams contributing the scalar self-energy in the
broken phase. In addition to the lines explained in 
figure~\ref{scalar-quartic-graphs}, in diagram~(f) we have dotted lines 
indicating ghost propagators.} 
 \label{scalar-selfenergy-graphs}
\end{figure}
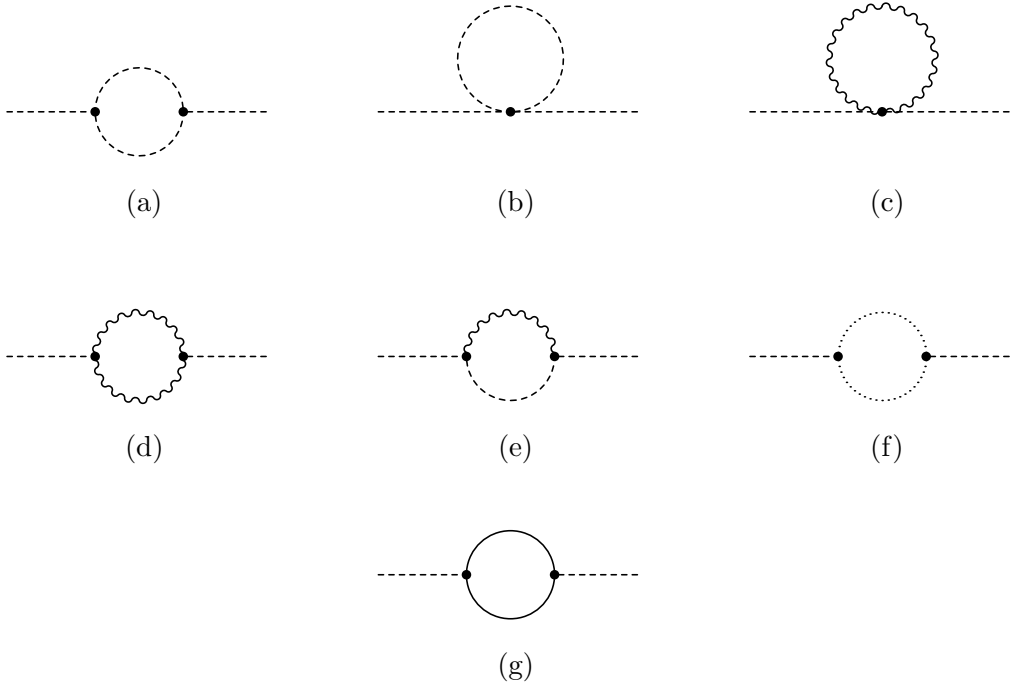
\noindent
Now we turn to the divergencies of the scalar two-point function. 
Having obtained the result for $\delta\tilde\lambda_{ijkl}$, 
we need $\delta\mu^2_{ij}$ and $\delta v_i$
for the full counterterm of the two-point function, 
equation~(\ref{2pf-ct}). The aim of the present section is not only 
to compute $\delta\mu^2_{ij}$ and $\delta v_i$
but our computations also serve as a consistency check
that the scalar self-energy can indeed be made finite by a suitable choice 
of these parameters.

In the presentation of the one-loop results for the divergencies we use 
the vector boson masses of equation~(\ref{mV}),
the definitions of the matrices $\Lambda$, $K$ and $K'$ given in 
appendix~\ref{app-MM}, and equations~(\ref{UMU}) and~(\ref{VMV}).
The divergencies refer to those of $-i \Pi_{bb'}(p^2)$, where $\Pi_{bb'}(p^2)$ is 
the self-energy matrix of the neutral scalar mass eigenfields $S^0_b$.

Now we list the momentum-independent divergencies belonging to the diagrams of 
figure~\ref{scalar-selfenergy-graphs}. In the individual results we indicate 
the nature of the particle (or particles) in the loop.
\\
Diagram~(a), charged scalars:
\begin{equation}\label{csl}
\frac{i}{16\pi^2}\,c_\infty \lambda_{ijmn} \lambda_{nmkl}
\left( v_i^* V_{jb} + V_{ib}^* v_j \right)
\left( v_k^* V_{lb'} + V_{kb'}^* v_l \right),
\end{equation}
diagram~(a), neutral scalars:
\begin{eqnarray}
&&
\frac{i}{16\pi^2}\,c_\infty \left\{
\left( V_{kb}^* V_{lb'} + V_{kb'}^* V_{lb} \right)
\left[ \tilde\lambda_{ilmn} \tilde\lambda_{kjnm} +
\lambda_{imkn} \left( \lambda_{mjnl} + \lambda_{njml} \right) \right] v_i^* v_j
\right. \nonumber 
\\ &&
+ \left. \left( V_{jb} V_{lb'} v_i^* v_k^* + V_{ib}^* V_{kb'}^* v_j v_l \right)
\tilde\lambda_{ijmn} \tilde\lambda_{klnm} \right\},
\label{nsl}
\end{eqnarray}
diagram~(b), charged scalars:
\begin{equation}\label{csb}
\frac{i}{16\pi^2}\,c_\infty 
\left( V_{ib}^* V_{jb'} + V_{ib'}^* V_{jb} \right)
\left[ \frac{1}{4}\,g^2 \xi_W\, \Lambda_{ij} + 
\lambda_{ijkl} \left( \mu^2 + \Lambda \right)_{lk} \right],
\end{equation}
diagram~(b), neutral scalars:
\begin{eqnarray}
&&
\frac{i}{16\pi^2}\,c_\infty \left\{ \frac{g^2 \xi_Z}{8 c_w^2} \left[
\left( V_{ib}^* V_{jb'} + V_{ib'}^* V_{jb} \right) \left( \Lambda + K' \right)_{ij}
- \lambda_{ijkl} \left( V_{jb} V_{lb'} v_i^* v_k^* + 
V_{ib}^* V_{kb'}^* v_j v_l \right) \right]
\right. \nonumber \\ && \left.
+ \tilde\lambda_{ijkl} 
\left( V_{ib}^* V_{jb'} + V_{ib'}^* V_{jb} \right)
\left( \mu^2 + \Lambda +  K' \right)_{lk} + 
\lambda_{ijkl} V_{jb} V_{lb'} K_{ik}^* +
\lambda_{ijkl} V_{ib}^* V_{kb'}^* K_{jl} \right\},
\label{nsb}
\end{eqnarray}
diagram~(c), $W^\pm$ and $Z$ bosons:
\begin{equation}\label{vbb}
\frac{i}{16\pi^2}\,c_\infty \delta_{bb'}
\left[ \frac{g^4 v^2}{8} \left( 3 + \xi_W^2 \right) +  
\frac{g^4 v^2}{16 c_w^4} \left( 3 + \xi_Z^2 \right) \right],
\end{equation}
diagram~(d), $W^\pm$ and $Z$ bosons:
\begin{equation}\label{vbl}
\frac{i}{16\pi^2}\,c_\infty 
\left( v_k^* V_{kb} + V_{kb}^* v_k \right)
\left( v_l^* V_{lb'} + V_{lb'}^* v_l \right)
\left[ \frac{g^4}{16} \left( 3 + \xi_W^2 \right) +
\frac{g^4}{32 c_w^4} \left( 3 + \xi_Z^2 \right) \right],
\end{equation}
diagram~(e), $W^\pm$ boson and charged scalars:
\begin{equation}
-\frac{i}{16\pi^2}\,c_\infty 
\left( V_{ib}^* V_{jb'} + V_{ib'}^* V_{jb} \right)
\left[ \frac{1}{16}\,g^4 \xi_W^2 \left( \delta_{ij} v^2 + v_i v_j^* \right) + 
\frac{1}{4}\,g^2 \xi_W \left( \mu^2 + \Lambda \right)_{ij} \right],
\label{wcl}
\end{equation}
diagram~(e), $Z$ boson and neutral scalars:
\begin{eqnarray}
&& \nonumber
\frac{i}{16\pi^2}\,c_\infty \left\{
-\frac{g^4 \xi_Z^2}{64 c_w^4} \left[
2\left( V_{ib}^* V_{jb'} + V_{ib'}^* V_{jb} \right) \delta_{ij} v^2 +
\left( v_k^* V_{kb} + V_{kb}^* v_k \right)
\left( v_l^* V_{lb'} + V_{lb'}^* v_l \right) \right] \right.
\\ && \left.
+\frac{g^2 \xi_Z}{8 c_w^2} \left[
V_{ib} V_{jb'} K_{ij}^* + V_{ib}^* V_{jb'}^* K_{ij} - 
\left( V_{ib}^* V_{jb'} + V_{ib'}^* V_{jb} \right) 
\left( \mu^2 + \Lambda + K' \right)_{ij} \right] \right\},
\label{znl}
\end{eqnarray}
diagram~(f), charged ghosts:
\begin{equation}\label{cgl}
-\frac{i}{16\pi^2}\,c_\infty 
\frac{g^4 v^2}{16} \, \xi_W^2 
\left( \omega_k^* V_{kb}\, \omega_l^* V_{lb'} +  
\omega_k V_{kb}^*\, \omega_l V_{lb'}^* \right),
\end{equation}
diagram~(f), neutral ghost:
\begin{equation}\label{ngl}
-\frac{i}{16\pi^2}\,c_\infty 
\frac{g^4 v^2}{16 c_w^4} \, \xi_Z^2 \,
\mbox{Re} \left(\omega_k^* V_{kb} \right)
\mbox{Re} \left(\omega_l^* V_{lb'} \right)
\end{equation}
diagram~(g), charged leptons:
\begin{eqnarray}
-\frac{i}{16\pi^2}\,c_\infty & \hspace{-3mm}\times\hspace{-3mm} &
\mbox{Tr} \left\{ 
\left( \Gamma_k^\dagger \Gamma_j M_\ell^\dagger M_\ell + 
\Gamma_j \Gamma_k^\dagger M_\ell M_\ell^\dagger \right)
\left( V_{jb}^* V_{kb'} + V_{jb'}^* V_{kb} \right) \right.
\nonumber \\
&& + \left.
\Gamma_k^\dagger M_\ell \Gamma_j^\dagger M_\ell\,  V_{jb'} V_{kb} +
\Gamma_k M_\ell^\dagger \Gamma_j M_\ell^\dagger\, V_{jb'}^* V_{kb}^* 
\right\},
\label{cll}
\end{eqnarray}
diagram~(g), neutrinos:
\begin{eqnarray}
-\frac{i}{16\pi^2}\,c_\infty & \hspace{-3mm}\times\hspace{-3mm} &
\mbox{Tr} \left\{ \left[
M_D^\dagger M_D \Delta_j^\dagger \Delta_k + 
\left( M_D M_D^\dagger +  M_R M_R^\dagger \right) \Delta_k \Delta_j^\dagger
\right] \left( V_{jb}^* V_{kb'} + V_{jb'}^* V_{kb} \right) \right.
\nonumber \\
&& + \left.
\Delta_k M_D^\dagger \Delta_j M_D^\dagger\,  V_{jb'} V_{kb} +
\Delta_k^\dagger M_D \Delta_j^\dagger M_D\,  V_{jb'}^* V_{kb}^* 
\right\}.
\label{nll}
\end{eqnarray}
In the last two equations we have exploited the mass relations for 
the leptons, as presented in section~\ref{scalar interactions}.

Some remarks concerning the $\xi$-dependence of the divergencies are in order.
In equations~(\ref{csb}) and~(\ref{nsb}) the linear $\xi$-dependence 
comes from the Goldstone bosons in the loop because, due to the $R_\xi$ gauge,
the Goldstone boson masses $M_{+1}^2 = M_1^2 = 0$ are replaced by
$\xi_W m_W^2$ and $\xi_Z m_Z^2$, respectively. The vector boson loops of 
equation~(\ref{vbb}) and~(\ref{vbl}) lead 
to a quadratic $\xi$-dependence, 
stemming from the $\xi$-dependence of the vector boson propagators.
Finally, the mixed vector boson--scalar loops have a linear $\xi$-dependence 
originating in the vector boson propagator, however, an additional 
factor $\xi$ comes into play in the case of Goldstone bosons in the loop.

\subsection{Determination of 
\texorpdfstring{$\delta \mu^2_{ij}$}{delta musq} and %
\texorpdfstring{$\delta v_k$}{delta vk}}
Having computed $\delta\tilde\lambda_{ijkl}$ in section~\ref{quartic}, we
are now in a position to determine $\delta \mu^2_{ij}$ and $\delta v_k$ 
from the divergencies of the scalar self-energy as presented 
in the previous subsection. 
In a graphical presentation, $\delta\mu^2_{ij}$ and $\delta v_k$ 
are to be computed from
\begin{fmffile}{scalar-self-divs-cts-labels}
	\fmfset{thin}{.7pt}
	\fmfset{dash_len}{1.5mm}
	\begin{align}
	\left(\quad
	\begin{gathered}
	\vspace{-4pt}
	\begin{fmfgraph*}(50,20)
	\fmfleft{i}
	\fmfright{o}
	\fmf{dashes}{i,v}
	\fmf{dashes}{o,v}
	 	\fmfv{label=\footnotesize{$S^0_b$},l.d=1}{i}
	 	\fmfv{label=\footnotesize{$S^0_{b'}$},l.d=1}{o}
	\fmfblob{17}{v}
	\end{fmfgraph*}
	\end{gathered}
	\quad
	\right)_{\!\! c_\infty}
	 + 
	\quad \;
	\begin{gathered}
	\vspace{-4pt}
	\begin{fmfgraph*}(50,20)
	\fmfleft{i}
	\fmfright{o}
	\fmf{dashes}{i,v}
	\fmf{dashes}{o,v}
	\fmfv{label=\footnotesize{$S^0_b$},l.d=1}{i}
	\fmfv{label=\footnotesize{$S^0_{b'}$},l.d=1}{o}
	\defotimes
	\fmfv{d.sh=otimes,d.f=empty}{v}
	\end{fmfgraph*}
	\end{gathered}
	\quad \;
	=
	\;0.
	\end{align}
\end{fmffile}%
It can be checked straightforwardly that the divergencies of 
the scalar self-energy given by equations~(\ref{csl})--(\ref{nll}) have the
same types of terms as those in the decomposition of 
$\delta\tilde\lambda_{ijkl}$ in equation~(\ref{decomp-lambda}). 
Therefore, $\delta\mu^2_{ij}$ can be decomposed in the same way:
\begin{equation}
\delta\mu^2_{ij} = 
\delta\mu^2_{ij}(S) + \delta\mu^2_{ij}(\ell^\pm) + 
\delta\mu^2_{ij}(\chi) + \delta\mu^2_{ij}(\xi^0) + 
\delta\mu^2_{ij}(\xi^1) + \delta\mu^2_{ij}(\xi^2).
\end{equation}

It will turn out that, after
insertion of $\delta\tilde\lambda_{ijkl}$ of equation~(\ref{4pf-dlambda})
into the counterterm of equation~(\ref{2pf-ct1}) and adding it to the 
divergencies of the scalar self-energy, 
the determination of 
$\delta \mu^2_{ij}$ and $\delta v_k$ is unique for the following reasons:
\begin{enumerate}
\item
As proven in~\cite{sperling}, $\delta v_k$ is a linear function in $\xi$.
\item
Therefore, with the exception of the terms proportional to $\xi^1$,
for the cancellation of the divergencies 
we only have the counterterm containing $\delta \mu^2_{ij}$ at our disposal.
\item
As we will see, both the divergencies proportional to $\xi^1$
and the counterterm induced by $\delta v_k$ are
linear combinations of the two linearly independent matrices 
$\delta_{bb'} M_{b'}^2$ and
$\left( V_{ib}^* V_{jb'} + V_{ib'}^* V_{jb} \right) \mu^2_{ij}$, while 
the counterterm induced by $\delta \mu^2_{ij}(\xi^1)$ is proportional to the 
second matrix. Therefore these two counterterms are linearly independent and 
a unique combination of the two cancels the divergencies.
\end{enumerate}
One might think that the usage of the scalar one-point function is appropriate 
to fix $\delta v_k$, but this does not offer any advantage because 
one would need $\delta \mu^2_{ij}$ anyway since it occurs 
in the counterterm---see equation~(\ref{1pf-ct}).

Inserting $\delta\tilde\lambda_{ijkl}$ of equation~(\ref{4pf-dlambda})
into the counterterm of equation~(\ref{2pf-ct1}), we find 
\begin{fmffile}{scalar-self-amps-ct}
	\fmfset{thin}{.7pt}
	\fmfset{dash_len}{1.5mm}
  \begin{equation}
	\left( \;
	\begin{gathered}
	\vspace{-4pt}
		\begin{fmfgraph*}(50,20)\fmfkeep{scalar-self-amps}
			\fmfleft{i}
			\fmfright{o}
			\fmf{dashes}{i,v}
			\fmf{dashes}{o,v}
			\fmfblob{17}{v}
		\end{fmfgraph*}
	\end{gathered}
	\;
	\right)_{\!\!c_\infty,\,X}
	+\quad
	\left( \;
	\begin{gathered}
	\vspace{-4pt}
	\begin{fmfgraph*}(50,20)\fmfkeep{scalar-self-ct}
		\fmfleft{i}
		\fmfright{o}
		\fmf{dashes}{i,v}
		\fmf{dashes}{o,v}
		\defotimes
		\fmfv{d.sh=otimes,d.f=empty}{v}
	\end{fmfgraph*}
	\end{gathered}
	\;
	\right)_{\!\! \delta\tilde\lambda(X)} =
%%%%
	0 \quad \text{for} \quad X = \ell^\pm, \xi^0.
\end{equation}
Moreover, 
\end{fmffile}%
\begin{equation}
\left( \;
\begin{gathered}
\vspace{-4pt}
\fmfreuse{scalar-self-amps}
\end{gathered}
\;
\right)_{\!\! \xi^2}
= \;
\left( \;
\begin{gathered}
\vspace{-4pt}
\fmfreuse{scalar-self-ct}
\end{gathered}
\;
\right)_{\!\! \delta\tilde\lambda(\xi^2)} =
0.
\end{equation}
Therefore, in these cases we obtain
\begin{equation}
\delta\mu^2_{ij}(\ell^\pm) = \delta\mu^2_{ij}(\xi^0) = \delta\mu^2_{ij}(\xi^2) = 0.
\end{equation}
However, in the cases of $X= S, \chi, \xi^1$  the counterterm parameter
$\delta\mu^2_{ij}(X)$ is non-trivial. From 
\begin{equation}
\left( \;
\begin{gathered}
\vspace{-4pt}
\fmfreuse{scalar-self-amps}
\end{gathered}
\;
\right)_{\!\! c_\infty, X}
+ \;
\left( \;
\begin{gathered}
\vspace{-4pt}
\fmfreuse{scalar-self-ct}
\end{gathered}
\;
\right)_{\!\! \delta\tilde\lambda(X)} 
+ \;
\left( \;
\begin{gathered}
\vspace{-4pt}
\fmfreuse{scalar-self-ct}
\end{gathered}
\;
\right)_{\!\! \delta\mu^2(X)}
= 0 \quad \text{for} \quad X = S, \chi,
\end{equation}
we compute
\begin{equation}
\delta\mu^2_{ij}(S) = 
\frac{2}{16\pi^2}\,c_\infty \left( 2\lambda_{ijkl} + \lambda_{ilkj} \right)
\mu^2_{lk}
\end{equation}
and
\begin{equation}
\delta\mu^2_{ij}(\chi) = 
-\frac{2}{16\pi^2}\,c_\infty \mbox{Tr} 
\left( M_R M_R^\dagger \Delta_j \Delta_i^\dagger \right).
\end{equation}
It is amusing to notice that the latter equation 
is the only instance where $M_R$, the mass matrix 
of the right-handed neutrino singlets, appears in a counterterm.

The linear $\xi$-terms need a special treatment and we will be very detailed
in their discussion. Our aim is to determine the remaining counterterm 
parameters $\delta\mu^2_{ij}(\xi^1)$ and $\delta v_k$ from the divergencies 
linear in $\xi$ of the scalar self-energy. 
In order to streamline the notation, we define
\begin{equation}\label{A1}
A_1 = \frac{c_\infty}{16 \pi^2} 
\left( \frac{g^2 \xi_W}{4} + \frac{g^2 \xi_Z}{8c_w^2} \right),
\end{equation}
where the index~1 indicates linearity in $\xi$. 
In terms of this quantity, the sum over all divergencies linear in $\xi$ 
of the scalar self-energy---see section~\ref{divergencies-neutral-scalar}---can 
be written as
\begin{equation}
\left( \;
\begin{gathered}
\vspace{-4pt}
\fmfreuse{scalar-self-amps}
\end{gathered}
\;
\right)_{\!\!c_\infty,\,\xi^1} =
-i \left( V_{ib}^* V_{jb'} + V_{ib'}^* V_{jb} \right) \mu^2_{ij} A_1.
\end{equation}
Now we turn to the counterterm of equation~(\ref{2pf-ct})
and discuss the various contributions linear in $\xi$.
It is easy to see from equations~(\ref{dlambda}) and~(\ref{4pf-dlambda}) 
that $\delta\tilde\lambda(\xi^1)_{ijkl}$, 
the part proportional to $\xi^1$ of $\delta\tilde\lambda_{ijkl}$, can 
be written in terms of $A_1$ as well:
\begin{equation}\label{xi-lambda}
\delta\tilde\lambda(\xi)_{ijkl} = -4A_1 \tilde\lambda_{ijkl}.
\end{equation}
Plugging this expression into the counterterm formula 
of equation~(\ref{2pf-ct1}) and using equation~(\ref{useful form}), 
we obtain
  \begin{equation}
  \left( 
    \begin{gathered}
    \vspace{-4pt}
    \fmfreuse{scalar-self-ct}
    \end{gathered}
    \right)_{\!\! \delta\tilde\lambda(\xi^1)} =
    \;
    4iA_1 \delta_{bb'} M_{b'}^2 
	-2i \left( V_{ib}^* V_{jb'} + V_{ib'}^* V_{jb} \right) \mu^2_{ij} A_1.
  \end{equation}
The remaining terms linear in $\xi$ in equation~(\ref{2pf-ct})
consist of the $\delta\mu^2_{ij}(\xi^1)$-part of
equation~(\ref{2pf-ct1}) and the counterterm induced by $\delta v_i$,
equation~(\ref{2pf-ct2}), and contain thus the parameters we want to 
determine.

Adding up all terms linear in $\xi$, divergence and the three 
counterterm contributions, we have
\begin{fmffile}{scalar-self-divs-cts}
\fmfset{thin}{.7pt}
\fmfset{dash_len}{1.5mm}
\begin{subequations}\label{scalar-div+cts}
\begin{align}
  \left(\;
    \begin{gathered}
    \vspace{-4pt}
      \begin{fmfgraph*}(50,20)
	\fmfleft{i}
	\fmfright{o}
	\fmf{dashes}{i,v}
	\fmf{dashes}{o,v}
	\fmfblob{17}{v}
      \end{fmfgraph*}
    \end{gathered}
    \;
    \right)_{\!\! c_\infty, \, \xi^1}
    & + 
    \;
    \left(\;
    \begin{gathered}
    \vspace{-4pt}
      \fmfreuse{scalar-self-ct}
    \end{gathered}
    \;
    \right)_{\!\!\xi^1}
    \nonumber
    \\[1em]
    =
    &
    -i \left( V_{ib}^* V_{jb'} + V_{ib'}^* V_{jb} \right) \mu^2_{ij} A_1
    \label{muA1}
    \\
    &
    +4iA_1 \delta_{bb'} M_{b'}^2 
    -2i \left( V_{ib}^* V_{jb'} + V_{ib'}^* V_{jb} \right) \mu^2_{ij} A_1 
    \label{4A1}
    \\
    &
    -\frac{i}{2} \left( V_{ib}^* V_{jb'} + V_{ib'}^* V_{jb} \right) 
    \delta\mu^2_{ij}(\xi^1)
    \\
    &\nonumber
    -\frac{i}{2} \tilde\lambda_{ijkl} 
    \left( \delta v_k^* v_l + v_k^* \delta v_l \right) 
    \left( V_{ib}^* V_{jb'} + V_{ib'}^* V_{jb} \right) 
    \\ &
    -\frac{i}{4} \tilde\lambda_{ijkl} \left[
    \left( \delta v_j v_l + v_j \delta v_l  \right) V_{ib}^* V_{kb'}^* +
    \left( \delta v_i^* v_k^* + v_i^* \delta v_k^* \right) V_{jb} V_{lb'}
    \right].
    \label{dv-ct}
    \end{align}
  \end{subequations}
\end{fmffile}%
We know that the sum of these terms must be zero.
Since in equations~(\ref{muA1}) and~(\ref{4A1}) these gauge parameters only 
occur in $A_1$ and taking into account that $\delta v_k$ is linear in 
$\xi_W$ and $\xi_Z$, 
we are lead to the ansatz $\delta v_k = c A_1 v_k$, 
where $c$ is a constant to be 
determined by the cancellation of the divergencies.
Plugging this ansatz into equation~(\ref{dv-ct}) 
(last two lines of equation~(\ref{scalar-div+cts}))
and using equation~(\ref{useful form}), 
after some computation these two lines are rewritten as
\begin{equation}\label{ddv}
-icA_1 \left[ 2 \delta_{bb'} M_{b'}^2 - 
\left( V_{ib}^* V_{jb'} + V_{ib'}^* V_{jb} \right) \mu^2_{ij} \right].
\end{equation}
Obviously, with $c = 2$ or
\begin{equation}\label{dv}
\delta v_k = 2 A_1 v_k
\end{equation}
the $A_1$-part of the counterterm induced by $\delta v_k$, 
equation~(\ref{ddv}), 
cancels the term $4iA_1\delta_{bb'} M_{b'}^2$ of equation~(\ref{4A1}). 

Having thus determined $\delta v_k$, we consider 
the sum of the remaining terms in equation~(\ref{scalar-div+cts}) 
which amounts to
\begin{equation}
-i \left( V_{ib}^* V_{jb'} + V_{ib'}^* V_{jb} \right) \mu^2_{ij} A_1
-\frac{i}{2} \left( V_{ib}^* V_{jb'} + V_{ib'}^* V_{jb} \right) 
\delta\mu^2_{ij}(\xi^1).
\end{equation}
Thus we find that
\begin{equation}\label{dmuxi}
\delta \mu^2_{ij}(\xi^1) = -2 A_1 \mu^2_{ij}
\end{equation}
together with $\delta v_k$ of equation~(\ref{dv}) induce counterterms which 
cancel the terms linear in $\xi$ in the scalar self-energy.
With this, we have finally determined the complete set of parameters, 
$\delta\mu^2_{ij}$, $\delta\tilde\lambda_{ijkl}$ and $\delta v_k$
that make the scalar self-energy finite.

\subsection{Finiteness of the scalar one-point function}
\label{finiteness-1pf}
\begin{figure}[ht]
\begin{fmffile}{vector-ghost-tadpole}
\fmfset{thin}{.7pt}
\fmfset{dash_len}{1.5mm}
\fmfset{wiggly_len}{2mm}
\fmfset{wiggly_slope}{75}
\fmfset{dot_len}{.8mm}
\fmfset{dot_size}{1.5thick}
\begin{center}
\begin{subfigure}[t]{.3\textwidth} \centering
\begin{fmfgraph*}(80,80)
\fmftop{i1}
\fmfbottom{b1,b2}
\fmf{phantom}{b1,v2,b2}
\fmffreeze
\fmf{phantom}{i1,v1,v2}
\fmf{dashes}{v2,v1}
\fmf{wiggly,tension=.5,left}{v1,i1,v1}
\end{fmfgraph*}
\caption{}
\end{subfigure}
\begin{subfigure}[t]{.3\textwidth} \centering
\begin{fmfgraph*}(80,80)
\fmftop{i1}
\fmfbottom{b1,b2}
\fmf{phantom}{b1,v2,b2}
\fmffreeze
\fmf{phantom}{i1,v1,v2}
\fmf{dashes}{v2,v1}
\fmf{dots,tension=.5,left}{v1,i1,v1}
\end{fmfgraph*}
\caption{}
\end{subfigure}
\end{center}
\end{fmffile}
\caption{The tadpole diagrams involving the vector bosons and 
the ghost fields.}
\label{tp-vb-gh}
\end{figure}
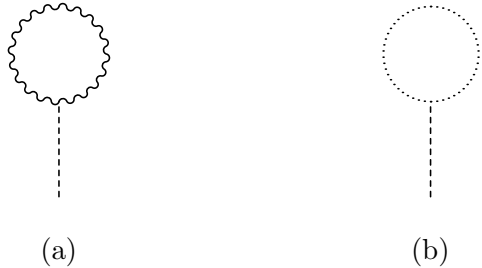
\noindent
As in the case of the counterterm of equation~(\ref{1pf-ct}),
we consider the ``truncated'' one-point function of $S^0_b$, where the external 
propagator $i/(-M_b^2)$ and the factor $\mc^{-\varepsilon/2}$ are 
removed. Also we emphasize that the scalar one-point function referring to
$S^0_1 \equiv G^0$ is 
zero---see discussion after equation~(\ref{TCt}) in section~\ref{VEV shift}. 
Thus we consider the truncated one-point functions
of $S^0_b$ with $b = 2, \ldots, 2n_H$.
The counterterms for the one-point function induced by 
$\delta \mu^2_{ij}$, $\delta\tilde\lambda_{ijkl}$ and $\delta v_i$ are
obtained by application of equation~(\ref{1pf-ct}).

\paragraph{Vector boson and ghost loops:}
The tadpole diagrams involving the vector boson and 
ghost loops---see figure~\ref{tp-vb-gh}---deserve a 
special treatment because the ghost loops cancel exactly that part of the 
vector boson loops deriving from the gauge-dependent part of 
the propagator in equation~(\ref{vbp1})~\cite{weinberg}.
Concretely, we are going to demonstrate that the $c_Z$ loop cancels the 
$\xi_Z$-dependent contribution of the $Z$ propagator and the
$c^+$ and $c^-$ loops cancel the $\xi_W$-dependent contribution of 
the $W$ propagator. 

First we consider the $Z$ and $c_Z$ loops in diagrams~(a) and~(b) of 
figure~\ref{tp-vb-gh}, respectively. According to the 
Lagrangians~(\ref{vb-s-non}) 
and~(\ref{cz}), 
we obtain the loop integrals
\begin{eqnarray}
&&
\int \frac{\dd^dk}{(2\pi)^d} \left\{
\frac{ig^2}{8c_w^2} \left( v_i^* V_{ib} + V_{ib}^* v_i \right)
\frac{k^2}{m_Z^2} \frac{-i}{k^2 - \xi_Z m_Z^2 + i\epsilon} 
\right. \nonumber \\ && \left.
+
(-i) \frac{g\xi_Z m_Z}{2c_w} \,\mbox{Re} \left( \omega_i V_{ib} \right)
\frac{-i}{k^2 - \xi_Z m_Z^2 + i\epsilon}
\right\}.
\label{int-z+cz}
\end{eqnarray}
Note that the minus sign in the numerator of the ghost propagator 
takes into account the anticommuting nature of the ghost fields.
Now we use 
\begin{equation}
\frac{k^2}{k^2 - \xi_Z m_Z^2} = 1 + \frac{\xi_Z m_Z^2}{k^2 - \xi_Z m_Z^2}
\quad \mbox{and} \quad 
\int \frac{\dd^dk}{(2\pi)^d}\, 1 = 0,
\end{equation}
the formula for the $Z$ mass of equation~(\ref{mV}),
the definition of $\omega_k$ in 
equation~(\ref{omega}), 
and the relation
\begin{equation}
v_i^* V_{ib} + V_{ib}^* v_i = 2v\,\mbox{Re} \left( \omega_i^* V_{ib} \right).
\end{equation}
It is then easy to see that the integral~(\ref{int-z+cz}) is zero.

In the case of the $W$ boson in diagram~(a) and the ghost fields $c^+$ and 
$c^-$ in diagram~(b) of figure~\ref{tp-vb-gh}, Lagrangians~(\ref{vb-s-non}) 
and~(\ref{c+-}) lead to the loop integral
\begin{eqnarray}
&&
\int \frac{\dd^dk}{(2\pi)^d} \left\{
\frac{ig^2}{4} \left( v_i^* V_{ib} + V_{ib}^* v_i \right)
\frac{k^2}{m_W^2} \frac{-i}{k^2 - \xi_W m_W^2 + i\epsilon} 
\right. \nonumber \\ && \left.
+ (-i)
\frac{g\xi_W m_W}{2} \left( \omega_i^* V_{ib} + \omega_i V_{ib}^* \right)
\frac{-i}{k^2 - \xi_Z m_W^2 + i\epsilon}
\right\}
\end{eqnarray}
With the same arguments as before we conclude that it is zero.

\paragraph{Divergencies of the Goldstone boson loops and 
the $\xi$-dependent terms:}
The $\xi$-dependent divergencies of the tadpoles come from the 
Goldstone-boson loops. The relevant Lagrangian is displayed in 
equation~(\ref{sgg}). Using the identity 
\begin{equation}
\mbox{Im} \left( V^\dagger V \right)_{1b} = -\frac{1}{2v}
\left( v_i^* V_{ib} + V_{ib}^* v_i \right),
\end{equation}
these are given by
\vspace{10pt}
\begin{fmffile}{tadpole-G0}
\fmfset{thin}{.7pt}
\fmfset{dash_len}{1.5mm}
\begin{equation}
  \left(\;\;
    \begin{gathered}
    \label{1pf-div(xi)}
    \vspace{-10pt}
      \begin{fmfgraph*}(25,30)
	\fmfbottom{b}
	\fmftop{t}
	\fmf{phantom}{b,v,t}
	\fmf{dashes}{b,v}
	\fmf{dashes,left}{v,t}
	\fmf{dashes,left}{t,v}
	\fmfv{label=\scriptsize{$G^\pm/G^0$},l.a=75,l.d=2.5}{t}
      \end{fmfgraph*}
    \end{gathered}
    \vphantom{\begin{minipage}[t][.6cm]{0pt}\end{minipage}}
    \;\;
    \right)_{\!\! c_\infty,\, \xi^1}
    = 
    \;
    \frac{i}{2} \left( v_i^* V_{ib} + V_{ib}^* v_i \right) M_b^2 A_1.
\end{equation}
\end{fmffile}%
The counterterm of equation~(\ref{1pf-ct1}) 
associated with $\delta\tilde\lambda_{ijkl}(\xi^1)$ of 
equation~(\ref{Vvc}) can be put into the form 
\begin{fmffile}{scalar-tadpole-ct-lambda}
\fmfset{thin}{.7pt}
\fmfset{dash_len}{1.5mm}
\vspace{5pt}
  \begin{equation}
  \label{dlxi}
  \left( \;
     \begin{gathered}
      \vspace{-8pt}
      \begin{fmfgraph*}(25,25)\fmfkeep{scalar-tadpole-ct}
	\fmfbottom{b}
	\fmftop{t}
	\fmf{dashes}{b,t}
	\defotimes
	\fmfv{d.sh=otimes,d.f=empty}{t}
      \end{fmfgraph*}
    \end{gathered}
    \vphantom{\begin{minipage}[t][.5cm]{0pt}\end{minipage}}
    \;
    \right)_{\!\! \delta\tilde{\lambda}(\xi^1)}
    =
    \frac{i}{2} \left( v_i^* V_{ib} + V_{ib}^* v_i \right) M_b^2 A_1 -
    i \left( v_i^* V_{jb} + V_{ib}^* v_j \right) \mu^2_{ij} A_1.
  \end{equation}
\end{fmffile}%
Now we add up all terms of the scalar one-point function proportional to 
$\xi^1$, \textit{i.e.}\ 
equations~(\ref{1pf-div(xi)}) and~(\ref{dlxi}) and the remaining 
counterterms of equation~(\ref{1pf-ct}), namely those with 
$\delta\mu^2_{ij}(\xi^1)$ and $\delta v_i$:
\begin{fmffile}{scalar-tadpole-condition}
\fmfset{thin}{.7pt}
\fmfset{dash_len}{1.5mm}
\vspace{5pt}
  \begin{subequations}\label{tadpole-div+cts}
  \begin{align}
  \left(\;
     \begin{gathered}
      \vspace{-8pt}
      \begin{fmfgraph*}(25,25)\fmfkeep{tadpole-divs}
	\fmfbottom{b}
	\fmftop{t}
	\fmf{dashes}{b,t}
	\fmfblob{17}{t}
      \end{fmfgraph*}
    \end{gathered}
    \vphantom{\begin{minipage}[t][.5cm]{0pt}\end{minipage}} %i
    \;
\right)_{\!\! c_\infty,\,\xi^1} %i
    +
\left(
    \;
    \begin{gathered}
      \vspace{-8pt}
      \fmfreuse{scalar-tadpole-ct}
    \end{gathered}
    \vphantom{\begin{minipage}[t][.5cm]{0pt}\end{minipage}}
    \;
    \right)_{\!\!\xi^1}
    =& \quad    
    \frac{i}{2} \left( v_i^* V_{ib} + V_{ib}^* v_i \right) M_b^2 A_1
    \label{tadpole-div}
    \\
    &
    +\frac{i}{2} \left( v_i^* V_{ib} + V_{ib}^* v_i \right) M_b^2 A_1 -
    i \left( v_i^* V_{jb} + V_{ib}^* v_j \right) \mu^2_{ij} A_1
    \label{tadpole-ct1}
    \\
    &
    -\frac{i}{2} \left( v_i^* V_{jb} + V_{ib}^* v_j \right) 
    \delta \mu^2_{ij}(\xi^1)
    \label{tadpole-ct2}
    \\
    &
    -\frac{i}{2} \left( \delta v_i^* V_{ib} + V_{ib}^* \delta v_i \right) M_b^2.
    \label{tadpole-ct3}
    \end{align}
  \end{subequations}
\end{fmffile}%
Taking $\delta v_i$ from equation~(\ref{dv}) and 
$\delta\mu^2_{ij}(\xi^1)$ 
from equation~(\ref{dmuxi}), the terms in equation~(\ref{tadpole-div+cts}) 
add up to zero. Note that the blob in equation~(\ref{tadpole-div+cts}) 
is identical with the Goldstone loops in equation~(\ref{1pf-div(xi)}) because,
as explained in the beginning of this subsection, the $\xi$-dependence of
vector boson propagators is cancelled by the ghost loops.

\paragraph{The remaining tadpole diagrams:}
A tedious but straightforward computation de\-mon\-strates that
the $\xi$-independent divergencies of the tadpole diagrams are cancelled 
by the counterterm~(\ref{1pf-ct1}) by plugging in the expressions for
$\delta\mu^2_{ij}(X)$ and 
$\delta\tilde\lambda_{ijkl}(X)$
with $X = S,\ell^\pm,\chi,\xi^0$.

Summarizing, we have found that the counterterms determined by 
$\overline{\mbox{MS}}$ renormalization of 
the scalar four-point function and 
the scalar self-energy 
make the scalar one-point function finite. 

\subsection{The VEV shift \texorpdfstring{$\Delta v_i$}{Delta vi} %
and the tadpoles}
\label{VEV shift}
Using equation~(\ref{columns}), we find that 
a finite VEV shift $\Delta v_i$ induces the term 
\begin{equation}\label{V-shift}
-\mc^{-\varepsilon/2} \frac{1}{2} M_b^2 \left( 
\Delta v_i^* V_{ib} + V_{ib}^* \Delta v_i \right) S^0_b 
\equiv -\mc^{-\varepsilon/2} \Delta t_b S^0_b
\end{equation}
in the scalar potential. As announced in section~\ref{introduction}, 
we will now show that it is 
possible to choose $\Delta v_i$ such that the scalar one-point 
function is zero at the one-loop level. 
There are three contributions to the truncated one-point 
function: the loop integrals $-iT_b$, the sum of the 
counterterms~(\ref{tadpole-ct1})--(\ref{tadpole-ct3}) 
denoted by $-i\mathcal{C}_b$, 
and the contribution of equation~(\ref{V-shift}). Thus we require,
in order to achieve a vanishing one-point function,
\begin{fmffile}{scalar-tadpole-condition-full}
\fmfset{thin}{.7pt}
\fmfset{dash_len}{1.5mm}
\vspace{5pt}
\begin{equation}\label{TCt}
    \begin{gathered}
      \fmfreuse{tadpole-divs}
    \end{gathered}
    \;
    +
    \;
    \begin{gathered}
      \fmfreuse{scalar-tadpole-ct}
    \end{gathered}
    \;
    +
    \;
     \begin{gathered}
      \begin{fmfgraph*}(25,25)\fmfkeep{scalar-tadpole-vev-shift}
	\fmfbottom{b}
	\fmftop{t}
	\fmf{dashes}{b,t}
	\fmfv{decor.shape=triangle,decor.filled=shaded,decor.size=12}{t}
      \end{fmfgraph*}
    \end{gathered}
    \;
    = 
    \;    
    -i\left(T_b + \mathcal{C}_b + \Delta t_b \right)\equiv 0
    \quad \mbox{or} \quad
    \Delta t_b = - T_b - \mathcal{C}_b.
  \end{equation}
\end{fmffile}%
The triangle represents the term $-i \Delta t_b$ induced by 
equation~(\ref{V-shift}).

Before we derive a formula for VEV shift $\Delta v_k$, 
let us dwell a little bit on the one-point function of 
$G^0$. That $G^0$ is an unphysical field is suggestive of its vanishing. 
This is indeed borne out by explicit one-loop considerations: 
all couplings of $G^0$ to bosons, including the ghost field $c_Z$, 
vanish, the $c^+$ and $c^-$ loops cancel each other exactly, and the fermion 
loops give zero when the trace is taken in flavour and Dirac 
space. Concerning the 
counterterms~(\ref{1pf-ct}) and taking into account equation~(\ref{dv}),
we see that all counterterms of the one-point function of $G^0$ 
contain the factor $v_i^* V_{i1} + V_{i1}^* v_i$.
Since $V_{i1} = iv_i/v$---see equation~(\ref{GUV}), this factor obviously 
vanishes.

To proceed further, some remarks are in order:
\begin{enumerate}
\renewcommand{\labelenumi}{\roman{enumi}.}
\item
From the considerations in section~\ref{finiteness-1pf} we know that 
the scalar one-point functions, \textit{i.e.}\ the quantities 
$T_b + \mathcal{C}_b = -\Delta t_b$ are finite, 
thus the $\Delta v_i$ are finite as well.\footnote{
 We remind the reader that we use 
 the symbol ``$\Delta$,'' occurring in $\Delta v_k$ and $\Delta t_b$, 
 for finite quantities.}
 Diagrammatically, this can be expressed as
\begin{fmffile}{scalar-tadpole-condition-finite}
\fmfset{thin}{.7pt}
\fmfset{dash_len}{1.5mm}
\vspace{5pt}
  \begin{equation*}
    \begin{gathered}
    \vspace{-8pt}
      \fmfreuse{scalar-tadpole-vev-shift}
    \end{gathered}
    \; = \;
    -\left(
    \begin{gathered}
    \vspace{-8pt}
      \fmfreuse{tadpole-divs}
    \end{gathered}
    \vphantom{\begin{minipage}[t][.5cm]{0pt}\end{minipage}}
    \right)_{\!\! \text{finite}},
  \end{equation*}
\end{fmffile}%
where the subscript ``finite'' indicates that all terms 
proportional to $c_\infty$ have been subtracted.
\item
Since $\Delta t_b$ belongs to the real scalar field $S^0_b$,
this quantity must be real.
\item 
In the quantity $\Delta t_b$,
the masses $M_b$ derive 
from the mass matrix of the neutral scalars 
where the Goldstone boson has zero mass, while the Goldstone mass-squared 
$\xi_Z m_Z^2$ derives from the $R_\xi$-gauge condition and occurs only 
in the propagator.
\end{enumerate}
We can summarize this discussion in the following way:
\begin{equation}
\Delta t_1 = 0, \quad \left( \Delta t_b \right)^* = \Delta t_b,
\quad \mbox{and} \quad 
\frac{1}{2} \left( \Delta v_i^* V_{ib} + V_{ib}^* \Delta v_i \right) = 
\frac{\Delta t_b}{M_b^2} 
\quad \mbox{for} \quad b=2,\ldots,2n_H.
\end{equation}

Eventually, our aim is to obtain $\Delta v_k$ from the $\Delta t_b$,
but there is the obstacle that $\Delta v_i^* V_{i1} + V_{i1}^* \Delta v_i$ 
is not determined, because it is multiplied by $M_1^2 = 0$ in $\Delta t_1$.
However, as we will shortly see, the only consistent value of this quantity is
\begin{equation}\label{stronger}
\Delta v_i^* V_{i1} + V_{i1}^* \Delta v_i = 0.
\end{equation}
Taking this relation into account,
the first two relations of equation~(\ref{orthV}) 
allow to derive the VEV shifts 
\begin{equation}\label{dvk}
\Delta v_k = \sum_{b=2}^{2n_H} \frac{\Delta t_{b}}{M_{b}^2}\, V_{kb}.
\end{equation}
This means that it is indeed possible to make the scalar 
one-point function vanish by a finite VEV shift $\Delta v_k$.

Let us check now that equation~(\ref{dvk}) is indeed consistent with 
equation~(\ref{stronger}). We plug the result for $\Delta v_k$ 
into equation~(\ref{stronger}) and utilize the third relation in 
equation~(\ref{orthV}). In this way we obtain
\begin{equation*}
\Delta v_i^* V_{i1} + V_{i1}^* \Delta v_i = 
\sum_{b=2}^{2n_H} \frac{\Delta t_{b}}{M_{b}^2} 
\left( V_{ib}^* V_{i1} + V_{i1}^* V_{ib} \right) = 
2\sum_{b=2}^{2n_H} \frac{\Delta t_{b}}{M_{b}^2} \, \delta_{b1} = 0,
\end{equation*}
which is the desired result.

Now we want to demonstrate that the insertion of all tadpole contributions,
including the tadpole counterterm, 
on a fermion line is equivalent to make the shift
$v_k \to v_k + \Delta v_k$ in the mass term of the 
respective fermion in the Lagrangian. We begin with the charged-lepton lines. 
The corresponding expression for these diagrams is 
\begin{fmffile}{fermion-tadpoles-ct}
\fmfset{thin}{.7pt}
\fmfset{dash_len}{1.5mm}
\fmfset{arrow_len}{2.5mm}
\vspace{5pt}
  \begin{equation}\label{cl-tp}
      \begin{gathered}
      \vspace{-4pt}
      \begin{fmfgraph*}(50,35)\fmfkeep{fermion-tadpole-amps}
	\fmfleft{i}
	\fmftop{t}
	\fmfright{o}
	\fmf{fermion}{o,v,i}
	\fmffreeze
	\fmf{dashes}{v,t}
	\fmfblob{17}{t}
      \end{fmfgraph*}
    \end{gathered}
    \;
    +
    \;
    \begin{gathered}
      \vspace{-4pt}
      \begin{fmfgraph*}(50,35)\fmfkeep{fermion-tadpole-ct}
	\fmfleft{i}
	\fmftop{t}
	\fmfright{o}
	\fmf{fermion}{o,v,i}
	\fmffreeze
	\fmf{dashes}{v,t}
	\defotimes
	\fmfv{d.sh=otimes,d.f=empty}{t}
      \end{fmfgraph*}
    \end{gathered}
    \; = \;
    -\frac{i}{\sqrt{2}} \mc^{\varepsilon/2} \sum_{b=2}^{2n_H} 
    \left[ G_b \gamma_L + G_b^\dagger \gamma_R \right] \times
    \frac{i}{-M_b^2} \times 
    (-i) \mc^{-\varepsilon/2} \left( T_b + \mathcal{C}_b \right).
   \end{equation}
\end{fmffile}%
Replacing $T_b + \mathcal{C}_b$ by $-\Delta t_b$, \textit{c.f.}\ 
equation~(\ref{TCt}), and using the expression for $\Delta t_b$ given
in equation~(\ref{V-shift}), we obtain
\begin{fmffile}{fermion-tadpoles-vev-shift}
\fmfset{thin}{.7pt}
\fmfset{dash_len}{1.5mm}
\fmfset{arrow_len}{2.5mm}
\vspace{5pt}
  \begin{equation}
      \begin{gathered}
      \vspace{-4pt}
      \begin{fmfgraph*}(50,35)\fmfkeep{fermion-tadpole-vev-shift}
	\fmfleft{i}
	\fmftop{t}
	\fmfright{o}
	\fmf{fermion}{o,v,i}
	\fmffreeze
	\fmf{dashes}{v,t}
	\fmfv{decor.shape=triangle,decor.filled=shaded,decor.size=12}{t}
      \end{fmfgraph*}
    \end{gathered}
    \; = \;
    -\frac{i}{\sqrt{2}} \sum_{b=2}^{2n_H} 
    \left[ G_b \gamma_L + G_b^\dagger \gamma_R \right]
    \times \frac{i}{-M_b^2} \times 
    \frac{i}{2} M_b^2 \left( 
    \Delta v_i^* V_{ib} + V_{ib}^* \Delta v_i \right). 
   \end{equation}
\end{fmffile}%
Here the scalar squared masses $M_b^2$ cancel. Taking into account 
equation~(\ref{stronger}), we can---after this cancellation---take 
the sum from $b=1$ to $2n_H$.\footnote{In this context, 
 equation~(\ref{stronger}) means that the Goldstone boson, \textit{i.e.}\ 
 $b=1$ does not contribute, which we know already from an earlier discussion in
 the present subsection. This is further evidence that  
 equation~(\ref{stronger}) is correct.}
Since $G_b$ contains the factor $V_{kb}^*$---see equation~(\ref{Gb}),
we apply the first two relations in equation~(\ref{orthV}). This 
gives the final form of equation~(\ref{cl-tp}), namely
\begin{fmffile}{fermion-vev-shift}
\fmfset{thin}{.7pt}
\fmfset{dash_len}{1.5mm}
\fmfset{arrow_len}{2.5mm}
\vspace{5pt}
  \begin{equation}\label{ell-shift}
      \begin{gathered}
      \vspace{-4pt}
      \begin{fmfgraph*}(50,35)\fmfkeep{fermion-vev-shift}
	\fmfleft{i}
	\fmfright{o}
	\fmf{fermion}{o,v,i}
	\fmfv{decor.shape=triangle,decor.filled=shaded,decor.size=12}{v}
      \end{fmfgraph*}
    \end{gathered}
    \; = \;
    -\frac{i}{\sqrt{2}} \left( W_R^\dagger \Gamma_k \Delta v_k^* W_L \gamma_L + 
    W_L^\dagger \Gamma_k^\dagger \Delta v_k W_R \gamma_R \right).
   \end{equation}
\end{fmffile}%
This expression has exactly the form of the contribution of a mass term to the
fermion self-energy, however, with $\Delta v_k$ instead
of $v_k$. We have thus proven the above statement. 

Now we turn to neutrino lines. We proceed as before and obtain the 
intermediate expression 
\begin{eqnarray}
\lefteqn{
2 \times \left( -\frac{i}{\sqrt{2}} \right) 
\mc^{\varepsilon/2} \sum_{b=2}^{2n_H} 
\left[ F_b \gamma_L + F_b^\dagger \gamma_R \right] \times
\frac{i}{-M_b^2} \times 
(-i) \mc^{-\varepsilon/2} \left( T_b + \mathcal{C}_b \right) }
\nonumber \\ &=&
-i\sqrt{2} \sum_{b=2}^{2n_H} 
\left[ F_b \gamma_L + F_b^\dagger \gamma_R \right]
\times \frac{i}{-M_b^2} \times 
\frac{i}{2} M_b^2 \left( 
\Delta v_i^* V_{ib} + V_{ib}^* \Delta v_i \right).
\end{eqnarray}
Note that the factor~2 comes about due to the 
Majorana nature of 
the neutrinos. Then we plug in $F_b$ of equation~(\ref{Fb}) and employ again 
equation~(\ref{orthV}). We finally arrive at the expression
\begin{equation}\label{nu-shift}
- \frac{i}{\sqrt{2}} \left\{ 
\left( U_R^\dagger \Delta_k U_L + U_L^T \Delta_k^T U_R^* \right) 
\Delta v_k \gamma_L +
\left( U_L^\dagger \Delta_k^\dagger U_R + U_R^T \Delta_k^* U_L^* \right) 
\Delta v_k^* \gamma_R \right\}
\end{equation}
This contribution to $-i \Sigma$ corresponds precisely to a term 
\begin{equation}
-\frac{1}{\sqrt{2}} \bar\chi \left( U_R^\dagger \Delta_k U_L \,
\Delta v_k \gamma_L + 
U_L^\dagger \Delta_k^\dagger U_R\, \Delta v_k^* \gamma_R \right) \chi
\end{equation}
in the Lagrangian---compare with the $M_D$ term in equation~(\ref{Lmass}) 
after utilizing equation~(\ref{mass-eigen}).
This proves that taking into account the tadpole diagrams on the neutrino lines 
corresponds to making a finite VEV shift in the Dirac-type 
neutrino mass term in the Lagrangian.

\section{Gauge-parameter independence of the one-loop fer\-mion masses}
\label{gauge-invariance}
\subsection{Two decompositions of the fermion self-energy}
\label{two decomp}
We denote by $\Sigma$ the \emph{renormalized} fermion self-energy. It 
can be decomposed as 
\begin{equation}\label{sigma1}
\Sigma(p) = \slashed{p} \left( \Sigma^{(A)}_L(p^2) \gamma_L + 
\Sigma^{(A)}_R(p^2) \gamma_R \right) + 
\Sigma^{(B)}_L(p^2) \gamma_L + \Sigma^{(B)}_R(p^2) \gamma_R.
\end{equation}
For definiteness we use the index $i$ 
for the fermion masses 
in this subsection though, in the light 
of our notation convention laid out in section~\ref{introduction}, for 
charged fermions we should be using $\alpha$ instead. For $n$ fermions the 
quantities $\Sigma^{(A)}_L$, $\Sigma^{(A)}_R$, $\Sigma^{(B)}_L$, $\Sigma^{(B)}_R$
are $n \times n$ matrices that fulfill the matrix relations
\begin{equation}\label{sigma-h}
\left( \Sigma^{(A)}_L \right)^\dagger = \Sigma^{(A)}_L, \quad
\left( \Sigma^{(A)}_R \right)^\dagger = \Sigma^{(A)}_R, \quad
\left( \Sigma^{(B)}_L \right)^\dagger = \Sigma^{(B)}_R.
\end{equation}
Strictly speaking these relations hold only for the dispersive 
part of the self-energy. 
In the case of Majorana fermions, one has in addition 
\begin{equation}\label{sigma-t}
\left( \Sigma^{(A)}_L \right)^T = \Sigma^{(A)}_R, \quad
\left( \Sigma^{(B)}_L \right)^T = \Sigma^{(B)}_L, \quad
\left( \Sigma^{(B)}_R \right)^T = \Sigma^{(B)}_R.
\end{equation}
If there are no degeneracies in the tree-level masses $m_i$, 
the diagonal elements of the coefficient matrices in equation~(\ref{sigma1}) 
allow to express---at lowest non-trivial order---the mass shifts as 
\begin{equation}\label{dm1}
\Delta m_i = \frac{1}{2} \left\{
m_i \left[ \left( \Sigma^{(A)}_L \right)_{ii} (m_i^2) + 
\left( \Sigma^{(A)}_R \right)_{ii} (m_i^2) \right] + 
\left( \Sigma^{(B)}_L \right)_{ii} (m_i^2) + 
\left( \Sigma^{(B)}_R \right)_{ii} (m_i^2) \right\}.
\end{equation}
Therefore, 
the pole masses, comprising tree-level plus radiative corrections, are 
given by
\begin{equation}
\mt{i} = m_i + \Delta m_i.
\end{equation}

As we will see, another useful decomposition of $\Sigma$ is given by
\begin{eqnarray}
\Sigma &=& A_L \gamma_L + A_R \gamma_R +
\left( \slashed{p} - \hat m \right) 
\left( B^{(r)}_L \gamma_L + B^{(r)}_R \gamma_R \right) + 
\left( B^{(l)}_L \gamma_L + B^{(l)}_R \gamma_R \right) 
\left( \slashed{p} - \hat m \right) 
\nonumber \\ && + 
\left( \slashed{p} - \hat m \right) 
\left( C_L \gamma_L + C_R \gamma_R \right)
\left( \slashed{p} - \hat m \right),
\label{sigma2}
\end{eqnarray}
where, for our purposes, $\hat m$ is either $\hat m_\nu$ or $\hat m_\ell$.
For simplicity of notation we have dropped the $p^2$-dependence in 
the coefficient matrices $A_{L,B}$, $B^{(r)}_{L,R}$, $B^{(l)}_{L,R}$, $C_{L,R}$.
Of course, one can convert equation~(\ref{sigma2}) into the form of 
equation~(\ref{sigma1}), in which case one obtains the identifications
\begin{subequations}\label{convert}
\begin{eqnarray}
\Sigma^{(A)}_L &=& B^{(r)}_L + B^{(l)}_R - C_L \hat m - \hat m C_R, \\
\Sigma^{(A)}_R &=& B^{(r)}_R + B^{(l)}_L - C_R \hat m - \hat m C_L, \\
\Sigma^{(B)}_L &=& A_L - \hat m B^{(r)}_L - B^{(l)}_L \hat m + p^2 C_R + 
\hat m C_L \hat m,
\\
\Sigma^{(B)}_R &=& A_R - \hat m B^{(r)}_R - B^{(l)}_R \hat m + p^2 C_L + 
\hat m C_R \hat m.
\end{eqnarray}
\end{subequations}
The nice feature of the form of equation~(\ref{sigma2}) is that the radiative 
mass shifts to the tree-level masses are simply given by
\begin{equation}\label{dm2}
\Delta m_i = \frac{1}{2} \left\{ 
\left( A_L(m_i^2) \right)_{ii} + \left( A_R(m_i^2) \right)_{ii} \right\}.
\end{equation}
Of course, this is to be expected but can also be checked explicitly 
by plugging the expressions of equation~(\ref{convert}) 
into equation~(\ref{dm1}).

At this point we want to stress that the discussion in the last paragraph 
holds also for any part of the fermion self-energy. If such a part is 
decomposed as in equation~(\ref{sigma2}), then the coefficient matrices 
$B^{(r)}_{L,R}$, $B^{(l)}_{L,R}$, $C_{L,R}$ of this part will not contribute 
to the physical mass shifts $\Delta m_i$. Therefore, any gauge 
dependence in $B^{(r)}_{L,R}$, $B^{(l)}_{L,R}$, $C_{L,R}$ is irrelevant for the 
masses. This will be utilized in the following. 
The $\xi$-independence of the one-loop neutrino masses for the model put 
forward in~\cite{neufeld} has recently been shown in~\cite{gajdosik}.

\subsection{Gauge-parameter cancellation in fermion self-energy loops}
\label{loops}
\begin{figure}[ht]
  \begin{fmffile}{fermion-loops}
   \fmfset{thin}{.7pt}
   \fmfset{dash_len}{1.5mm}
   \fmfset{wiggly_len}{2mm}
   \fmfset{wiggly_slope}{75}
   \fmfset{dot_len}{.8mm}
   \fmfset{dot_size}{1.5thick}
  \begin{center} 
    \begin{subfigure}[t]{.3\textwidth} \centering
      \begin{fmfgraph*}(100,80)
	\fmfleft{i}
	\fmfright{o}
	\fmf{plain,tension=5}{i,v1}
	\fmf{plain,tension=5}{v2,o}
	\fmf{plain}{v1,v2}
	\fmf{wiggly,left,label=\footnotesize{$W^\pm/Z^0$}}{v1,v2}
      \end{fmfgraph*}
    \setlength{\abovecaptionskip}{-20pt}
    \caption{}
    \end{subfigure}
     \begin{subfigure}[t]{.3\textwidth} \centering
      \begin{fmfgraph*}(100,80)
	\fmfleft{i}
	\fmfright{o}
	\fmf{plain,tension=5}{i,v1}
	\fmf{plain,tension=5}{v2,o}
	\fmf{plain}{v1,v2}
	\fmf{dashes,left,label=\footnotesize{$G^\pm/G^0$}}{v1,v2}
      \end{fmfgraph*}
    \setlength{\abovecaptionskip}{-20pt}
    \caption{}
    \end{subfigure}
    \begin{subfigure}[t]{.3\textwidth} \centering
      \begin{fmfgraph*}(100,80)
	\fmftop{i1}
	\fmfleft{b1}
	\fmfright{b2}
	\fmf{plain}{b1,v2,b2}
	\fmffreeze
	\fmf{phantom}{i1,v1,v2}
 	\fmf{dashes,label=\footnotesize{$S^0_b$}}{v2,v1}
	\fmf{dashes,tension=.5,left}{v1,i1,v1}
	\fmflabel{\footnotesize{$G^\pm/G^0$}}{i1}
      \end{fmfgraph*}
    \setlength{\abovecaptionskip}{-20pt}
    \caption{}
    \end{subfigure}
  \end{center}
  \end{fmffile}
\caption{Loop contributions to the fermion self-energies that 
depend on $\xi$. In diagram~(c) only physical neutral scalars 
contribute, \textit{i.e.}\ $b=2,\ldots,2n_H$.} 
\label{fermion-loops}
\end{figure}
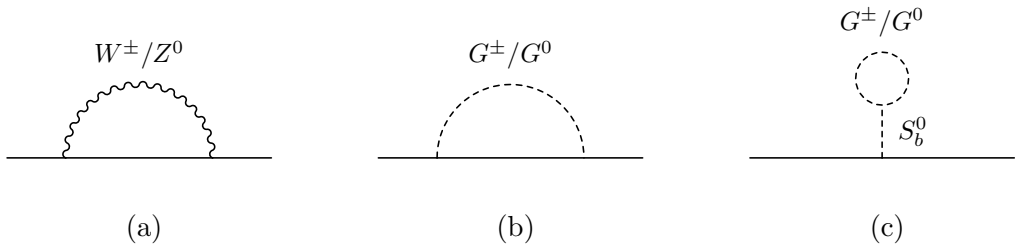
\noindent
The diagrams in figure~\ref{fermion-loops} are those loop diagrams 
which have $\xi$-dependent boson propagators. 
Using equation~(\ref{vbp1}) for the vector boson propagator, 
the gauge dependence resides in 
\begin{equation}\label{vbp-xi}
-\frac{k_\mu k_\nu}{m_V^2}\, \frac{1}{k^2 - \xi_V m_V^2 + i\epsilon}
\quad \mbox{with} \quad V = Z,W.
\end{equation}
In this subsection we will demonstrate that, 
when we include only this part of the vector boson propagator in diagram~(a),
then the sum of the diagrams in figure~\ref{fermion-loops} has the form 
of equation~(\ref{sigma2}) with vanishing $A_L$ and $A_R$. In other words,
of the diagrams in figure~\ref{fermion-loops} only diagram~(a) with 
the vector boson propagator 
\begin{equation}
\frac{1}{k^2 - m_V^2 + i\epsilon} \left( -g_{\mu\nu} + 
\frac{k_\mu k_\nu}{m_V^2} \right)
\end{equation}
contributes to $\Delta m_i$ and 
the sum over the one-loop diagrams gives a $\xi$-independent 
contribution to $\Delta m_i$, as it must be on physical grounds.
We will prove this separately for charged leptons and neutrinos 
and for charged and neutral boson exchange.
The discussion presented here does not apply to photon exchange. 
This case will be treated separately at the end of this subsection.

We first consider the contribution of diagram~(a) of 
figure~\ref{fermion-loops}, with the vector boson propagator of
equation~(\ref{vbp-xi}), to the fermion self-energies
$-i \Sigma_f(p)$ ($f=\chi,\ell$). 
In order to streamline the discussion, we have to introduce some notation.
We define 
\begin{equation}\label{aa}
\ca \equiv \ca_L \gamma_L + \ca_R \gamma_R
\quad \mbox{and} \quad 
\tilde\ca \equiv \ca_L \gamma_R + \ca_R \gamma_L
\end{equation}
such that the coupling matrices of the vector bosons $V$ to the fermions $f$
have the structure $\gamma^\mu \ca$, where $\ca$ also includes 
the flavour part. The matrices $\ca$ can be read off from the respective
Lagrangians. Here is a list of all matrices $\ca$ under 
discussion, with $f$ being the incoming fermion:
\begin{alignat}{4}
\ca &= -\frac{g}{4c_w} F_{LR} & \quad &\mbox{for} \quad & &V = Z,& \;&f = \chi,
\label{Anuz}
\\ 
\ca &= -\frac{g}{\sqrt{2}} W_L^\dagger U_L \gamma_L
& &\mbox{for} & &V = W,& &f = \chi,
\label{Anuw}
\\
\ca_c &= +\frac{g}{\sqrt{2}} \left( W_L^\dagger U_L \right)^* \gamma_R
& &\mbox{for} & &V = W,& &f = \chi,
\label{Anuw_c}
\\
\ca &= -\frac{g}{c_w} 
\left[ \left( s_w^2 - \frac{1}{2} \right) \gamma_L + s_w^2 \gamma_R \right] 
& \quad &\mbox{for} \quad & &V = Z,& \;&f = \ell,
\label{Aellz}
\\
\ca &= -\frac{g}{\sqrt{2}} U_L^\dagger W_L \gamma_L
& &\mbox{for} & &V = W,& &f = \ell.
\label{Aellw}
\end{alignat}
The matrix $F_{LR}$ is defined in equation~(\ref{FLR}).
The expression $\ca_c$ occurs in the Lagrangian of equation~(\ref{Lccb}),
its usage will be explained below.
Moreover, we stipulate that $\hat m$ is the 
diagonal mass matrix of the fermions with momentum $p$ on the external line,
while $\um$ denotes the diagonal mass matrix of the fermions with momentum 
$p-k$ on the internal line of diagram~(a) in figure~\ref{fermion-loops}.
Eventually, with the abbreviation 
\begin{equation}
\mathcal{P} \equiv \frac{1}%
{\slashed{p} - \slashed{k} - \um + i\epsilon},
\end{equation}
we find for the contribution to $-i\Sigma_f(p)$, 
pertaining to the propagator of equation~(\ref{vbp-xi}), the expression
\begin{equation}\label{sw}
-\frac{1}{m^2_V} \int \frac{\dd^d k}{(2\pi)^d}\, 
\frac{1}{k^2 - \xi_V m^2_V + i\epsilon}\, 
\slashed{k} \ca^\dagger \mathcal{P} \slashed{k} \ca.
\end{equation}

Since we are dealing with Majorana neutrinos, 
it has to be taken into account 
that $\chi$ cannot only be Wick-contracted with $\bar\chi$ but also with
$\chi$, and the analogue applies to $\bar\chi$. 
Dealing with one-loop computations, this complication arises only in the case
of the self-energy $\Sigma_\chi(p)$.
In this context a very convenient identity is given by~\cite{denner}
\begin{equation}\label{fof}
\bar f_1 \mathcal{O} f_2 = \overline{(f_2)^c}\, C \mathcal O^T C^{-1} (f_1)^c,
\end{equation}
where $f_1$ and $f_2$ are any vectors of fermion fields and 
$\mathcal{O}$ is product of an arbitrary matrix in Dirac space times a matrix
in flavour space or a sum over matrices of this type;
the superscript $c$ indicates the charge-conjugated field and $C$ 
is the charge-conjugation matrix, which acts only in Dirac space.
Thus if one contracts $\chi$ with an interaction term 
$\bar f_1 \mathcal{O} f_2$ where $f_2 = \chi$,\footnote{For 
  simplicity of notation we assume that 
  $\mathcal{O}$ also contains the boson fields.} 
because of $\chi^c = \chi$ one can simply take advantage
of this identity and the aforementioned contraction 
becomes the ordinary contraction of $\chi$ with $\bar\chi$, however, at the
expense of transforming $\mathcal{O}$ into $C \mathcal O^T C^{-1}$.
Actually, in our one-loop computations two cases\footnote{As a side remark, for
 neutral-scalar vertex corrections a third case occurs 
 which requires the Lagrangian of equation~(\ref{LS0c}).} 
cover all possible situations:
\begin{enumerate}
\renewcommand{\labelenumi}{\roman{enumi}.}
\item
In the couplings of $S^0_b$ and $Z$---see equations~(\ref{LS0}) 
and~(\ref{Lnc}), respectively---we have $f_1 = f_2 = \chi$ and 
our respective coupling matrices are defined in 
such a way that they are invariant under the 
transformation of equation~(\ref{fof}):
\begin{equation}
C \left( \gamma^\mu F_{LR} \right)^T C^{-1} = \gamma^\mu F_{LR},
\quad
C \left( F_b \gamma_L + F_b^\dagger \gamma_R  \right)^T C^{-1} = 
F_b \gamma_L + F_b^\dagger \gamma_R.
\end{equation}
Consequently, in our formalism, 
a contraction of $\chi$ with an interaction term 
of the type $\bar\chi \mathcal{O} \chi$ simply gives a factor of~2.
\item
In the case of charged-boson interactions, we display
both versions of the Lagrangian interaction density, the common one with the
charged-lepton fields $\ell$ and, using equation~(\ref{fof}), the one with the
fields $\ell^c$---see equations~(\ref{LS+}) and~(\ref{Lcc}), in order to
clearly spell out both contractions of the external $\chi$ or $\bar\chi$.
\end{enumerate}

For $W^\pm$ exchange, \textit{i.e.}\ diagram~(a) of 
figure~\ref{fermion-loops}, the second case applies. 
In the light of the discussion presented here
there are two contributions to $\Sigma_\chi(p)$ to be taken into account, 
stemming from $\ca$ of equation~(\ref{Anuw}) and $\ca_c$ of 
equation~(\ref{Anuw_c}).

Now we return to a discussion of equation~(\ref{sw}). We
follow~\cite{weinberg} and make the decomposition
\begin{equation}\label{decomp}
\slashed{k} \ca^\dagger \mathcal{P} \slashed{k} \ca = 
-{\tilde \ca}^\dagger \slashed{k} \ca + A' + B',
\end{equation}
where $\tilde \ca$ is defined in equation~(\ref{aa}), and
\begin{eqnarray}
A' &=&
-\frac{1}{2} \left( \slashed p - \hat m \right) \ca^\dagger \ca 
-\frac{1}{2} {\tilde \ca}^\dagger \tilde\ca \left( \slashed p - \hat m \right)   
\nonumber
\\ && 
-\frac{1}{2} \hat m \ca^\dagger \ca 
-\frac{1}{2} {\tilde \ca}^\dagger \tilde\ca \hat m 
+ {\tilde \ca}^\dagger \um \ca,
\label{A}
\\
B' &=&
\hphantom{+}
\left( \slashed p - \hat m \right) \ca^\dagger \mathcal{P} \tilde\ca
\left( \slashed p - \hat m \right) 
\nonumber \\ && + 
\left( \slashed p - \hat m \right) \ca^\dagger \mathcal{P}
\left( \tilde\ca \hat m - \um \ca \right)
+
\left( \hat m \ca^\dagger - {\tilde\ca}^\dagger \um \right)
\mathcal{P} \tilde\ca \left( \slashed p - \hat m \right) 
\nonumber \\ && + 
\left( \hat m \ca^\dagger - {\tilde\ca}^\dagger \um \right)
\mathcal{P}
\left( \tilde\ca \hat m - \um \ca \right).
\label{B}
\end{eqnarray}
Obviously, the first term on the right-hand side of equation~(\ref{decomp})
drops out when performing the integration in equation~(\ref{sw})
and all terms in equations~(\ref{A}) and~(\ref{B}) that have 
$\slashed{p} - \hat m$ do not contribute to $\Delta m_i$ or $\Delta m_\alpha$.
Therefore, it useful to introduce the definitions
\begin{equation}\label{A'B'}
A \equiv
-\frac{1}{2} \hat m \ca^\dagger \ca 
-\frac{1}{2} {\tilde \ca}^\dagger \tilde\ca \hat m 
+ {\tilde \ca}^\dagger \um \ca,
\quad
B \equiv 
\left( \hat m \ca^\dagger - {\tilde\ca}^\dagger \um \right)
\mathcal{P}
\left( \tilde\ca \hat m - \um \ca \right),
\end{equation}
which refer to the last line in equation~(\ref{A}) and in equation~(\ref{B}),
respectively. 

When we use in the following the notions A-term and B-term, 
we mean the contribution of $A$ and $B$, respectively, 
to the loop integral of equation~(\ref{sw}).
We will now prove the following~\cite{weinberg}:
\begin{enumerate}
\item
The A-term contribution of equation~(\ref{A'B'}) to diagram~(a) of
figure~\ref{fermion-loops} is exactly cancelled by the sum over all physical
neutral scalars $S^0_b$ in diagram~(c).
\item
The B-term contribution of equation~(\ref{A'B'}) to diagram~(a) of
figure~\ref{fermion-loops} is exactly cancelled by diagram~(c).
\end{enumerate}
These cancellations occur separately for both neutrino and charged-leptons on
the external lines and for both neutrino and charged-leptons on
the inner lines. Therefore, in total there are eight cancellations.

\paragraph{Neutrinos and the cancellation of the A-term:}
Firstly we consider $Z$ exchange and neutrinos on the internal line of 
diagram~(a) in figure~\ref{fermion-loops}.
With equation~(\ref{Anuz}) and 
using some formulas of section~\ref{scalar interactions}, we find in this case
\begin{equation}
A = -\frac{g^2}{32 c_w^2} \left[ 
\hat m_\nu \left( U_L^\dagger U_L \gamma_L + U_L^T U_L^* \gamma_R \right) +
\left( U_L^\dagger U_L \gamma_R + U_L^T U_L^* \gamma_L \right) \hat m_\nu
\right].
\end{equation}
Taking into account a factor~4 from the Majorana contractions, we obtain 
for the loop integral of equation~(\ref{sw}) the A-term 
\begin{eqnarray}
&& \nonumber
4 \times \frac{1}{m_Z^2} 
\int \frac{\dd^d k}{(2\pi)^d}\, 
\frac{1}{k^2 - \xi_Z m^2_Z + i\epsilon} 
\\ && \times
\frac{g^2}{32 c_w^2} \left[ 
\hat m_\nu \left( U_L^\dagger U_L \gamma_L + U_L^T U_L^* \gamma_R \right) +
\left( U_L^\dagger U_L \gamma_R + U_L^T U_L^* \gamma_L \right) \hat m_\nu
\right].
\end{eqnarray}
The expression for diagram~(c) of figure~\ref{fermion-loops} is given by
\begin{equation}
\sum_{b=2}^{2n_H} (-i \sqrt{2}) \left( F_b \gamma_L + F_b^\dagger \gamma_R \right)
\times \frac{i}{-M_b^2} \times \frac{i}{2v} M_b^2 \times
\mbox{Im} \left( V^\dagger V \right)_{1b}
\int \frac{\dd^d k}{(2\pi)^d}\, 
\frac{i}{k^2 - \xi_Z m^2_Z + i\epsilon}.
\end{equation}
Note the factor $1/2$ on the vertex of $S^0_b$ coupling to $G^0$---see
equation~(\ref{sgg}). 
Since here $M_b^2$ cancels, we can use 
equation~(\ref{VIm}) to perform the summation 
\begin{equation}
\sum_{b=2}^{2n_H} F_b \,\mbox{Im} \left( V^\dagger V \right)_{1b} =
-\frac{1}{\sqrt{2}v} \left( 
\hat m_\nu U_L^\dagger U_L + U_L^T U_L^* \hat m_\nu \right).
\end{equation}
To obtain this expression, we have also utilized equation~(\ref{URLM}).
Note that in this sum we can include $b=1$ because 
$\mbox{Im} \left( V^\dagger V \right)_{1b} = 0$.
Finally, we end up with the expression
\begin{equation}\label{nu-c}
-\frac{1}{2v^2} \left[ 
\left( \hat m_\nu U_L^\dagger U_L + U_L^T U_L^* \hat m_\nu \right) \gamma_L + 
\left( U_L^\dagger U_L \hat m_\nu + \hat m_\nu U_L^T U_L^* \right) \gamma_R
\right]
\int \frac{\dd^d k}{(2\pi)^d}\, 
\frac{1}{k^2 - \xi_Z m^2_Z + i\epsilon}.
\end{equation}
for diagram~(c).
Because of 
\begin{equation}\label{mzv}
\frac{g^2}{m_Z^2 c_w^2} = \frac{4}{v^2}
\end{equation}
it exactly cancels the A-term.

Secondly we consider $W$ exchange and charged leptons on the internal line of 
diagram~(a) in figure~\ref{fermion-loops}.
Here we obtain
\begin{equation}
A = -\frac{g^2}{4} \left( \hat m_\nu U_L^\dagger U_L \gamma_L + 
U_L^\dagger U_L \hat m_\nu \gamma_R \right).
\end{equation}
However, due to the Majorana nature, we also have the contribution from 
$\ca_c$ of equation~(\ref{Anuw_c}), leading to
\begin{equation}
A_c = -\frac{g^2}{4} \left( \hat m_\nu U_L^T U_L^* \gamma_R + 
U_L^T U_L^* \hat m_\nu \gamma_L \right).
\end{equation}
The full A-term is, therefore,
\begin{eqnarray}
&& \nonumber
\frac{1}{m_W^2} 
\int \frac{\dd^d k}{(2\pi)^d}\, 
\frac{1}{k^2 - \xi_W m^2_W + i\epsilon} 
\\ && \times
\frac{g^2}{4} \left[ 
\left( \hat m_\nu U_L^\dagger U_L \gamma_L + U_L^\dagger U_L \hat m_\nu \gamma_R 
\right) +
\left( \hat m_\nu U_L^T U_L^* \gamma_R + U_L^T U_L^* \hat m_\nu \gamma_L
\right) \right].
\end{eqnarray}
As for diagram~(c) we can 
take over 
the previous result,
equation~(\ref{nu-c}), with minor modifications:
\begin{equation}
-\frac{1}{v^2} \left[ 
\left( \hat m_\nu U_L^\dagger U_L + U_L^T U_L^* \hat m_\nu \right) \gamma_L + 
\left( U_L^\dagger U_L \hat m_\nu + \hat m_\nu U_L^T U_L^* \right) \gamma_R
\right]
\int \frac{\dd^d k}{(2\pi)^d}\, 
\frac{1}{k^2 - \xi_W m^2_W + i\epsilon}.
\end{equation}
Note there is no factor $1/2$ on the vertex of $S^0_b$ coupling to 
$G^\pm$---see equation~(\ref{sgg}). 
With 
\begin{equation}\label{mwv}
\frac{g^2}{m^2_W} = \frac{4}{v^2} 
\end{equation}
we see that also here the tadpoles cancel the A-term.

\paragraph{Neutrinos and the cancellation of the B-term:}
Firstly we discuss $Z$ and $\chi$ exchange in the loop diagram~(a) of 
figure ~\ref{fermion-loops}. 
%%%%
For the computation of $B$ we need
\begin{equation}\label{T}
\tilde\ca \hat m_\nu -  \hat m_\nu \ca = -\frac{g}{4c_w}
\left( T \gamma_R - T^\dagger \gamma_L \right)
\quad \mbox{with} \quad 
T = U_L^\dagger U_L \hat m_\nu + \hat m_\nu U_L^T U_L^*.
\end{equation}
Since here $\ca$ is hermitian, we have
\begin{equation}
\hat m_\nu \ca^\dagger - {\tilde\ca}^\dagger \hat m_\nu = 
-\left( \tilde\ca \hat m_\nu -  \hat m_\nu \ca \right).
\end{equation}
Thus the B-term is given by
\begin{equation}
-\frac{4}{m_Z^2} \int \frac{\dd^d k}{(2\pi)^d}\, 
\frac{1}{k^2 - \xi_Z m^2_Z + i\epsilon} \times (-1) \times
\frac{g^2}{16 c_w^2} \left( T \gamma_R - T^\dagger \gamma_L \right)
\mathcal{P} 
\left( T \gamma_R - T^\dagger \gamma_L \right).
\end{equation}
Turning to diagram~(b) of figure~\ref{fermion-loops}, the $G^0$-coupling matrix 
$F_1$ of equation~(\ref{F1G1}) can be written with the help of $T$ as
\begin{equation}\label{F1}
F_1 = \frac{i}{\sqrt{2}v} T^\dagger.
\end{equation}
This leads to the following expression for diagram~(b):
\begin{equation}
-\int \frac{\dd^d k}{(2\pi)^d}\, 
\frac{1}{k^2 - \xi_Z m^2_Z + i\epsilon} \times 
\frac{1}{v^2} \left( T^\dagger \gamma_L - T \gamma_R \right)
\mathcal{P} 
\left( T^\dagger \gamma_L - T \gamma_R \right).
\end{equation}
Using again equation~(\ref{mzv}) we see that the expression for diagram~(b) 
exactly cancels the B-term.

Secondly we discuss $W^\pm$ and charged-fermion exchange in the loop 
diagram~(a) of figure ~\ref{fermion-loops}. In the light of the discussion 
concerning Majorana fermions, we have to take into account both
$W^+$ and $\ell^-$ exchange and $W^-$ and $\ell^+$ exchange. Defining 
for simplicity of notation the $n_L \times (n_L + n_R)$ matrix 
\begin{equation}\label{VL}
V_L = W_L^\dagger U_L,
\end{equation}
for the quantity $B$ of equation~(\ref{A'B'}) 
we require the expressions 
\begin{subequations}
\begin{eqnarray}
\tilde\ca \hat m_\nu -  \hat m_\ell \ca &=& -\frac{g}{\sqrt{2}} 
\left( V_L \hat m_\nu \gamma_R - \hat m_\ell V_L \gamma_L \right),
\\
\hat m_\nu \ca^\dagger - {\tilde\ca}^\dagger \hat m_\ell &=& -\frac{g}{\sqrt{2}} 
\left( \hat m_\nu V_L^\dagger \gamma_L - V_L^\dagger \hat m_\ell \gamma_R \right),
\\
\tilde\ca_c \hat m_\nu -  \hat m_\ell \ca_c &=& +\frac{g}{\sqrt{2}} 
\left( V_L^* \hat m_\nu \gamma_L - \hat m_\ell V_L^* \gamma_R \right),
\\
\hat m_\nu \ca_c^\dagger - {\tilde\ca}_c^\dagger \hat m_\ell &=& +\frac{g}{\sqrt{2}} 
\left( \hat m_\nu V_L^T \gamma_R - V_L^T \hat m_\ell \gamma_L \right).
\end{eqnarray}
\end{subequations}
Then, the corresponding B-term is given by
\begin{eqnarray}
&& \nonumber
-\frac{1}{m_W^2} \int \frac{\dd^d k}{(2\pi)^d}\, 
\frac{1}{k^2 - \xi_W m^2_W + i\epsilon} \times
\frac{g^2}{2} 
\\
&& \nonumber \times \left\{ 
\left( \hat m_\nu V_L^\dagger \gamma_L - V_L^\dagger \hat m_\ell \gamma_R \right)
\mathcal{P} 
\left( V_L \hat m_\nu \gamma_R - \hat m_\ell V_L \gamma_L \right) \right.
\\
&& + \left.
\left( \hat m_\nu V_L^T \gamma_R - V_L^T \hat m_\ell \gamma_L \right)
\mathcal{P} 
\left( V_L^* \hat m_\nu \gamma_L - \hat m_\ell V_L^* \gamma_R \right) \right\}.
\end{eqnarray}
In equation~(\ref{LS+}) we have formulated the couplings of $G^\pm$ to the
fermions with the help of the matrices of equation~(\ref{R1L1}). With the
matrix $V_L$ they are simply 
\begin{equation}
R_1 = \frac{\sqrt{2}}{v}\,V_L \hat m_\nu, 
\quad
L_1 = \frac{\sqrt{2}}{v}\, \hat m_\ell V_L. 
\end{equation}
Just as diagram~(a) of figure~\ref{fermion-loops}, diagram~(b) has two
contributions as well---see the Lagrangians of equations~(\ref{LS+a})
and~(\ref{LS+b}), given by
\begin{eqnarray}
&& \nonumber
\int \frac{\dd^d k}{(2\pi)^d}\, 
\frac{1}{k^2 - \xi_W m^2_W + i\epsilon} \times
\frac{2}{v^2} 
\\
&& \nonumber \times \left\{ 
\left( \hat m_\nu V_L^\dagger \gamma_L - V_L^\dagger \hat m_\ell \gamma_R \right)
\mathcal{P} 
\left( V_L \hat m_\nu \gamma_R - \hat m_\ell V_L \gamma_L \right) \right.
\\
&& + \left.
\left( \hat m_\nu V_L^T \gamma_R - V_L^T \hat m_\ell \gamma_L \right)
\mathcal{P} 
\left( V_L^* \hat m_\nu \gamma_L - \hat m_\ell V_L^* \gamma_R \right) \right\}.
\end{eqnarray}
Obviously, using equation~(\ref{mwv}), diagram~(c) cancels the B-term.

\paragraph{Charged leptons and the cancellation of the A-term:}
For $Z$ and neutrino exchange 
$A$ is simply given by
\begin{equation}
A = -\frac{g^2}{8 c_w^2}\,\hat m_\ell
\end{equation}
and the A-term is thus
\begin{equation}
\frac{1}{m_Z^2} \int \frac{\dd^d k}{(2\pi)^d}\, 
\frac{1}{k^2 - \xi_Z m^2_Z + i\epsilon} \times
\frac{g^2}{8 c_w^2} \, \hat m_\ell.
\end{equation}
The sum over the tadpoles in diagram~(c) of figure~\ref{fermion-loops} now
reads
\begin{equation}
\sum_{b=2}^{2n_H} \frac{-i}{\sqrt{2}} \left( G_b \gamma_L + G_b^\dagger
\gamma_R \right)
\times \frac{i}{-M_b^2} \times \frac{i}{2v} M_b^2 \times
\mbox{Im} \left( V^\dagger V \right)_{1b}
\int \frac{\dd^d k}{(2\pi)^d}\, 
\frac{i}{k^2 - \xi_Z m^2_Z + i\epsilon}.
\end{equation}
With 
\begin{equation}
\sum_{b=2}^{2n_H} G_b \,\mbox{Im} \left( V^\dagger V \right)_{1b} =
-\frac{\sqrt{2}}{v}\, \hat m_\ell
\end{equation}
the tadpole contributions are thus
\begin{equation}
-\frac{1}{2v^2}\,\hat m_\ell \, 
\int \frac{\dd^d k}{(2\pi)^d}\, 
\frac{i}{k^2 - \xi_Z m^2_Z + i\epsilon}.
\end{equation}
Applying equation~(\ref{mzv}), we see that they exactly cancel the A-term.

Though for charged $W^\pm$ and charged-lepton 
exchange $\ca$ of equation~(\ref{Aellw}) 
looks very different from that of equation~(\ref{Aellz}),  
the result for $A$ is quite similar:
\begin{equation}
A = -\frac{g^2}{4}\,\hat m_\ell.
\end{equation}
It gives the A-term 
\begin{equation}
\frac{1}{m_W^2} \int \frac{\dd^d k}{(2\pi)^d}\, 
\frac{1}{k^2 - \xi_W m^2_W + i\epsilon} \times
\frac{g^2}{4} \, \hat m_\ell,
\end{equation}
while the tadpole contributions are 
\begin{equation}
-\frac{1}{v^2}\,\hat m_\ell \, 
\int \frac{\dd^d k}{(2\pi)^d}\, 
\frac{i}{k^2 - \xi_W m^2_W + i\epsilon}.
\end{equation}
Both contributions exactly cancel each other, when taking into account
equation~(\ref{mwv}).

\paragraph{Charged leptons and the cancellation of the B-term:}
Firstly we consider $Z$ and $\ell$ exchange in diagram~(a) of
figure~\ref{fermion-loops}. 
According to equation~(\ref{A'B'}), the expressions 
\begin{equation}
\tilde\ca \hat m_\ell - \hat m_\ell \ca = \frac{g}{2 c_w} \, 
\hat m_\ell \left( \gamma_R - \gamma_L \right)
\quad \mbox{and} \quad
\hat m_\ell \ca^\dagger - {\tilde\ca}^\dagger \hat m_\ell = -\frac{g}{2 c_w} \, 
\hat m_\ell \left( \gamma_R - \gamma_L \right)
\end{equation}
leads to $B$ and thus to the B-term
\begin{equation}
\frac{1}{m_Z^2} \int \frac{\dd^d k}{(2\pi)^d}\, 
\frac{1}{k^2 - \xi_Z m^2_Z + i\epsilon} \times
\frac{g^2}{4 c_w^2} \, \hat m_\ell \left( \gamma_L - \gamma_R \right) 
\mathcal{P} m_\ell \left( \gamma_L - \gamma_R \right).
\end{equation}
With $G_1$ of equation~(\ref{F1G1}) and the Lagrangian of
equation~(\ref{LS0}), diagram~(b) of figure~\ref{fermion-loops} gives
\begin{equation}
-\int \frac{\dd^d k}{(2\pi)^d}\, 
\frac{1}{k^2 - \xi_Z m^2_Z + i\epsilon} \times
\frac{1}{v^2} \, \hat m_\ell \left( \gamma_L - \gamma_R \right) 
\mathcal{P} m_\ell \left( \gamma_L - \gamma_R \right),
\end{equation}
which exactly cancels the B-term.

Secondly we consider $W$ and $\chi$ exchange in diagram~(a) of
figure~\ref{fermion-loops}. Here we have
\begin{subequations}
\begin{eqnarray}
\tilde\ca \hat m_\ell - \hat m_\nu \ca &=& -\frac{g}{\sqrt{2}} 
\left( V_L^\dagger \hat m_\ell \gamma_R - 
\hat m_\nu V_L^\dagger \gamma_L \right),
\\ 
\hat m_\ell \ca^\dagger - {\tilde\ca}^\dagger \hat m_\nu &=& -\frac{g}{\sqrt{2}} 
\left( \hat m_\ell V_L \gamma_L - 
V_L \hat m_\nu \gamma_R \right) 
\end{eqnarray}
and the B-term
\end{subequations}
\begin{eqnarray}
&& \nonumber
-\frac{1}{m_W^2} \int \frac{\dd^d k}{(2\pi)^d}\, 
\frac{1}{k^2 - \xi_W m^2_W + i\epsilon} 
\\ &&
\times \frac{g^2}{2} 
\left( \hat m_\ell V_L \gamma_L - 
V_L \hat m_\nu \gamma_R \right) 
\mathcal{P}
\left( V_L^\dagger \hat m_\ell \gamma_R - 
\hat m_\nu V_L^\dagger \gamma_L \right).
\end{eqnarray}
Considering diagram~(b) of figure~\ref{fermion-loops}, we need 
the Lagrangian of equation~(\ref{LS+a}) and the matrices 
$R_1$ and $L_1$ of equation~(\ref{R1L1}).
The expression for this diagram is then
\begin{eqnarray}
&& \nonumber
\int \frac{\dd^d k}{(2\pi)^d}\, 
\frac{1}{k^2 - \xi_W m^2_W + i\epsilon} 
\\ &&
\times \frac{2}{v^2} 
\left( V_L \hat m_\nu \gamma_R  - 
\hat m_\ell V_L \gamma_L \right) 
\mathcal{P} 
\left( \hat m_\nu V_L^\dagger \gamma_L - 
V_L^\dagger \hat m_\ell \gamma_R \right),
\end{eqnarray}
which precisely
cancels the B-term.

\paragraph{Photon exchange:}
This is only possible for charged leptons. Moreover, diagrams~(b) and~(c)
in figure~\ref{fermion-loops} do not exist in this case.
However, here $\ca = e \bone$ and 
the decomposition of equation~(\ref{decomp}) is simply given by 
\begin{equation}
\slashed{k} \mathcal{P} \slashed{k} = -\slashed{k} - 
\left( \slashed{p} - \hat m_\ell \right) + 
\left( \slashed{p} - \hat m_\ell \right) \mathcal{P}
\left( \slashed{p} - \hat m_\ell \right).
\end{equation}
This demonstrates that the part of the photon propagator proportional to 
$\xi_A$---\textit{c.f.}\ equation~(\ref{vbp2})---does not contribute 
to $\Delta m_\alpha$.

\subsection{Yukawa coupling renormalization and mass counterterms}
\label{Counterterms}
\begin{figure}[ht]
  \begin{fmffile}{yukawa-vertex}
   \fmfset{thin}{.7pt}
   \fmfset{dash_len}{1.5mm}
   \fmfset{wiggly_len}{2mm}
   \fmfset{wiggly_slope}{75}
   \fmfset{dot_len}{.8mm}
   \fmfset{dot_size}{1.5thick}
  \begin{center} 
    \begin{subfigure}[t]{.24\textwidth} \centering
      \begin{fmfgraph*}(60,80)
	\fmftop{t1,t2}
	\fmfbottom{b}
	\fmf{plain}{t1,vtl,vb}
	\fmf{plain}{t2,vtr,vb}
	\fmf{dashes,tension=1.5}{vb,b}
	\fmffreeze
	\fmf{wiggly,left=.5}{vtl,vtr}
      \end{fmfgraph*}
    \caption{}
    \end{subfigure}
     \begin{subfigure}[t]{.24\textwidth} \centering
      \begin{fmfgraph*}(60,80)
	\fmftop{t1,t2}
	\fmfbottom{b}
	\fmf{plain}{t1,vtl,vb}
	\fmf{plain}{t2,vtr,vb}
	\fmf{dashes,tension=1.5}{vb,b}
	\fmffreeze
	\fmf{dashes,left=.5}{vtl,vtr}
      \end{fmfgraph*}
    \caption{}
    \end{subfigure}
    \begin{subfigure}[t]{.24\textwidth} \centering
      \begin{fmfgraph*}(60,80)
	\fmftop{t1,t2}
	\fmfbottom{b}
	\fmf{phantom}{t1,vtl,vb}
	\fmf{phantom}{t2,vtr,vb}
	\fmf{dashes,tension=1.5}{vb,b}
	\fmffreeze
	\fmf{plain}{t1,vtl,vtr,t2}
	\fmf{dashes}{vtl,vb}
	\fmf{wiggly}{vtr,vb}
      \end{fmfgraph*}
    \caption{}
    \end{subfigure}
    \begin{subfigure}[t]{.24\textwidth} \centering
      \begin{fmfgraph*}(60,80)
	\fmftop{t1,t2}
	\fmfbottom{b}
	\fmf{phantom}{t1,vtl,vb}
	\fmf{phantom}{t2,vtr,vb}
	\fmf{dashes,tension=1.5}{vb,b}
	\fmffreeze
	\fmf{plain}{t1,vtl,vtr,t2}
	\fmf{wiggly}{vtl,vb}
	\fmf{dashes}{vtr,vb}
      \end{fmfgraph*}
    \caption{}
    \end{subfigure}
  \end{center}
  \end{fmffile}
\caption{Vertex corrections to the couplings of neutral scalars to fermions.}
\label{vertex-corrections}
\end{figure}
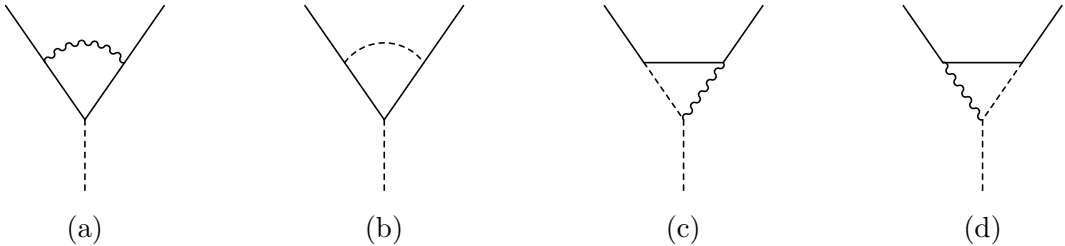
\paragraph{Yukawa coupling renormalization:}
Vertex corrections to the $S^0_b$ coupling to neutrinos can effectively 
be written as counter\-terms to the Yukawa coupling matrices $\Delta_k$ or can
be computed in the unbroken theory as a correction to the $\varphi^0_k$ 
vertex. The result is
\begin{equation}\label{dDelta}
\delta\Delta_k = -2A_1 \Delta_k - \frac{1}{16\pi^2}\, c_\infty 
\Delta_j \Gamma_k^\dagger \Gamma_j,
\end{equation}
where the first term stems from diagrams~(c) and~(d) in
figure~\ref{vertex-corrections} 
and the second one from diagram~(b) with charged-scalar exchange.
Note that the contributions of diagram~(a) with neutral and charged vector
boson exchange and of diagram~(b) with neutral scalar exchange 
are zero separately.

Now we discuss vertex corrections of the $S^0_b$ coupling to leptons. 
Those can be subsumed as $\delta \Gamma_k$. The result is 
\begin{eqnarray}
\delta\Gamma_k &=& -2A_1 \Gamma_k - 
\frac{1}{16\pi^2}\, c_\infty \Gamma_j \Delta_k^\dagger \Delta_j 
\nonumber \\ && -
\frac{g^2}{16\pi^2 c_w^2}\, c_\infty \left( 3 + \xi_Z \right) 
s_w^2 \left( s_w^2 - \frac{1}{2} \right) \Gamma_k -
\frac{e^2}{16\pi^2}\, c_\infty \left( 3 + \xi_A \right) \Gamma_k, 
\label{dGamma}
\end{eqnarray}
where the two terms in the first line originate, just as before,
from diagrams~(c) and~(d) in
figure~\ref{vertex-corrections} and from diagram~(b) with charged-scalar 
exchange. As before, diagram~(a) with $W^\pm$ exchange and diagram~(b) with
$S^0_{b'}$ exchange give vanishing contributions. However, $Z$ and photon
exchange in diagram~(a) are non-zero, leading to the two terms in the second
line.

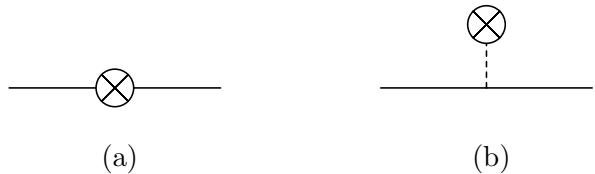
\begin{figure}[ht]
  \begin{fmffile}{fermion-counterterms}
   \fmfset{thin}{.7pt}
   \fmfset{dash_len}{1.5mm}
   \fmfset{wiggly_len}{2mm}
   \fmfset{wiggly_slope}{75}
   \fmfset{dot_len}{.8mm}
   \fmfset{dot_size}{1.5thick}
  \begin{center} 
    \begin{subfigure}[t]{.3\textwidth} \centering
      \begin{fmfgraph*}(80,80)
	\fmfleft{i}
	\fmfright{o}
	\fmf{plain}{i,v,o}
	\defotimes
	\fmfv{d.sh=otimes,d.f=empty}{v}
      \end{fmfgraph*}
    \setlength{\abovecaptionskip}{-20pt}
    \caption{}
    \end{subfigure}
     \begin{subfigure}[t]{.3\textwidth} \centering
      \begin{fmfgraph*}(80,80)
	\fmftop{i1}
	\fmfleft{b1}
	\fmfright{b2}
	\fmf{plain}{b1,v,b2}
	\fmffreeze
	\fmf{phantom}{v,vx,i1}
	\fmf{phantom,tension=2}{i1,vx}
	\fmf{dashes}{v,vx}
	\defotimes
	\fmfv{d.sh=otimes,d.f=empty}{vx}
      \end{fmfgraph*}
    \setlength{\abovecaptionskip}{-20pt}
    \caption{}
    \end{subfigure}
  \end{center}
  \end{fmffile}
\caption{Counterterms on fermion lines.} \label{counterterms}
\end{figure}
\paragraph{Mass counterterms:}
There are two types of counterterms for the fermion self-energies. 
The first type, depicted in diagram~(a) of figure~\ref{counterterms}, 
originates in $\delta v_k$, $\delta \Delta_k$ and $\delta M_R$
for neutrinos, while for charged leptons it stems from 
$\delta v_k$ and $\delta \Gamma_k$. Both have also wave-function
renormalization counterterms. Defining
\begin{equation}\label{dM}
\delta M_\ell \equiv \frac{1}{\sqrt{2}} \sum_k 
\left(  \delta v_k^\ast\, \Gamma_k + v_k^\ast\, \delta \Gamma_k \right)
\quad \mathrm{and} \quad
\delta M_D \equiv \frac{1}{\sqrt{2}} \sum_k 
\left( \delta v_k\, \Delta_k + v_k\, \delta \Delta_k \right),
\end{equation}
diagram~(a) thus refers to the terms 
\begin{equation}\label{ct-sigma-ell}
i \slashed{p}\, \delta^{(\ell)}  -i 
\left[ W_R^\dagger\, \delta M_\ell\, W_L \gamma_L +
W_L^\dagger\, \delta M_\ell^\dagger\, W_R \gamma_R \right]
\quad \mbox{with} \quad 
\delta^{(\ell)} = \delta^{(\ell)}_L \gamma_L + \delta^{(\ell)}_R \gamma_R
\end{equation}
in the case of charged leptons and to 
\begin{subequations}\label{ct-sigma-nu}
\begin{eqnarray}
&& \hphantom{-}
i \slashed{p} \left(  \delta^{(\chi)} \gamma_L + 
\left(\delta^{(\chi)} \right)^* \gamma_R \right)
\label{ct-sigma-nu-a} \\ &&
-i \left[ \left( U_R^\dagger\, \delta M_D\, U_L + 
U_L^T\, \delta M_D^T\, U_R^* \right) \gamma_L + 
\left( U_L^\dagger \delta M_D^\dagger U_R + 
U_R^T\, \delta M_D^*\, U_L^* \right) \gamma_R \right]
\label{ct-sigma-nu-b} \\ &&
-i \left[ U_R^\dagger\, \delta M_R\, U_R^*\, \gamma_L + 
U_R^T\, \delta M_R^*\, U_R\, \gamma_R \right]
\label{ct-sigma-nu-c}
\end{eqnarray}
\end{subequations}
for neutrinos.
The second type of counterterm is symbolized by diagram~(b) of
figure~\ref{counterterms} and connects the tadpole counterterm to the fermion
line. 

In order to discuss the $\xi$-dependence of the counterterms it is expedient to
have the explicit $\xi$-dependence of the wave-function renormalization 
matrices as well. For neutrinos the result is 
\begin{equation}\label{delta-chi}
\delta^{(\chi)}(\xi) = -2A_1 U_L^\dagger U_L, 
\end{equation}
whereas for charged lepton we find the more involved result
\begin{equation}
\delta^{(\ell)}(\xi) = -\frac{1}{16\pi^2}\,c_\infty \left\{
\frac{g^2}{2} \xi_W \gamma_L + 
\frac{g^2}{c_w^2} \xi_Z 
\left[ \left( s_w^2 - \frac{1}{2} \right)^2 \gamma_L + s_w^4 \gamma_R \right]
+ e^2 \xi_A \right\}.
\end{equation}\label{delta-ell}
Using the definition of $A_1$, equation~(\ref{A1}), we can recast this
equation into 
\begin{equation}
\delta^{(\ell)}(\xi) = -2A_1 \gamma_L 
-\frac{1}{16\pi^2}\,c_\infty \left\{
\frac{g^2}{c_w^2} \xi_Z 
\left[ \left( s_w^4 - s_w^2 \right) \gamma_L + s_w^4 \gamma_R \right]
+ e^2 \xi_A \right\}.
\end{equation}

\paragraph{Gauge-parameter independence of mass counterterms:}
Let us take stock of the $\xi$-dependence in the counterterms.
\begin{enumerate}
\renewcommand{\labelenumi}{(\alph{enumi})}
\item
In the counterterms pertaining to diagram~(a) of figure~\ref{counterterms},
$\xi$-dependence occurs in
\begin{enumerate}
\renewcommand{\labelenumii}{(a-\roman{enumii})}
\item
$\delta v_k$---see equation~(\ref{dv})---of $\delta M_\ell$ and $\delta M_D$,
\item
the $A_1$-term of $\delta\Delta_k$ and $\delta\Gamma_k$,
\item
the residual $\xi$-dependence of $\delta\Gamma_k$ not contained in $A_1$,
\item
the $A_1$-term of $\delta^{(\chi)}(\xi)$ and of $\delta^{(\ell)}(\xi)$ and
\item
the residual $\xi$-dependence of 
$\delta^{(\ell)}(\xi)$ not contained in $A_1$.
\end{enumerate}
\item
Finally, the $\xi$-dependence that resides in the tadpole counterterms 
symbolized by diagram~(b) of figure~\ref{counterterms}
has to be accounted for in the present discussion.\footnote{Note that in 
section~\ref{loops} we have already treated the tadpole loops and their
$\xi$-dependence--see figure~\ref{fermion-loops}, but that discussion did not 
include the $\xi$-dependence of the tadpole counterterms.}
The relevant expression can be read off from  
equation~(\ref{tadpole-div+cts}), where the first line is the divergence of the
tadpole loops. Thus the sum of the other three lines, given by
\begin{equation}\label{t-ct}
  \left(
  \begin{gathered}
  \vspace{-8pt}
  \fmfreuse{scalar-tadpole-ct}
  \end{gathered}
  \vphantom{\begin{minipage}[t][.5cm]{0pt}\end{minipage}}
  \right)_{\xi^1}
  \; = \;
  -\frac{i}{2} \left( v_i^* V_{ib} + V_{ib}^* v_i \right) M_b^2 A_1,
\end{equation}
are the $\xi$-dependent tadpole counterterms to be taken into account here.
\end{enumerate}
In the following we will show that the $\xi$-dependent counterterms listed
here do not contribute to $\Delta m_i$ and $\Delta m_\alpha$.

With equations~(\ref{dv}), (\ref{dDelta}) and~(\ref{dGamma})
it is obvious that the terms stemming from ~(a-i) and~(a-ii) 
cancel each other in both $\delta M_\ell$ and $\delta M_D$.

Next we consider the terms originating in~(a-iv) and diagram~(b).
In the case of neutrinos the contribution of diagram~(b)
of figure~\ref{counterterms} to $-i\Sigma_\chi(p)$ gives
\begin{fmffile}{neutrino-tadpole-ct}
\fmfset{thin}{.7pt}
\fmfset{dash_len}{1.5mm}
\fmfset{arrow_len}{2.5mm}
\vspace{5pt}
\begin{equation}
\begin{gathered}
 \vspace{-4pt}
  \begin{fmfgraph*}(50,35)\fmfkeep{neutrino-tadpole-ct}
	\fmfleft{i}
	\fmftop{t}
	\fmfright{o}
	\fmf{plain}{i,v,o}
	\fmffreeze
	\fmf{dashes}{v,t}
	\defotimes
	\fmfv{d.sh=otimes,d.f=empty}{t}
      \end{fmfgraph*}
\end{gathered}
\;=\;
\sum_{b=2}^{2n_H} (-i \sqrt{2}) \left( F_b \gamma_L + F_b^\dagger \gamma_R \right)
\times \frac{i}{-M_b^2} \times 
\left(-\frac{i}{2}\right) \left( v_i^* V_{ib} + V_{ib}^* v_i \right) M_b^2 A_1.
\end{equation}
\end{fmffile}%
Since the expression in equation~(\ref{t-ct}) is zero for 
$b=1$---\textit{c.f.}\ equation~(\ref{GUV})---and the scalar masses cancel,
we can include $b=1$ in the sum. Moreover, taking
advantage of the first two relations of equation~(\ref{orthV}) and
using
\begin{equation}
\left( U_R^\dagger \Delta_k U_L + U_L^T \Delta_k^T U_R^* \right) v_k = 
-2vi F_1,
\end{equation}
which follows from  equations~(\ref{Fb}) and~(\ref{F1G1}), the contribution 
of diagram~(b) of figure~\ref{counterterms} finally has the form
\vspace{5pt}
\begin{equation}
\begin{gathered}
 \vspace{-4pt}
 \fmfreuse{neutrino-tadpole-ct}
\end{gathered}
 \; = \;
i A_1 \left[ 
\left( \hat m_\nu U_L^\dagger U_L + U_L^T U_L^* \hat m_\nu \right) \gamma_L + 
\left( U_L^\dagger U_L \hat m_\nu + \hat m_\nu U_L^T U_L^* \right) \gamma_R
\right].
\end{equation}
Adding this to the $A_1$-term of $\delta^{(\chi)}(\xi)$ and introducing the
abbreviation $\mathcal{Z} \equiv U_L^\dagger U_L$, we arrive at
\begin{eqnarray}
&& -i A_1 \left[ 2 \slashed{p} \left( \mathcal{Z} \gamma_L + \mathcal{Z}^T 
\gamma_R \right) - 
\left( \hat m_\nu \mathcal{Z} + \mathcal{Z}^T \hat m_\nu \right) \gamma_L + 
\left( \mathcal{Z} \hat m_\nu + \hat m_\nu \mathcal{Z}^T \right) 
\gamma_R \right] 
\nonumber \\
&=& 
-i A_1 \left[ \left( \slashed{p} - \hat m_\nu \right) \mathcal{Z} \gamma_L +
\mathcal{Z}^T \gamma_L \left( \slashed{p} - \hat m_\nu \right) + 
\left( \slashed{p} - \hat m_\nu \right) \mathcal{Z}^T \gamma_R + 
\mathcal{Z} \gamma_R \left( \slashed{p} - \hat m_\nu \right) \right]. 
\hphantom{xxx}
\end{eqnarray}
According to section~(\ref{two decomp}), we can read off from this expression
that it does not contribute to $\Delta m_i$.
With this we have concluded the discussion of counterterms to 
the neutrino self-energy.

Now we proceed in an analogous fashion in the case of charged leptons.
Skipping here all details, 
the $A_1$-term of $\delta^{(\ell)}(\xi)$ together with the contribution of 
diagram~(b) of figure~\ref{counterterms} leads to
\begin{equation}
-i A_1 \left( 2\slashed{p } \gamma_L - \hat m_\ell \right) = 
-i A_1 \left[ 
\left( \slashed{p } - \hat m_\ell \right) \gamma_L + 
\gamma_R \left( \slashed{p } - \hat m_\ell \right) \right]
\end{equation}
in $-i \Sigma_\ell(p)$. Therefore, this does not contribute 
to $\Delta m_\alpha$.

It remains to consider the terms~(a-iii) and~(a-v), 
which refer solely to charged leptons. These are 
\begin{eqnarray}
&&
-\frac{i}{16\pi^2}\,c_\infty \left\{
\frac{g^2}{c_w^2} \xi_Z 
\left[ \left( s_w^4 - s_w^2 \right) \gamma_L + s_w^4 \gamma_R \right]
+ e^2 \xi_A \right\} \slashed{p} 
\nonumber \\ && + 
\frac{i}{16\pi^2}\,c_\infty \left[ 
\frac{g^2}{c_w^2} \xi_Z s_w^2 \left( s_w^2 - \frac{1}{2} \right) + 
e^2 \xi_A \right] \hat m_\ell.
\end{eqnarray}
Obviously, the $\xi_A$-terms combine to give $\slashed{p} - \hat m_\ell$.
That the $\xi_Z$-terms can also be decomposed into expressions having
external factors $\slashed{p} - \hat m_\ell$, 
can be concluded from the discussion in appendix~\ref{app-on-shell}. 
Explicitly, one such  decomposition is given by
\begin{eqnarray}
&& \nonumber
\left[ \left( s_w^4 - s_w^2 \right) \gamma_L + s_w^4 \gamma_R \right] 
\slashed{p} - s_w^2 \left( s_w^2 - \frac{1}{2} \right) \hat m_\ell 
\\ &=&
\left( s_w^4 - \frac{1}{2} s_w^2 \right) 
\left( \slashed{p} - \hat m_\ell \right) \gamma_L - 
\frac{1}{2} s_w^2 \gamma_R 
\left( \slashed{p} - \hat m_\ell \right) + 
s_w^4 \left( \slashed{p} - \hat m_\ell \right) \gamma_R.
\end{eqnarray}

Summarizing, we have found that all $\xi$-dependent counterterms to the 
neutrino or charged-lepton self-energies have external factors 
$\slashed{p} - \hat m_\nu$ or
$\slashed{p} - \hat m_\ell$, respectively. Therefore, in our renormalization
scheme there are no $\xi$-dependent counterterms for the one-loop radiative
masses $\Delta m_i$ ($i = 1,\ldots,n_L+n_R$)
and $\Delta m_\alpha$ ($\alpha = e,\mu,\tau$).

\section{Finiteness of the fermion self-energies}
\label{finiteness}
\begin{figure}[ht]
 \begin{fmffile}{fermion-loops-counterterm}
 \fmfset{thin}{.7pt}
 \fmfset{dash_len}{1.5mm}
 \fmfset{wiggly_len}{2mm}
 \fmfset{wiggly_slope}{75}
 \fmfset{dot_len}{.8mm}
 \fmfset{dot_size}{1.5thick}
 \begin{center}
 \begin{subfigure}[t]{.3\textwidth} \centering
 \begin{fmfgraph*}(100,80)
 \fmfleft{i}
 \fmfright{o}
 \fmf{plain,tension=5}{i,v1}
 \fmf{plain,tension=5}{v2,o}
 \fmf{plain}{v1,v2}
 \fmf{wiggly,left}{v1,v2}
 \end{fmfgraph*}
 \setlength{\abovecaptionskip}{-20pt}
 \caption{}
 \end{subfigure}
 \begin{subfigure}[t]{.3\textwidth} \centering
 \begin{fmfgraph*}(100,80)
 \fmfleft{i}
 \fmfright{o}
 \fmf{plain,tension=5}{i,v1}
 \fmf{plain,tension=5}{v2,o}
 \fmf{plain}{v1,v2}
 \fmf{dashes,left}{v1,v2}
 \end{fmfgraph*}
 \setlength{\abovecaptionskip}{-20pt}
 \caption{}
 \end{subfigure}
 \begin{subfigure}[t]{.3\textwidth} \centering
 \begin{fmfgraph*}(80,80)
 \fmfleft{i}
 \fmfright{o}
 \fmf{plain}{i,v,o}
 \defotimes
 \fmfv{d.sh=otimes,d.f=empty}{v}
 \end{fmfgraph*}
 \setlength{\abovecaptionskip}{-20pt}
 \caption{}
 \end{subfigure}
 \end{center}
 \end{fmffile}
\caption{Fermion self-energy diagrams.}
\label{fermion-selfenergy-diagrams}
\end{figure}
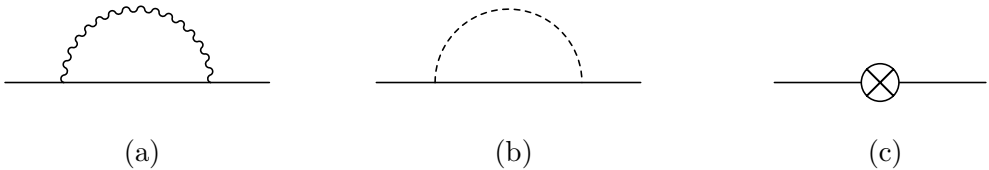
Here we want to demonstrate that in our renormalization scheme the 
fermion self-energies are 
finite, while the explicit formulas are given later in 
section~\ref{self-energies}. 
The corresponding diagrams are displayed in
figure~\ref{fermion-selfenergy-diagrams}.
Since the wave-function renormalization 
matrices can always be chosen such that the divergent terms proportional 
to $\slashed{p}$ cancel, it is sufficient to consider $\Sigma^{(B)}_{L,R}$,
defined in equation~(\ref{sigma1}), of the fermion self-energies. Moreover,
due to equation~(\ref{sigma-h}) we only need to examine $\Sigma^{(B)}_L$.
According to our renormalization scheme, 
diagram~(c) of figure~\ref{fermion-selfenergy-diagrams}
is induced by the renormalization of the Yukawa coupling matrices and the VEVs.
The only genuine mass renormalization we have refers to the renormalization 
of the mass matrix $M_R$. However, as we will see shortly, at one-loop level 
it is not needed.
\subsection{Neutrinos}
\label{neutrinos}
With $\delta v_k$ of equation~(\ref{dv}) and 
$\delta \Delta_k$ of equation~(\ref{dDelta}), 
the part of the counterterm of equation~(\ref{ct-sigma-nu}) that 
is supposed to make 
$-i \left( \Sigma^{(B)}_{\nu L} \gamma_L + \Sigma^{(B)}_{\nu R} \gamma_R \right)$
finite reads
\begin{fmffile}{neutrino-ct-no-p}
	\fmfset{thin}{.7pt}
	\fmfset{dash_len}{1.5mm}
	\fmfset{arrow_len}{2.5mm}
	\begin{align}
	\left( \;
	\begin{gathered}
	\vspace{-4pt}
	\begin{fmfgraph*}(50,20)
	\fmfleft{i}
	\fmfright{o}
	\fmf{plain}{o,v,i}
	\defotimes
	\fmfv{d.sh=otimes,d.f=empty}{v}
	\end{fmfgraph*}
	\end{gathered}
	\; \right)_{\!\! \slashed{p}=0}
	\; = \;
	&
	\frac{ic_\infty}{16 \pi^2} \left\{
	\left( U_R^\dagger \Delta_j M_\ell^\dagger \Gamma_j U_L + 
	U_L^T \Gamma_j^T M_\ell^* \Delta_j^T U_R^*  \right) \gamma_L +
	\right. \nonumber \\
	 & + \left. 
	\left( U_L^\dagger \Gamma_j^\dagger M_\ell \Delta_j^\dagger U_R +
	U_R^T \Delta_j^* M_\ell^T \Gamma_j^* U_L^* \right) \gamma_R \right\}
\nonumber \\
&-i \left[ U_R^\dagger\, \delta M_R\, U_R^*\, \gamma_L + 
U_R^T\, \delta M_R^*\, U_R\, \gamma_R \right].
	\label{cdDelta}
	\end{align}
\end{fmffile}%

Beginning with the $Z$ contribution to the neutrino self-energy, 
it is easy to see that equation~(\ref{d1}) makes it finite. 
Concerning neutral scalar exchange, it also 
does not give a 
divergence because of $V_{kb} V_{lb} = 0$---see equation~(\ref{orthV}).
As for $W^\pm$ exchange, here the chiral projector is the reason that 
both divergent and finite contributions to $\Sigma^{(B)}_{\nu L}$ vanish.
However, the charged-scalar exchange has a non-vanishing divergence, which 
agrees, apart from the sign, with the counterterms in 
the first and second line of equation~(\ref{cdDelta}). Therefore, 
the $\xi$-independent Yukawa counterterm in equation~(\ref{dDelta}) 
indeed cancels the divergence of $\Sigma^{(B)}_{\nu L}$. Consequently, 
the $\delta M_R$ counterterm of equation~(\ref{ct-sigma-nu-c}) must 
be zero because it is not needed. In other words, 
at the one-loop level we find
\begin{equation}
\delta M_R = 0.
\end{equation}
For explicit formulas for the $Z$ and scalar contributions to 
$\Sigma_{\nu L}^{(B)}$ see equation~(\ref{sigma-nu-ell}) and equations thereafter.

\subsection{Charged leptons}
We have to plug $\delta v_k$ of equation~(\ref{dv}) and 
$\delta \Gamma_k$ of equation~(\ref{dGamma}) into the $\slashed{p}$-independent
counterterm of equation~(\ref{ct-sigma-ell}). In this way, 
we obtain the complete counterterm pertaining to  
$-i \left( \Sigma^{(B)}_{\ell L} \gamma_L + \Sigma^{(B)}_{\ell R} \gamma_R \right)$ 
as
\begin{fmffile}{fermion-ct-no-p}
	\fmfset{thin}{.7pt}
	\fmfset{dash_len}{1.5mm}
	\fmfset{arrow_len}{2.5mm}
	\begin{align}
	\left( \;
	\begin{gathered}
	\vspace{-4pt}
	\begin{fmfgraph*}(50,20)
	\fmfleft{i}
	\fmfright{o}
	\fmf{plain}{o,vl,v,vr,i}
	\fmf{fermion,tension=2}{o,vr}
	\fmf{fermion,tension=2}{vl,i}
	\defotimes
	\fmfv{d.sh=otimes,d.f=empty}{v}
	\end{fmfgraph*}
	\end{gathered}
	\; \right)_{\!\! \slashed{p}=0}
	\; = \;
	& \nonumber
	\frac{ic_\infty}{16 \pi^2} \left\{
	\left[ \frac{g^2}{c_w^2} \left( 3 + \xi_Z \right) 
	s_w^2 \left( s_w^2 - \frac{1}{2} \right) 
	+ e^2 \left( 3 + \xi_A \right) \right] \hat m_\ell \right.
	\\
	 & + \left.
	W_R^\dagger \Gamma_j M_D^\dagger \Delta_j W_L \gamma_L + 
	W_L^\dagger \Delta_j^\dagger M_D \Gamma_j^\dagger W_R \gamma_R 
	\vphantom{\frac{g^2}{c_w^2}} \right\}.
	\end{align}
\end{fmffile}%
It is straightforward to check that the $\xi_Z$, $\xi_A$ and $M_D$ terms in this
formula cancel the divergences of the $Z$, photon and charged-lepton loops
in $\Sigma^{(B)}_{\ell L}$. As in the case of neutrinos, the $W^\pm$ loop is zero 
and the sum over the neutral-lepton loops is finite because of the second 
relation in equation~(\ref{orthV}).
Therefore, we find that $\delta v_k$ and $\delta \Gamma_k$ indeed make 
the self-energy of the charged leptons finite.
For explicit formulas for the gauge boson and scalar contributions to 
$\Sigma_{\ell L}^{(B)}$ see equation~(\ref{sigma-nu-ell}) 
and equations thereafter.

\section{One-loop fermion self-energy formulas in Feynman gauge}
\label{one-loop}
In this section, our goal is to present formulas for the fermion 
self-energies that allow to compute the one-loop radiative corrections to
the tree-level masses. Since we have proven that these corrections
are $\xi$-independent, we can resort to a specific gauge. In order to have 
the most simple form of the vector boson 
propagators---see equation~(\ref{vbp2}), we choose the Feynman gauge with
\begin{equation}
\xi_W = \xi_Z = \xi_A = 1
\quad \mbox{and} \quad
M_{+1}^2 = m_W^2, \quad M_1^2 = m_Z^2.
\end{equation}
Consequently, in our formulas for the lepton self-energies, summations 
over scalar contributions always include the Goldstone bosons.
At the end we use the self-energies to investigate radiative 
corrections to the seesaw mechanism.

\subsection{Self-energies}
\label{self-energies}
The mass formula, equation~(\ref{dm1}), 
can be adapted to the type of fermion 
by taking into account equations~(\ref{sigma-h}) and~(\ref{sigma-t}).
In this way we find 
the radiative fermion mass corrections 
\begin{equation}\label{dm-nu}
\Delta m_i = 
m_i \left( \Sigma^{(A)}_{\nu L} \right)_{ii} (m_i^2) +
\mbox{Re} \left( \Sigma^{(B)}_{\nu L} \right)_{ii} (m_i^2) 
\quad (i=1,\ldots,n_L + n_R)
\end{equation}
and
\begin{equation}\label{dm-ell}
\Delta m_\alpha =  
\frac{m_\alpha}{2} \left[ 
\left( \Sigma^{(A)}_{\ell L} \right)_{\alpha\alpha} (m_\alpha^2) + 
\left( \Sigma^{(A)}_{\ell R} \right)_{\alpha\alpha} (m_\alpha^2) \right] + 
\mbox{Re}
\left( \Sigma^{(B)}_{\ell L} \right)_{\alpha\alpha} (m_\alpha^2)
\quad (\alpha = e,\mu,\tau)
\end{equation}
for neutrinos and charged leptons, respectively.

Because we are dealing with Majorana neutrinos, 
according to equation~(\ref{sigma-t}) we only need 
$\Sigma^{(A)}_{\nu L}$ and $\Sigma^{(B)}_{\nu L}$ for $\Sigma_\nu$, while 
for $\Sigma_\ell$, the charged-lepton self-energy, 
$\Sigma^{(A)}_{\ell L}$, $\Sigma^{(A)}_{\ell R}$ and $\Sigma^{(B)}_{\ell L}$ are
required for the determination of the total self-energy. 
The self-energies can be formulated with the two
functions\footnote{A conversion to other widely used
loop functions is given in appendix~\ref{conversion}.}
\begin{equation}
\cf_0(r,s,t) = \int_0^1 \dd x \ln \frac{\Delta(r,s,t)}{\mc^2}
\quad \mbox{and} \quad
\cf_1(r,s,t) = \int_0^1 \dd x\, x\ln \frac{\Delta(r,s,t)}{\mc^2},
\end{equation}
where
\begin{equation}
\Delta(r,s,t) = xr + (1-x)s - x(1-x) t.
\end{equation}
In order to write the $(n_L + n_R) \times (n_L + n_R)$ matrices 
$\Sigma^{(A)}_{\nu L}$ and $\Sigma^{(B)}_{\nu L}$ in a compact way, we define the 
following diagonal matrices:
\begin{subequations}
\begin{eqnarray}
\widehat\cf_{0,a,\nu} &=& \diag 
\left( \cf_0(M^2_{+a},m^2_1, p^2), \ldots,  \cf_0(M^2_{+a},m^2_{n_L+n_R}, p^2) 
\right),
\\
\widehat\cf_{0,b,\nu} &=& \diag 
\left( \cf_0(M^2_b,m^2_1, p^2), \ldots,  \cf_0(M^2_b,m^2_{n_L+n_R}, p^2) \right),
\\
\widehat\cf_{0,W,\nu} &=& \diag 
\left( \cf_0(m_W^2,m^2_1, p^2), \ldots,  \cf_0(m_W^2,m^2_{n_L+n_R}, p^2) \right),
\\
\widehat\cf_{0,Z,\nu} &=& \diag 
\left( \cf_0(m_Z^2,m^2_1, p^2), \ldots,  \cf_0(m_Z^2,m^2_{n_L+n_R}, p^2) \right).
\end{eqnarray}
\end{subequations}
In addition, we need the analogous matrices with $\cf_1$.
For the charged leptons, we have $n_L \times n_L = 3 \times 3$ matrices
\begin{equation}
\widehat\cf_{0,a,\ell} = \diag 
\left( \cf_0(M^2_{+a},m^2_e, p^2), 
\cf_0(M^2_{+a},m^2_\mu, p^2), 
\cf_0(M^2_{+a},m^2_\tau, p^2) \right),
\end{equation}
etc. In the case of charged fermions, photon exchange occurs as well, 
therefore, we also introduce the matrices 
$\widehat\cf_{0,A,\ell}$ and $\widehat\cf_{1,A,\ell}$, which have no analogue in 
the neutrino sector.

In the case of $\Sigma^{(B)}_{\nu L}$ and $\Sigma^{(B)}_{\ell L}$ we distinguish 
between that part stemming from the diagrams of 
figure~\ref{fermion-selfenergy-diagrams} and those from the VEV 
shift---see section~\ref{VEV shift}. The first one we indicate by 
the superscript ``proper'' and the second one by ``shift'':
\begin{equation}\label{sigma-nu-ell}
\Sigma^{(B)}_{\nu L} = 
\Sigma^{(B,\,\mathrm{proper})}_{\nu L} + \Sigma^{(B,\,\mathrm{shift})}_{\nu L}
\quad \mbox{and} \quad
\Sigma^{(B)}_{\ell L} = 
\Sigma^{(B,\,\mathrm{proper})}_{\ell L} + \Sigma^{(B,\,\mathrm{shift})}_{\ell L}.
\end{equation}

For the neutrinos, the result for the self-energy is
\begin{eqnarray}
16 \pi^2\, \Sigma^{(A)}_{\nu L} &=&
L_a^\dagger \widehat\cf_{1,a,\ell} L_a + R_a^T \widehat\cf_{1,a,\ell} R_a^* + 
2 F_b^\dagger \widehat\cf_{1,b,\nu} F_b 
\nonumber \\ && + \frac{g^2}{2} V_L^\dagger \left( \bone + 2 \widehat\cf_{1,W,\ell} 
\right) V_L + 
\frac{g^2}{4c_w^2} \left( U_L^\dagger U_L +  2 U_L^\dagger U_L 
\widehat\cf_{1,Z,\nu} U_L^\dagger U_L \right)\!,
\hphantom{xxx}
\\ \label{sigmaBproper}
16 \pi^2\, \Sigma^{(B,\,\mathrm{proper})}_{\nu L} &=& -
R_a^\dagger \hat m_\ell \widehat\cf_{0,a,\ell} L_a - 
L_a^T \hat m_\ell \widehat\cf_{0,a,\ell} R_a^* + 
2 F_b \hat m_\nu \widehat\cf_{0,b,\nu} F_b 
\nonumber \\ && +
\frac{g^2}{c_w^2} U_L^T U_L^* \hat m_\nu \widehat\cf_{0,Z,\nu} U_L^\dagger U_L.
\end{eqnarray}
The corresponding 
coupling matrices are found in 
sections~\ref{scalar interactions} and~\ref{current interactions}.
In the $W^\pm$ term, the matrix $V_L$ of equation~(\ref{VL}) occurs, 
which will also be relevant for the charged-lepton self-energy.
In the computation of the $Z$ contribution, we have employed 
\begin{equation}
F_{LR}^2 = U_L^\dagger U_L \gamma_L + U_L^T U_L^* \gamma_R 
\end{equation}
and
\begin{equation}
F_{RL} \hat m_\nu F_{LR} = 0
\quad \mbox{with} \quad
F_{RL} = U_L^\dagger U_L \gamma_R - U_L^T U_L^* \gamma_L.
\end{equation}
These relations follow from equation~(\ref{d1}).
The matrix $F_{RL}$ occurs in the $Z$ contribution because of changing 
the position of the Dirac matrices:
$\gamma^\mu F_{LR} = F_{RL} \gamma^\mu$.

Turning to charged leptons, the self-energy is given by
\begin{eqnarray}
16 \pi^2\, \Sigma^{(A)}_{\ell L} &=&
R_a \widehat\cf_{1,a,\nu} R_a^\dagger + 
\frac{1}{2}\,G_b^\dagger \widehat\cf_{1,b,\ell} G_b +
e^2 \left( \bone +  2 \widehat\cf_{1,A,\ell} \right)
\nonumber \\ && + 
\frac{g^2}{2} \left( \bone + 2 V_L \widehat\cf_{1,W,\nu} V_L^\dagger
\right) + 
\frac{g^2}{c_w^2} \left( \bone +  2 \widehat\cf_{1,Z,\ell} \right)
\left( s_w^2 - \frac{1}{2} \right)^2\!,
\\
16 \pi^2\, \Sigma^{(A)}_{\ell R} &=&
L_a \widehat\cf_{1,a,\nu} L_a^\dagger + 
\frac{1}{2}\,G_b \widehat\cf_{1,b,\ell} G_b^\dagger +
e^2 \left( \bone +  2 \widehat\cf_{1,A,\ell} \right)
\nonumber \\ && + 
\frac{g^2}{c_w^2} \left( \bone +  2 \widehat\cf_{1,Z,\ell} \right) s_w^4,
\\
16 \pi^2\, \Sigma^{(B,\,\mathrm{proper})}_{\ell L} &=&
- L_a \hat m_\nu \widehat\cf_{0,a,\nu} R_a^\dagger + 
\frac{1}{2}\,G_b \hat m_\ell \widehat\cf_{0,b,\ell} G_b - 
2 e^2 \hat m_\ell \left( \bone +  2 \widehat\cf_{0,A,\ell} \right)
\nonumber \\ && - 
\frac{2 g^2}{c_w^2} \hat m_\ell \left( \bone +  2 \widehat\cf_{0,Z,\ell} \right)
s_w^2 \left( s_w^2 - \frac{1}{2} \right).
\end{eqnarray}

The parts of the fermion self-energies induced by the VEV shift 
are read off from 
equations~(\ref{nu-shift}) and~(\ref{ell-shift}), respectively:
\begin{equation}
\Sigma^{(B,\,\mathrm{shift})}_{\nu L} = \frac{1}{\sqrt{2}} 
\left( U_R^\dagger \Delta_k U_L + U_L^T \Delta_k^T U_R^* \right) 
\Delta v_k
\quad \mbox{and} \quad
\Sigma^{(B,\,\mathrm{shift})}_{\ell L} = 
\frac{1}{\sqrt{2}} W_R^\dagger \Gamma_k W_L  \Delta v_k^*.
\end{equation}
The shifts $\Delta v_k$ are related to the $\Delta t_b$ 
of equation~(\ref{V-shift}) via equation~(\ref{dvk}), while the 
$\Delta t_b$ are given by the sum over the finite parts of the tadpole 
diagrams---\textit{c.f.}\ equation~(\ref{TCt}).

The individual results obtained from the tadpole diagrams are
\begin{subequations}
\begin{alignat}{2}
&\Delta t_b^{(W)} &&=\; \frac{1}{16 \pi^2} 
\left( v_j^* V_{jb} + V_{jb}^* v_j \right)
\frac{g^2 m_W^2}{4} \left( 1 -  3 \ln \frac{m_W^2}{\mc^2} \right),
\\
&\Delta t_b^{(Z)} &&=\; \frac{1}{16 \pi^2} 
\left( v_j^* V_{jb} + V_{jb}^* v_j \right)
\frac{g^2 m_Z^2}{8 c_w^2} \left( 1 -  3 \ln \frac{m_Z^2}{\mc^2} \right),
\\
&\Delta t_b^{(S^\pm)} &&=\; \frac{1}{16 \pi^2} \lambda_{ijkl}  
\left( v_i^* V_{jb} + V_{ib}^* v_j \right) U_{ka}^* U_{la} M_{+a}^2 
\left( 1 - \ln \frac{M_{+a}^2}{\mc^2} \right),
\\
&\Delta t_b^{(S^0)} &&=\; \frac{1}{16 \pi^2} \left[ \tilde\lambda_{ijkl}  
\left( v_i^* V_{jb} + V_{ib}^* v_j \right) V_{kb'}^* V_{lb'} +
\lambda_{ijkl} \left( v_i^* V_{jb'} V_{kb}^* V_{lb'} + 
V_{ib'}^* v_j V_{kb'}^* V_{lb} \right) \right]
\nonumber \\ &&&\hphantom{=}\;\; \times
\frac{M_{b'}^2}{2} \left( 1 - \ln \frac{M_{b'}^2}{\mc^2} \right),
\\
&\Delta t_b^{(\ell)} &&=\; -\frac{\sqrt{2}}{16 \pi^2} 
\, \mbox{Tr} \left[ {\hat m_\ell}^3 
\left( \bone - \ln \frac{{\hat m_\ell}^2}{\mc^2} \right) 
\left( G_b + G_b^\dagger \right) \right],
\\
\label{dt-chi}
&\Delta t_b^{(\chi)} &&=\; -\frac{\sqrt{2}}{16 \pi^2} 
\, \mbox{Tr} \left[ {\hat m_\nu}^3 
\left( \bone - \ln \frac{{\hat m_\nu}^2}{\mc^2} \right) 
\left( F_b + F_b^\dagger \right) \right],
\end{alignat}
\end{subequations}
where the superscripts indicate the particles in the loop. In the 
$W^\pm$ and $Z$ contributions the respective ghost loops are contained.

\subsection{Seesaw mechanism}
Our computation of the fermion self-energies did not assume anything 
about the scales of the neutrino masses. However, in the seesaw 
mechanism~\cite{minkowski,yanagida,glashow,gell-mann,mohapatra} one stipulates 
that the $M_R$ is non-singular and the eigenvalues of $M_R^\dagger M_R$ 
are of a scale $m_R^2$ such that $m_R$ is much larger than all entries in $M_D$.
With this assumption, there are $n_L$ light neutrinos and $n_R$ 
heavy neutrinos with approximate mass matrices
\begin{equation}
M_\mathrm{light} = - M_D^T M_R^{-1} M_D 
\quad \mbox{and} \quad
M_\mathrm{heavy} = M_R. 
\end{equation}
The matrix 
$\mathcal{U}$, as occurring in equation~(\ref{rgets}), is approximated by 
\begin{equation}\label{Useesaw}
\renewcommand{\arraystretch}{1.2}
\mathcal{U} \simeq \left( 
\begin{array}{cc}
\bone_{n_L} & \left( M_R^{-1} M_D \right)^\dagger \\
-M_R^{-1} M_D & \bone_{n_R}
\end{array} \right)
\left( \begin{array}{cc}
S_\mathrm{light} & 0_{n_L \times n_R} \\
0_{n_R \times n_L} & S_\mathrm{heavy} 
\end{array} \right)
\end{equation}
with
\begin{subequations}
\begin{eqnarray}
S_\mathrm{light}^T M_\mathrm{light} S_\mathrm{light} & \simeq &
\diag \left( m_1, \ldots, m_{n_L} \right),
\\
S_\mathrm{heavy}^T M_\mathrm{heavy} S_\mathrm{heavy} & \simeq &
\diag \left( m_{n_L+1}, \ldots, m_{n_L + n_R} \right).
\end{eqnarray}
\end{subequations}

As demonstrated in~\cite{lavoura}, the dominant radiative corrections to the 
seesaw mechanism reside in the left upper corner of the 
$(n_L + n_R) \times (n_L + n_R)$ Majorana neutrino mass matrix
of equation~(\ref{rgets}):
\begin{equation}
\left( \begin{array}{cc}
\left( M_L \right)_{\mbox{\scriptsize 1-loop}} & M_D^T \\ M_D & M_R 
\end{array} \right).
\end{equation}
Since in the mHDSM gauge symmetry forbids Majorana mass terms of the $\nu_L$,
there is a zero at tree level in the left upper corner 
of this mass matrix and no counterterms are allowed for 
$\left( M_L \right)_{\mbox{\scriptsize 1-loop}}$. Therefore, this radiative 
$n_L \times n_L$ mass matrix must be finite. Moreover, 
the light neutrino mass matrix is modified to 
\begin{equation}
M_\mathrm{light} = \left( M_L \right)_{\mbox{\scriptsize 1-loop}}
- M_D^T M_R^{-1} M_D.
\end{equation}

Examination of the neutrino self-energy and using the seesaw approximation 
of $\mathcal{U}$ of equation~(\ref{Useesaw}) leads to the conclusion that
$\left( M_L \right)_{\mbox{\scriptsize 1-loop}}$ is determined by the 
contributions of the neutral scalars and the $Z$ boson to 
$\Sigma^{(B,\,\mathrm{proper})}_{\nu L}$ of 
equation~(\ref{sigmaBproper})~\cite{lavoura}.
As discussed in section~\ref{neutrinos}, these are indeed finite 
without any renormalization. Moreover, the dominant corrections are 
induced by heavy neutrino exchange and, therefore, we can neglect light
neutrino masses in $\cf_0$. In this way, we can define an effective neutrino 
mass matrix given by 
\begin{eqnarray}
\left( M_L \right)_{\mbox{\scriptsize 1-loop}}
&=& \frac{1}{16 \pi^2} U_L^* \left[
2 F_b\, \hat m_\nu \cf(M_b^2,{\hat m_\nu}^2) F_b + \frac{g^2}{c_w^2} 
U_L^T U_L^*\, \hat m_\nu \cf(m_Z^2,{\hat m_\nu}^2) U_L^\dagger U_L \right] U_L^\dagger
\nonumber \\
&=&
\frac{1}{16 \pi^2} \left[ 
\frac{1}{2}\, \Delta_k^T V_{kb} U_R^* \hat m_\nu \cf(M_b^2,{\hat m_\nu}^2) 
U_R^\dagger \Delta_l V_{lb} + 
\frac{g^2}{c_w^2} U_L^* \hat m_\nu \cf(m_Z^2,{\hat m_\nu}^2) U_L^\dagger \right]
\hphantom{xxx}
\label{ML}
\end{eqnarray}
with 
\begin{equation}\label{Fv1}
\cf(a,b) \equiv \cf_0(a,b,0) = \int_0^1 \dd x \ln 
\frac{\left[ x a + (1 - x) b \right]}{\mc^2} = 
-\ln \mc^2 - 1 + \frac{a \ln a - b \ln b}{a - b}.
\end{equation}
Obviously, the function has the symmetry $\cf(a,b) = \cf(b,a)$.
Since $\left( M_L \right)_{\mbox{\scriptsize 1-loop}}$ must be given in the same 
basis as the mass matrix of equation~(\ref{rgets}), we have performed 
the corresponding basis transformation, given by $U_L^*$ 
to the left and $U_L^\dagger$ to the right in the first line of 
equation~(\ref{ML}).

To proceed further, we convert $\cf$ in the two forms
\begin{equation}\label{Fv2}
\cf(a,b) = -\ln \mc^2 - 1 + \ln a + \frac{\ln \frac{a}{b}}{\frac{a}{b} - 1} =
-\ln \mc^2 - 1 + \ln a + \frac{\frac{b}{a} \ln \frac{b}{a}}{\frac{b}{a} - 1}.
\end{equation}
Next we define the diagonal matrices
\begin{equation}
\hat r_b \equiv \frac{{\hat m_\nu}^2}{M_b^2} 
\quad \mbox{and} \quad
\hat r_Z \equiv \frac{{\hat m_\nu}^2}{m_Z^2}.
\end{equation}
Then we can write the neutral-scalar contribution to equation~(\ref{ML}) as
\begin{equation}\label{cS0}
\frac{1}{2}\, \Delta_k^T V_{kb} U_R^* \hat m_\nu \left[ 
- \left( \ln \mc^2 + 1 \right) \bone + 
\ln {\hat m_\nu}^2 + \frac{\ln \hat r_b}{\hat r_b - \bone}
\right] U_R^\dagger \Delta _l V_{lb}.
\end{equation}
Summing in this equation from $b=1$ to $b=2n_H$, 
the second relation of equation~(\ref{orthV}) tells us that only the 
last term in the square brackets contributes.
Similarly, the $Z$ contribution can be formulated as 
\begin{equation}\label{cZ}
\frac{g^2}{c_w^2} U_L^* \hat m_\nu \left[
- \left( \ln \mc^2 + 1 - \ln m_Z^2 \right) \bone + 
\frac{\hat r_Z \ln \hat r_Z}{\hat r_Z - \bone} \right] U_L^\dagger.
\end{equation}
In this case, because of equation~(\ref{d1}), again only the last term in the
square brackets contributes. Now we consider 
the Goldstone boson contribution to equation~(\ref{cS0})  
separately. 
Since $\hat r_1 = \hat r_Z$, it is suggestive to add it to the $Z$ 
contribution. Indeed, 
using equations~(\ref{GUV}) and~(\ref{URLM}), 
we obtain
\begin{equation}\label{gs}
U_R^\dagger \Delta_k V_{k1} = \frac{i}{v}\,U_R^\dagger \Delta_k v_k =
\frac{i \sqrt{2}}{v}\,U_R^\dagger M_D = 
\frac{i \sqrt{2}}{v}\,\hat m_\nu U_L^\dagger.
\end{equation}
Plugging this into the part with $b=1$ of equation~(\ref{cS0}), we find 
that the $G^0$ contribution differs from the $Z$ contribution solely by 
the numerical factor $-1/4$. Eventually, we arrive at the result~\cite{lavoura}
\begin{equation}
\left( M_L \right)_{\mbox{\scriptsize 1-loop}} = 
\frac{1}{32 \pi^2} \sum_{b=2}^{2n_H} 
\Delta_k^T V_{kb}\, U_R^*\, \hat m_\nu\frac{\ln \hat r_b}{\hat r_b - \bone}
U_R^\dagger\, \Delta_l V_{lb} + 
\frac{3g^2}{64 \pi^2 m_W^2} \,
U_L^*\, {\hat m_\nu}^3 \frac{\ln \hat r_Z}{\hat r_Z - \bone}
U_L^\dagger.
\end{equation}
For $n_H = 1$ it agrees with the result in~\cite{pilaftsis}.
We observe that the Goldstone plus $Z$ contribution to 
$\left( M_L \right)_{\mbox{\scriptsize 1-loop}}$ is 
universal, \textit{i.e.}\ independent of $n_H$.\footnote{We thank A.~Pilaftsis
for drawing our attention to this fact.} 
After employing once more equation~(\ref{URLM}), we find that it is 
of the same order of 
magnitude as that of the physical neutral scalars. 
Moreover, the light neutrino 
masses are completely negligible in these one-loop corrections, which amounts 
to setting
\begin{equation}
U_R^* = \left( 0, S_\mathrm{heavy} \right).
\end{equation}
It has been pointed out 
in~\cite{neufeld,pilaftsis,lavoura,yaguna} 
that numerically these corrections can be sizeable.

\section{Conclusions}
\label{conclusions}
Extensions of the scalar sector play an important role in lepton mass and 
mixing models. However, predictions of such models have mostly been 
computed at tree level and their stability under radiative corrections 
has not been tested.

In this paper we have considered an important class of such models, the mHDSM,
which has an arbitrary number $n_H$ of Higgs doublets and 
an arbitrary number $n_R$ of right-handed neutrino singlets with Majorana 
mass terms. Using the $R_\xi$ gauge for the quantization of the
mHDSM, we have proposed a simple renormalization scheme 
which gives a straightforward recipe for the computation of radiative 
corrections. The idea is that all
counterterms are induced by the parameters of the unbroken theory, with the
exception of the VEV renormalization $\delta v_k$ ($k = 1,\ldots,n_H$).
Since all masses in the mHDSM are obtained by spontaneous gauge-symmetry
breaking, masses are derived quantities and there is no mass renormalization in
our scheme. The removal of the corresponding divergencies is procured 
by the renormalization of VEVs and Yukawa couplings.

In addition to the infinite counterterm parameters $\delta v_k$,
there are the well-known finite VEV shifts $\Delta v_k$ 
induced by the finite parts of tadpole diagrams, which guarantee that
beyond tree level the VEVs of the physical neutral scalars are still
vanishing. 

We have demonstrated, by determination of all counterterm
parameters, that at the one-loop level our renormalization 
scheme is capable of removing all divergences and we have elucidated 
the prescription for computing the VEV shifts $\Delta v_k$. 
Moreover, we have shown analytically for the one-loop fermion self-energies
that including the VEV shifts is equivalent to the insertion of all 
tadpole diagrams on the fermion line.

As an application of the renormalization scheme we have presented the full
fermion mass corrections at the one-loop level. In this context,
we have identified all the mechanisms and performed analytically the necessary 
computations to show that these corrections are $\xi$-independent,
in both the loop diagrams and the counterterms. In the case of loop diagrams
we have closely followed ref.~\cite{weinberg}. 
We have also demonstrated
that in the seesaw limit the radiative corrections to the seesaw mechanism
computed in ref.~\cite{lavoura} derive from our much more general framework.

We conclude with a speculation about the importance of tadpole contributions 
to one-loop fermion masses. Tadpoles induce VEV shifts $\Delta v_k$ via
equation~(\ref{dvk}). The potentially largest contribution 
comes from heavy neutrinos in the tadpole loop, \textit{i.e.}\ from 
$\Delta t_b^{(\chi)}$, equation~(\ref{dt-chi}), because this quantity roughly
scales with the third power in the seesaw scale $m_R$. 
%%%%
A very crude estimate 
suggests that a scale $m_R \gg 10^3$\,TeV induces huge VEV shifts, much larger 
than the electroweak scale, while for 
$m_R \lesssim 10^3$\,TeV
the induced VEV shifts are below 10\,GeV. Whether 
this apparent incompatibility of a generic mHDSM (or the SM), \textit{i.e.}\
without any suppression mechanism for $\Delta t_b^{(\chi)}$, with 
large seesaw scales is physical or an artefact of our renormalization scheme 
remains to be investigated.

\section*{Acknowledgments}
M.~L. is supported by the Austrian Science Fund (FWF), Project
No.\ P28085-N27 and in part by the FWF Doctoral Program No.\ W1252-N27 
Particles and Interactions. 
The authors are very grateful to F.~Jegerlehner and H.~Neufeld for 
many stimulating discussions and support. They are especially 
indebted M.~Sperling for clarifying the role of VEV renormalization 
in the $R_\xi$ gauge. Moreover, M.~L.\ thanks D.\ Lechner and C.\ Lepenik 
for further helpful discussions.

\newpage

\appendix
\section{The scalar mass matrices}
\label{app-MM}
\setcounter{equation}{0}
\renewcommand{\theequation}{A.\arabic{equation}}
In this section we discuss the tree-level scalar masses.
The scalar potential is given by
\begin{equation}\label{V}
V(\phi) = \mu^2_{ij} \phi_i^\dagger \phi_j +
\lambda_{ijkl} \phi_i^\dagger \phi_j \phi_k^\dagger \phi_l
\end{equation}
with
\begin{equation}\label{indices}
\mu^2_{ij} = \left( \mu^2_{ji} \right)^*, \quad
\lambda_{ijkl} = \lambda_{klij} = \lambda_{jilk}^*.
\end{equation}
We allow for an arbitrary number $n_H$ of scalar doublets $\phi_k$
($k=1,2, \ldots, n_H$)
and use the notation
\begin{equation}\label{VEV}
\phi_k = \left( \begin{array}{c}
\varphi_k^+ \\ \varphi_k^0 \end{array} \right),
\quad \mathrm{with} \quad
\left\langle 0 \left| \varphi_k^0 \right| 0 \right\rangle
= \frac{v_k}{\sqrt{2}}\, .
\end{equation}
We then write
\begin{equation}
\varphi^0_k = \frac{1}{\sqrt{2}} \left( v_k + \rho_k + i\sigma_k \right),
\quad \mbox{hence} \quad
\left\langle 0 \left| \rho_k \right| 0 \right\rangle = 
\left\langle 0 \left| \sigma_k \right| 0 \right\rangle = 0,
\quad
\rho_k^\dagger = \rho_k, \quad \sigma_k^\dagger = \sigma_k.
\end{equation}
The quadratic terms in the scalar potential are written as
\begin{equation}
V_\mathrm{mass} = \sum_{i,j} 
\varphi^-_i \left( \mathcal{M}^2_+ \right)_{ij}\, \varphi^+_j
+ \frac{1}{2} \left[
A_{ij}\, \rho_i \rho_j 
+ B_{ij}\, \sigma_i \sigma_j 
+ 2 C_{ij}\, \rho_i \sigma_j 
\right].
\label{ABC}
\end{equation}
The mass matrix of the charged scalars~\cite{grimus},
\begin{equation}
\mathcal{M}^2_+ = \mu^2 + \Lambda\, ,
\quad \mbox{where} \quad
\Lambda_{ij} = \sum_{k,l} \lambda_{ijkl}\, v_k^* v_l\,,
\label{M+}
\end{equation}
is complex and hermitian, with $\Lambda$ being hermitian as well.
The matrices $A$ and $B$ are real and symmetric;
$C$ is real but otherwise arbitrary.
All matrices defined so far are $n_H \times n_H$ matrices.

The mass matrix of the neutral real scalar fields $\rho_i$ and $\sigma_j$ 
may then be formulated as the real $2n_H \times 2n_H$ matrix
\begin{equation}\label{M02}
\mathcal{M}^2_0 = 
\left( \begin{array}{cc} A & C \\ C^T & B \end{array} \right).
\end{equation}
Defining complex $n_H \times n_H$ matrices
\begin{equation}
K_{ik} = \sum_{j,l} \lambda_{ijkl}\, v_j v_l
\quad \mbox{and} \quad
K'_{il} = \sum_{j,k} \lambda_{ijkl}\, v_j v_k^*,
\end{equation}
where $K$ is symmetric and $K'$ is hermitian.
The matrices $A$, $B$, $C$ are obtained as~\cite{grimus} 
\begin{subequations}\label{ABCK}
\begin{eqnarray}
A &=& \mbox{Re} \left( \mu^2 + \Lambda + K^\prime \right)
+ \mbox{Re}\, K\, , \\
B &=& \mbox{Re} \left( \mu^2 + \Lambda + K^\prime \right)
- \mbox{Re}\, K\, , \\
C &=& - \mbox{Im} \left( \mu^2 + \Lambda + K^\prime \right)
+ \mbox{Im}\, K\, ,
\end{eqnarray}
\end{subequations}
respectively.

\section{The diagonalization matrices of the charged and neutral 
scalar mass terms} 
\label{app-UV}
\setcounter{equation}{0}
\renewcommand{\theequation}{B.\arabic{equation}}
Let ${\hat M}_+^2$ be the diagonal matrix of the squares
of charged scalar masses.
Furthermore, we denote the unitary $n_H \times n_H$ matrix
which diagonalizes $\mathcal{M}^2_+$ 
by $U$. Thus we have
\begin{equation}
U^\dagger \mathcal{M}^2_+ U = {\hat M}_+^2.
\end{equation}
Inverting this relation, we obtain
\begin{equation}\label{UMU}
U {\hat M}_+^2 U^\dagger = \mu^2 + \Lambda.
\end{equation}

Let ${\hat M}_0^2$ be the diagonal matrix of the squares
of the neutral scalar masses and $\tilde V$ the orthogonal 
$2n_H \times 2n_H$ matrix $\tilde V$ that diagonalizes the mass matrix of 
the neutral scalars, \textit{i.e.}
\begin{equation}\label{diag0}
{\tilde V}^T \mathcal{M}^2_0 \tilde V = {\hat M}_0^2.
\end{equation}
Without loss of generality we can write 
\begin{equation}\label{tildeV}
\tilde V = \left(
\begin{array}{c}
\mathrm{Re}\,V \\ \mathrm{Im}\,V 
\end{array} \right),
\end{equation}
where $V$ is a complex $n_H \times 2n_H$ matrix. 
In terms of $V$, orthogonality of $\tilde V$ reads~\cite{lavoura,osland,bento}
\begin{equation}\label{orthV}
V_{jb}^* V_{kb} = 2 \delta_{jk}, 
\quad
V_{jb} V_{kb} = 0,
\quad
V_{jb}^* V_{jb'} + V_{jb'}^* V_{jb} = 2\delta_{bb'}. 
\end{equation}
Notice that 
\begin{equation}
V_{jb}^* V_{jb'} = \delta_{bb'} + i \mathrm{Im} \left( V^\dagger V \right)_{bb'}
\end{equation}
is in general not diagonal, but it is easy to see that
$\mathrm{Im} \left( V^\dagger V \right)_{bb'}$ is antisymmetric 
in the indices $b,b'$~\cite{bento}.
A further relation, useful in one-loop computations, is
\begin{equation}\label{VIm}
V_{kb'}\,\mathrm{Im} \left( V^\dagger V \right)_{b'b} = -i V_{kb}.
\end{equation}

The matrices $U$ and $V$ allow to write the Higgs doublets $\phi_k$ 
in terms of the mass eigenfields $S^+_a$ and $S^0_b$: 
\begin{equation}\label{scalar-mass-eigenfields}
\phi_k = \left(
\begin{array}{c}
U_{ka} S^+_a \\ \frac{1}{\sqrt{2}} \left( v_k + V_{kb} S^0_b \right)
\end{array} \right).
\end{equation}

There is no straightforward analogue to equation~(\ref{UMU}).
But rewriting equation~(\ref{diag0}) as
\begin{equation}
\mathcal{M}^2_0 = 
\tilde V {\hat M}_0^2 {\tilde V}^T = 
\left( \begin{array}{cc}
\mathrm{Re}\,V {\hat M}_0^2\, \mathrm{Re}\,V^T & 
\mathrm{Re}\,V {\hat M}_0^2\, \mathrm{Im}\,V^T \\
\mathrm{Im}\,V {\hat M}_0^2\, \mathrm{Re}\,V^T & 
\mathrm{Im}\,V {\hat M}_0^2\, \mathrm{Im}\,V^T 
\end{array} \right) 
\end{equation}
and using subsequently equation~(\ref{M02}) leads to
\begin{subequations}
\begin{eqnarray}
V {\hat M}_0^2 V^T &=& 
A - B + iC + iC^T,
\\
V {\hat M}_0^2 V^\dagger &=&
A + B - iC + iC^T.
\end{eqnarray}
\end{subequations}
Eventually, application of equation~(\ref{ABCK}) gives the useful
results~\cite{bento} 
\begin{subequations}\label{VMV}
\begin{eqnarray}
V \hat{M}_0^2 V^T &=& 2 K, \\
V \hat{M}_0^2 V^\dagger &=& 2 \left( \mu^2 + \Lambda + K' \right).
\end{eqnarray}
\end{subequations}

We denote the masses of the charged and neutral scalars by 
$M_{+a}$ ($a = 1, \ldots, n_H$) and $M_b$ ($b = 1, \ldots, 2n_H$), 
respectively. Obviously, the $a$-th column of $U$ is an eigenvector of  
$\mathcal{M}^2_+$ with eigenvalue $M_{+a}^2$. 
By definition, the $b$-th column of $\tilde V$ 
is an eigenvector of $\mathcal{M}^2_0$ with eigenvalue $M_b^2$, which  
in terms of the columns of $V$ reads~\cite{grimus}
\begin{equation}\label{columns}
\left( \mu^2 + \Lambda +  K' \right)_{ij} V_{jb} + K_{ij} V_{jb}^* = 
M_b^2 V_{ib}.
\end{equation}
Taking into account equation~(\ref{orthV}), this equation can be cast
into the very useful form
\begin{equation}\label{useful form}
\frac{1}{2} \left( \mu^2_{ij} + \tilde\lambda_{ijkl} v_k^* v_l \right)
\left( V_{ib}^* V_{jb'} + V_{ib'}^* V_{jb} \right) +
\frac{1}{2} \lambda_{ijkl} \left( v_i^* v_k^* V_{jb} V_{lb'} + 
v_j v_l V_{ib}^* V_{kb'}^* \right) = \delta_{bb'} M_b^2.
\end{equation}

The mass matrices $\mathcal{M}^2_+$ and $\mathcal{M}^2_0$ also contain 
one eigenvalue zero each, referring to the Goldstone bosons.
Allocating the indices $a=1$ and $b=1$ to the charged and the neutral 
Goldstone boson, respectively, 
the corresponding columns in $U$ and $V$ are given by~\cite{grimus}
\begin{equation}\label{GUV}
U_{k1}  = \frac{v_k}{v} 
\quad \mbox{and} \quad
V_{k1} = i\frac{v_k}{v},
\end{equation}
respectively.

\section{On-shell contributions to the fermion self-energies}
\label{app-on-shell}
\setcounter{equation}{0}
\renewcommand{\theequation}{C.\arabic{equation}}
Here we confine ourselves to one fermion field. Suppose a contribution 
$\sigma(p)$ to the total (renormalized) self-energy $\Sigma(p)$ is given by
\begin{equation}
\sigma(p) = \slashed{p} \left( a_L \gamma_L + a_R \gamma_R \right)
- b_L \gamma_L - b_R \gamma_R.
\end{equation}
What is the condition that $\sigma(p)$ vanishes on-shell?
With this we mean that it has the form
\begin{equation}\label{sigma}
\sigma(p) = 
\left( c_L \gamma_L + c_R \gamma_R \right) \left( \slashed{p} - m \right) + 
\left( \slashed{p} - m \right) \left( d_L \gamma_L + d_R \gamma_R \right).
\end{equation}
It is easy to prove that for $m \neq 0$ this is the case if and only if
\begin{equation}\label{cond-on-shell}
\frac{1}{m} \left( b_L + b_R \right) = a_L + a_R.
\end{equation}
Notice, however, that the coefficients $c_{L,R}$ and $d_{L,R}$ are not
uniquely determined by $a_{L,R}$ and $b_{L,R}$, because the shift 
$c'_{L,R} = c_{L,R} + s$ and $d'_{L,R} = d_{L,R} - s$
with an arbitrary (real) $s$ does not change $\sigma(p)$ of 
equation~(\ref{sigma}).

\section{Conversion to scalar Feynman integrals}
\label{conversion}
The loop functions $\cf_0$ and $\cf_1$ of section~\ref{one-loop} 
can easily be converted into the well-known Feynman integral representations 
$B_0$ and $B_1$ as defined in~\cite{Passarino:1978jh,Bohm:2001yx} 
and numerically evaluated using freely available computer algebra software
such as \textit{LoopTools}~\cite{Hahn:1998yk}. 
The conversion reads
\begin{subequations}
 \begin{eqnarray}
 \cf_0(m_1^2, m_2^2, p^2) &=& c_\infty - B_0(p^2; m_1, m_2),\\
 \cf_1(m_1^2, m_2^2, p^2) &=& \frac{c_\infty}{2} 
 - \left[ B_0(p^2; m_1, m_2) + B_1(p^2; m_1, m_2)\right],
\end{eqnarray}
\end{subequations}
where $B_1(p^2; m_1, m_2)$ is related to the vector two-point
integral $B^\mu(p; m_1, m_2)$ via
\begin{equation}
 B_1(p^2; m_1, m_2) = \frac{1}{p^2}\, p_\mu B^\mu(p; m_1, m_2), \quad p^2 \neq 0.
\end{equation}
The loop-integrals referred to are
\begin{subequations}
 \begin{eqnarray}
 B_0(p^2; m_1, m_2) &=& \frac{16\pi^2 }{i} \, \mathcal{M}^{4-d}\!\!
 \int \! \frac{\mathrm{d}^d k}{(2\pi)^d}
 \frac{1}{\left[ k^2 - m_1^2 \right] 
 \left[ \left(p + k\right)^2 - m_2^2\right]}, \\
 B^\mu(p; m_1, m_2) &=& \frac{16\pi^2 }{i} \, \mathcal{M}^{4-d}\!\! 
 \int \! \frac{\mathrm{d}^d k}{(2\pi)^d}
 \frac{k^\mu}{\left[ k^2 - m_1^2 \right] 
 \left[ \left(p + k\right)^2 - m_2^2\right]}.
 \end{eqnarray}
\end{subequations}

\end{document}